\documentclass[aps,prd,twocolumn,superscriptaddress,nofootinbib,10pt]{revtex4-1}

\usepackage[utf8]{inputenc}
\usepackage{amssymb}
\usepackage{amsmath}
\usepackage{amsfonts}
\usepackage{graphicx}
\usepackage{color}
\usepackage{xspace}
\usepackage{comment}
\usepackage{hyperref}
\usepackage[normalem]{ulem}
\usepackage[section]{placeins}
\usepackage{afterpage}

\usepackage{float}
\usepackage{slashed}
\usepackage{appendix}

\usepackage{multirow,rotating}
\usepackage[dvipsnames]{xcolor}

\newcommand{\GeV}{\,{\mathrm{GeV}}}
\newcommand{\TeV}{\,{\mathrm{TeV}}}

\begin{document}
	
    \title{Layered cavern surface tracker at future electron-positron colliders}
	
	\author{Ye Lu}
	\email{ye.lu@whut.edu.cn (Corresponding author)}
	\affiliation{Department of Physics, School of Physics and Mechanics, Wuhan University of Technology,\\ 430070 Wuhan, Hubei, China}
	
	\author{Ying-nan Mao}
	\email{ynmao@whut.edu.cn (Corresponding author)}
	\affiliation{Department of Physics, School of Physics and Mechanics, Wuhan University of Technology,\\ 430070 Wuhan, Hubei, China}

	\author{Kechen Wang}
	\email{kechen.wang@whut.edu.cn (Corresponding author)}
	\affiliation{Department of Physics, School of Physics and Mechanics, Wuhan University of Technology,\\ 430070 Wuhan, Hubei, China}
	
	\author{Zeren Simon Wang}
	\email{wzs@hfut.edu.cn (Corresponding author)}
        \affiliation{School of Physics, Hefei University of Technology, Hefei 230601, China}	
	
	\begin{abstract}       
            We propose a detector concept, LAYered CAvern Surface Tracker (LAYCAST), to be installed on the ceiling and the wall of the cavern hosting the main experiment of future electron-positron colliders such as CEPC and FCC-ee.
            With detailed and realistic considerations of the design of such a new experiment, the proposed detector is dedicated to extending the sensitivity reach of the main detector to various theoretical scenarios of long-lived particles (LLPs).
            We study carefully four such 
            scenarios involving a light scalar boson $X$, the heavy neutral lepton $N$, the lightest neutralino $\tilde{\chi}^0_1$ in the R-parity-violating supersymmetry, and the axion-like particle $a$. 
            Long-lived light scalar bosons are considered to be produced from the Standard-Model (SM) Higgs boson's decay ($h \to X X$) at the center-of-mass energy $\sqrt{s} =$ 240 GeV, while the other three types of LLPs are produced either from $Z$-boson decays (viz. $Z \to \nu\, N, ~\tilde{\chi}^0_1\, \tilde{\chi}^0_1 $) or direct scattering process ($ e^- e^+ \to ~\gamma\, a$) at $\sqrt{s} =$ 91.2 GeV, where $\gamma$ and $\nu$ denote the SM photon and neutrino, respectively.
            With Monte-Carlo simulations, we derive the sensitivities of the proposed experiment to these LLPs and the corresponding signal-event numbers.
            We also provide a dedicated estimate of a potentially important SM background from long-lived neutral kaons in hadronic $Z$ decays, and show that it is strongly suppressed by the combined requirements of the main detector and LAYCAST.
            Our findings show that LAYCAST can probe large new parameter space beyond both current bounds and the expected reach of the main experiments at CEPC and FCC-ee.
            Comparison with existing works in similar directions is also made.
 	\end{abstract}


	\vskip10mm
	
	\maketitle
	\flushbottom

\section{Introduction}
\label{sec:intro}

In various theories beyond the Standard Model (BSM), long-lived particles (LLPs) arise naturally for different reasons such as feeble couplings to Standard-Model (SM) particles and a heavy particle mediating their decays.
They are often proposed as solutions to certain issues in the SM such as dark matter and non-vanishing neutrino masses, and are thus well motivated.
Examples include heavy neutral leptons (HNLs)~\cite{Shrock:1980vy,Shrock:1980ct,Shrock:1981wq} (see also the recent review~\cite{Abdullahi:2022jlv}), dark photons~\cite{Okun:1982xi,Galison:1983pa,Holdom:1985ag,Boehm:2003hm,Pospelov:2008zw}, and dark Higgs bosons~\cite{OConnell:2006rsp,Wells:2008xg,Bird:2004ts,Pospelov:2007mp,Krnjaic:2015mbs,Boiarska:2019jym}.
LLPs can be searched for at various terrestrial facilities such as beam-dump experiments, $B$-factories, high-energy lepton or hadron colliders, and neutrino experiments.
Given their relatively long lifetime, signatures of displaced objects are the main targets of searches for LLPs, including displaced vertex (DV), displaced leptons, and displaced jets.
In particular, the currently leading collider machine, LHC, has conducted multiple LLP searches at its main experiments such as ATLAS~\cite{ATLAS:2022rme,ATLAS:2017tny,ATLAS:2020xyo,ATLAS:2020wjh,ATLAS:2022vhr} and CMS~\cite{CMS:2023mny,CMS:2024trg}.

Furthermore, a series of auxiliary far-detector (FD) experiments have been proposed to be operated in the vicinity of the different interaction points (IPs) at the LHC, dedicated mainly to searches for LLPs.
They are supposed to be external detectors with a distance of about $5-500$ meters away from their corresponding IP in the forward or transverse directions.
They are usually equipped with simple tracking technologies, targeting DVs formed by LLPs that may decay inside their fiducial volumes.
These concepts include FASER~\cite{Feng:2017uoz,FASER:2019aik,FASER:2018eoc} and FASER2~\cite{Feng:2022inv}, MATHUSLA~\cite{Chou:2016lxi,Curtin:2018mvb,MATHUSLA:2020uve}, MoEDAL-MAPP1 and MAPP2~\cite{Pinfold:2019zwp,Pinfold:2019nqj}, ANUBIS~\cite{ANUBIS:2019qdc}, as well as CODEX-b~\cite{Gligorov:2017nwh,Aielli:2019ivi}.
Additional dedicated detector concepts in complementary configurations include milliQan~\cite{milliQan:2018yeo} and the Forward Physics Facility (FPF) program and its proposed FORMOSA detector~\cite{FPF:2022fxm,FORMOSA:2021gll}.
In particular, FASER has been constructed and is currently under operation, and the collaboration has published the experiment's first results excluding certain new parameter space of the dark photons~\cite{FASER:2023tle}.
Phenomenological studies performed in e.g.~Refs.~\cite{Helo:2018qej,Dercks:2018eua,Dercks:2018wum,Hirsch:2020klk,Dreiner:2020qbi,Li:2023dbs,Lu:2023cet,Kling:2018wct,Dienes:2023uve,Kling:2022uzy,Aielli:2019ivi,Deppisch:2023sga,Frank:2019pgk,Sen:2021fha,Jodlowski:2019ycu,No:2019gvl,Berlin:2018jbm}, have shown that these relatively cheap external detectors can probe large parts of the parameter space of a number of LLP models that would, however, be inaccessible by the LHC main experiments.
We refer to Refs.~\cite{Antel:2023hkf,Alimena:2019zri,Lee:2018pag,Curtin:2018mvb,Beacham:2019nyx} for reviews on LLP theories and searches.

After the completion of the LHC programs in the mid-2030s, next-generation colliders are planned including Circular Electron Positron Collider (CEPC)~\cite{CEPCStudyGroup:2018rmc,CEPCAcceleratorStudyGroup:2019myu,CEPCStudyGroup:2023quu} in China and Future Circular Collider-ee (FCC-ee)~\cite{FCC:2018evy} at CERN.
Running as $Z$-boson or Higgs-boson factories, these experiments allow unprecedented electroweak and Higgs precision measurements~\cite{CEPCStudyGroup:2018ghi,An:2018dwb,FCC:2018byv,An:2018dwb,Grojean:2022blp,Guo:2022wti,Ruan:2014xxa,Ge:2016tmm}.
Moreover, given the comparatively clean environment at the electron-positron colliders, lower levels of background events and stronger sensitivities for searches for hidden particles including LLPs are expected.
Such searches have been conducted in some past lepton-collider experiments such as DELPHI~\cite{DELPHI:1996qcc}.
Also, phenomenological studies have been performed for the CEPC and FCC-ee main experiments, estimating their expected sensitivity reach to multiple LLP scenarios~\cite{Cheung:2019qdr,Wang:2019orr,Alipour-Fard:2018lsf,Cao:2023smj,Suarez:2021hpn,Antusch:2015mia,Zhang:2024bld,Blondel:2022qqo,Urquia-Calderon:2023dkf,Ovchynnikov:2023wgg,Chrzaszcz:2021nuk,Bauer:2018uxu}.

Similar to the LHC, for CEPC and FCC-ee, studies also exist for evaluating the physics potential of external detectors installed far from the IP~\cite{Wang:2019xvx,Chrzaszcz:2020emg,Tian:2022rsi, Schafer:2022shi}.
Ref.~\cite{Wang:2019xvx} (see also Ref.~\cite{Tian:2022rsi}) considered a list of far-detector concepts and estimated their sensitivities to several LLP benchmark scenarios.
These experiments would all be operated with a macroscopic distance away from the IP, thus allowing to make use of the space in between for purposes of shielding background events and thus achieving vanishing background levels in general.
The LLPs should travel towards a far detector and decay inside its fiducial volume.
In comparison, Ref.~\cite{Chrzaszcz:2020emg} proposed a detector called HErmetic CAvern TrackEr (HECATE) which consist of resistive plate chambers (RPCs) or scintillator plates, constructed from extruded scintillating bars.
The detector would be installed on the cavern wall, ceiling, as well as floor, of the CEPC/FCC-ee main experiment's hall and thus enclose the hall with a $4\pi$ solid-angle coverage.
LLPs are supposed to decay between the main detector (MD) and the inner surface of experimental hall and their decay products are detected by such new detector.
The effective decay volume in this case is the space inside the cavern excluding that occupied by the MD of the collider.
The authors consider the cavern as a spherical hermetic detector in numerical computation, and 
use an HNL that mixes exclusively with the muon neutrino as a benchmark model to showcase the proposal's potential in probing LLPs.
All these proposed auxiliary detectors are found to be sensitive to parameter regions in various LLP models that neither the LHC's main and far detectors, nor the main detectors of electron-positron colliders can probe.

Inspired by the previous proof-of-principle study~\cite{Chrzaszcz:2020emg}, in this work, 
we propose a similar tracker detector to be installed on the wall and ceiling of the main cavern 
at future electron-positron colliders such as CEPC and FCC-ee.
The fiducial volume is taken to be the space between the MD and the cavern's surface, similar to that of HECATE. 
However, the difference is that we consider a more refined geometry and integration assumptions, tailored to realistic cavern constraints.
Firstly, the tracker is not installed on the floor of the cavern because there will be load-bearing components for the MD and other facilities on the floor.
Therefore, the detector is not hermetic, and LLPs can escape the cavern through the floor.
We, thus, name the proposal as LAYered CAvern Surface Tracker (LAYCAST).
Besides, we also take into account other specific and realistic factors, e.g.~the actual planned geometries of the MD and the caverns.\footnote{
The detector materials are supposed to be hanging from the walls and the roof of the cavern, where elements including fire fighting protocols, cranes, cabling, and platforms are planned to be implemented. The detector materials should avoid these elements.
Similarly, the entrance to the cavern should be left clear of detector materials.
All these realistic factors will imply somewhat reduced coverage and lead to weaker sensitivities.
In this study, we do not take this into account and assume that this should affect the results to a minor extent.
}
We aim to achieve more realistic considerations of the new detector's fiducial volume and acceptance, and hence more accurate sensitivities to various LLPs.

We perform Monte-Carlo (MC) simulations to evaluate sensitivity reaches of the proposed detector to four LLP benchmark models: a light scalar from exotic Higgs decays, the HNL that mixes with the electron neutrino only, the lightest neutralino in the R-parity-violating (RPV) supersymmetry (SUSY), as well as axion-like particles that couple to photons and $Z$-bosons.
We also compare our estimated sensitivity results with existing bounds on the models, and sensitivity reaches of proposed LHC far detectors, or those of lepton-collider's main and proposed far detectors.

We note that in principle the same idea could also be implemented at future linear lepton colliders including International Linear Collider (ILC)~\cite{ILC:2007bjz,Fujii:2017vwa} (see also Refs.~\cite{Sakaki:2020mqb,Asai:2021xtg,Asai:2021xtg,Nojiri:2022xqn,Asai:2023dzs} for LLP studies at a proposed ILC beam-dump experiment) and Compact Linear Collider (CLIC)~\cite{CLIC:2018fvx}.
Geometries of the detector need to be adjusted according to the cavern sizes of the ILC and CLIC.

We organize the paper as follows.
In Sec.~\ref{sec:detector-setup} we elaborate on the experimental setup of LAYCAST at the CEPC/FCC-ee cavern.
We then introduce the theoretical models in Sec.~\ref{sec:model} for which we will perform sensitivity studies.
In Sec.~\ref{sec:simulation}, we detail the MC simulation procedures for the signal events and explain how we compute the estimated signal-event numbers.
The numerical results are presented and discussed in Sec.~\ref{sec:results}.
We conclude the paper in Sec.~\ref{sec:conclusions} with a summary.
In addition, in Appendix~\ref{appendix:cavern_geometries} we show the effect of the main cavern's size by considering the past and updated designs at the CEPC.
Appendix~\ref{appendix:IP} shows an analysis of the effect of IP's position relative to the cavern floor (corresponding to a support for the MD, of different heights).
Furthermore, Appendix~\ref{appendix:underside} illustrates briefly the importance of the floor of the cuboid fiducial volume on the expected sensitivities.
Appendix~\ref{app:KL_background} presents a dedicated estimate of the SM $K_L$ background and illustrates how the combined MD$\times$FD requirements suppress it.
Finally, Appendix~\ref{appendix:NS_main_detector} includes a discussion on the impact of the background level on the discovery sensitivities.

\section{Experimental setup}
\label{sec:detector-setup}

\begin{figure}[t]
	\centering
	\includegraphics[width=\columnwidth]{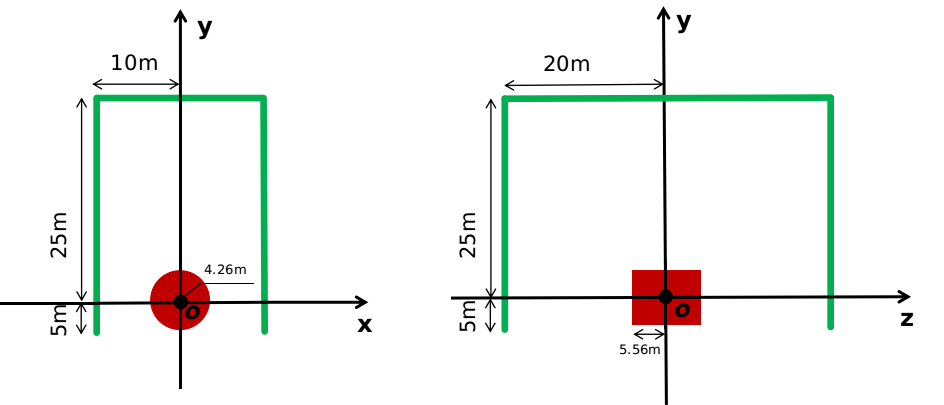}
	\caption{The front and side cross-section views of the proposed experiment.}
	\label{fig:FD902}
\end{figure}

We show schematic diagrams of LAYCAST in Fig.~\ref{fig:FD902}, where the origin of the coordinates corresponds to the IP; the $z$-axis is in parallel with the beam axis with the positive direction being the moving direction of the incident electron beam; the $y$-axis is the vertical direction with the positive side pointing upwards; and the $x$-axis lies in the horizontal plane and is perpendicular to the beam axis.
The left and right plots show section views of the experimental setup in the $xOy$ and $yOz$ planes, respectively.
In the plots, the red cylinder enclosing the coordinate origin represents the MD of CEPC/FCC-ee.
The green area depicts the proposed new detector, LAYCAST, in the shape of a thin layer to be instrumented on the cavern surface.
The shape of the experimental hall is simplified into a cuboid.
Considering that the floor of the experiment hall cannot be installed owing to load bearing and other reasons, the LAYCAST would be mounted on the roof surface and four vertical walls of the experimental hall. 
The LLPs produced at the IP, if decaying inside the MD, can potentially be observed therein via the decay products.
If they traverse the MD and decay before reaching the cavern's inner surface, they may be detected by the LAYCAST experiment.

For quantitative evaluation, we consider the shape of the cavern to be simply a cuboid 
with dimensions (L$\times$W$\times$H) of $40\text{ m}\times 20\text{ m}\times 30\text{ m}$~\cite{CEPC_cavern_old_slides}~\footnote{The surfaces of cavern might be curved in reality, which would change the fiducial volume.
However, with similar distances from the IP, the change in the fiducial volume between a curved surface and a flat one is expected to be small. 
The cuboid setup then provides a valid approximation in such case.},
excluding the MD's volume.
The MD is cylindrical and has a radius of 4.26 m and a half-length of 5.56 m, following the latest design of the CEPC detector~\cite{CEPC_detector_slides}.
For various reasons including the load bearing, the collision point and thus (the center of) the MD are roughly 5 m or above the floor, held by a support~\footnote{In the latest design of the CEPC collider~\cite{CEPCStudyGroup:2023quu}, the main cavern is planned to have the dimensions $50\text{ m}\times 30\text{ m}\times 30\text{ m}$, with the beam-pipe channel above the cavern floor by 12.15 m. We numerically verified that these changes relative to the setup considered here would lead to only almost negligible improvement on the sensitivity results presented in this work (see Appendix~\ref{appendix:cavern_geometries}).
For the FCC-ee infrastructure, the experiment caverns are supposed to have dimensions $66\text{ m}\times 35\text{ m}\times 35\text{ m}$~\cite{FCC:2018evy}; we also expect only rather minor improvement for the results in this work with this geometrical setup.\label{footnote:cavern_geometry}}.
Since the layered tracker is not installed on the floor of the cuboid fiducial volume, 
LLP decay products that reach the roof surface and the four vertical walls of the experimental hall are taken into account.
In order to quantify the effect of both the relative position of the IP from the cavern floor and omission of the floor on the detector sensitivities, we study numerically and present the results and discussion in Appendix~\ref{appendix:IP} and Appendix~\ref{appendix:underside}, respectively.
In general, we find these effects rather minor.

We note that the detector configurations investigated in Ref.~\cite{Wang:2019xvx} are supposed to be located in a specific direction relative to the IP and have a macroscopic distance from the IP.
We, thus, can add shielding between the IP and new detectors to reduce the background events stemming from the IP.
In contrast, for the proposal in this work, the new detector is capable of accepting LLPs from most directions, so only the MD serves as the main background veto and it is not easy to instrument more shielding. 
We thus propose no shielding between the MD and the LAYCAST detector.
Therefore, in general, the design of such a cavern-surface detector brings the advantage of a large solid-angle coverage which can lead to better acceptances for the LLPs' signal, but it potentially has higher levels of background events, compared with the designs in Ref.~\cite{Wang:2019xvx}.

Now we discuss the potential background sources, mainly following the discussion given in Ref.~\cite{Chrzaszcz:2020emg}.
The MD acts as a veto for most SM particles from the primary vertex.
The remaining primary background sources consist of cosmic-induced particles and neutrinos coming from the IP.
The former can be effectively rejected by directional cuts and timing information which would be enabled if at least two separated layers of detector material are implemented~\cite{MATHUSLA:2020uve,MATHUSLA:2022sze}, so that particles stemming from the IP and those from the outside can be distinguished.
For proper tracking, though, at least four layers of detector materials would be required.
To support the layers, the mechanical structure should be designed as modular and lightweight material with high rigidity such as aluminum should be used.

At the CEPC/FCC-ee IP, neutrinos can be produced from certain hard processes, and subsequently interact with detector materials leading to production of charged particles in charged-current interactions.
If such interactions take place inside the MD, the produced charged particles can be vetoed by various components of the MD.
Even if the neutrinos interact with the outer layers of the MD, an additional layer for tracking could be installed just outside the MD for veto purpose.
For more detail, we refer to the discussions below and Ref.~\cite{Chrzaszcz:2020emg}.
We clarify the neutrino-veto logic as follows: an additional veto layer placed just outside the MD can be designed to be thin and low-mass, so that its neutrino-interaction probability is strongly reduced compared to that in dense outer MD materials. Moreover, charged-current neutrino interactions in dense material typically yield multi-track and hadronic activity that can be rejected using topology and timing once multiple LAYCAST layers are available.

From an experimental-integration perspective, LAYCAST can be operated as a stand-alone subsystem with time-stamped readout synchronized to the machine clock. Candidate displaced activity can then be associated with objects reconstructed in the MD offline, avoiding assumptions about a dedicated hardware-trigger coupling between the two detectors. The large lever arm between the MD and the cavern surface can further aid in track matching and cosmic rejection. A quantitative assessment of potential improvements in SM muon momentum resolution would require a dedicated magnetic-field and material model and is beyond the scope of this phenomenological study.

In addition, long-lived neutral kaons from hadronic $Z$ decays can traverse the main detector and decay in the cavern, producing displaced photons that reach LAYCAST and thus form an additional SM background~\cite{MATHUSLA:2020uve}.
In Appendix~\ref{app:KL_background}, we estimate this $K_L$ background and show that the combined MD$\times$FD requirements strongly suppress it; we illustrate this with the benchmark $e^+e^-\to Z\to a\gamma$ and an ultra-long-lived ALP $a$.
Other displaced signatures, such as $\ell^+\ell^-$ pairs, are expected to be similarly clean once a similar strategy including timing/pointing and a ``DV-in-air'' requirement is imposed: prompt dileptons are removed by pointing/timing; conversion-induced displaced $e^+e^-$ vertices are vetoed by $m_{ee}\simeq 0$, small opening angle, and material-aligned vertices~\cite{CMS:2010nua}; and accidental displaced $\mu^+\mu^-$-vertex candidates from cosmic (or machine-induced) muons are rejected using multi-layer timing and directionality (4D vertexing)~\cite{MATHUSLA:2018bqv,ANUBIS:2019qdc}.

While a detailed analysis of these background events is beyond the scope of the present work, for the numerical results we will show contour curves corresponding to more than one values of the signal-event number in order to showcase the effect of background events on the sensitivity reach.
Additionally, the MD at CEPC or FCC-ee may also suffer from background events for LLP searches; we study the effect of the background level on the sensitivity reach and provide a brief discussion in Appendix~\ref{appendix:NS_main_detector}.

We assume to use the plastic scintillator as the tracker material, for its excellent capabilities of timing and spatial measurements.
The cost of such a layered tracker detector stems primarily from the detector materials i.e.~scintillators.
The secondary contribution is from readout electronics.
In Ref.~\cite{Chrzaszcz:2020emg}, the total cost of the proposed detector HECATE is estimated to be below 5 MCHF per layer.
Considering the same scintillator model (EJ-200) and and the same type of readout electronics (based on the Sci-Fi detector from LHCb~\cite{LHCb:2014uqj}) and re-scaling the surface area from HECATE to LAYCAST, our proposal should cost below 3.6 MCHF per layer.

Finally, Since we will compare numerical results of LAYCAST with those of the MD at the CEPC/FCC-ee and representative far-detector setups (``FD1'', ``FD3'', and ``FD6'') considered in Ref.~\cite{Wang:2019xvx}, we briefly describe their geometries here.
For the fiducial volume of MD, the geometry of baseline detector setup in Ref.~\cite{Wang:2019orr} is adopted for estimating its sensitivity results for all four LLP scenarios.
The three far-detector setups in Ref.~\cite{Wang:2019xvx} employ a cuboid fiducial volume, centered around the $\theta=\pi/2$ direction, where $\theta$ is the polar angle with respect to the positive $z$-axis.
FD1 has a volume of 5000 m$^3$ with dimensions (L$\times$W$\times$H) 50 m$\times$10 m$\times$10 m, and is located 5 m or 10 m horizontally from the IP.
Both FD3 and FD6 are placed on the ground with a vertical distance of 50 m or 100 m from the IP, and be centrally located in both the $x$- and $z$-dimensions.
FD3 (FD6) has a volume of $8\times 10^5$ m$^3$ ($8\times 10^7$ m$^3$) with dimensions 200 m$\times$200 m$\times$20 m (1000 m$\times$1000 m$\times$80 m).

\section{Theory models}
\label{sec:model}

In this section, we briefly review the theoretical scenarios for which we estimate the sensitivity reach of the proposed experiment to the corresponding LLPs.

\subsection{Exotic Higgs decays into light scalars}\label{subsec:modelHiggs}

Ever since the discovery of the SM Higgs boson, $h$, at the LHC in 2012~\cite{Aad:2012tfa,Chatrchyan:2012xdj}, it has been one of the utmost tasks in high-energy physics to measure the properties of this Higgs boson including its mass and couplings to various particles.
While the LHC experiments such as ATLAS and CMS have published search results in this direction~\cite{ATLAS:2023dak,ATLAS:2023xpr,ATLAS:2023tnc,ATLAS:2018mme,Ordek:2022ppj,Trevisani:2022die,Tackmann:2022tcv}, there is a consensus that the next-generation electron-positron colliders running at the center-of-mass (CM) energy around $240-250$ GeV as Higgs-boson factories should be able to perform more accurate precision measurements on the SM Higgs boson~\cite{CEPCStudyGroup:2018ghi,An:2018dwb,FCC:2018byv,An:2018dwb,Grojean:2022blp,Guo:2022wti,Ruan:2014xxa,Ge:2016tmm}.
Such measurements include the Higgs boson decays into a pair of light scalar bosons (dark Higgs), $X$'s, that arise in the minimal extension of the Higgs sector of the SM where a singlet scalar field is appended to the model and mixes with the SM Higgs boson (SM$+X$)~\cite{OConnell:2006rsp,Wells:2008xg,Bird:2004ts,Pospelov:2007mp,Krnjaic:2015mbs,Boiarska:2019jym}.
CEPC and FCC-ee can search for this dark Higgs boson which may either decay promptly~\cite{Liu:2016zki,Ma:2020mjz,Gao:2019ypl,Liu:2017lpo}, be long-lived~\cite{Alipour-Fard:2018lsf,Cheung:2019qdr}, or stay invisible~\cite{Liu:2017lpo,Tan:2020ufz,Wang:2020dtb,Wang:2020tap}.
Further, sensitivity reach of external far detectors to this scenario has been estimated in Ref.~\cite{Wang:2019xvx}.
Since we will work with model-independent parameters as in Ref.~\cite{Wang:2019xvx} such as the dark-Higgs mass $m_X$ and proper decay length $c\tau_X$, as well as the exotic Higgs decay branching ratio Br$(h\to X X)$, we choose not to explicitly show the Lagrangian of the SM$+X$ model here.

\subsection{Heavy neutral leptons}\label{subsec:modelHNL}

The observation of neutrino oscillations~\cite{Super-Kamiokande:1998kpq,deSalas:2017kay} has confirmed the non-vanishing neutrino masses, which are in disagreement with the SM.
To provide an explanation, the most common approach is to assume the existence of electroweak singlet fermions, called sterile neutrinos, and to apply seesaw mechanisms~\cite{Minkowski:1977sc,Yanagida:1979as,Gell-Mann:1979vob,Mohapatra:1979ia,Schechter:1980gr,Wyler:1982dd,Mohapatra:1986bd,Bernabeu:1987gr,Akhmedov:1995ip,Akhmedov:1995vm,Malinsky:2005bi}.
These sterile neutrinos are usually called ``heavy neutral leptons'', $N$, in the context of collider physics.
They mix with the active neutrinos and can participate in SM weak interactions suppressed by the mixing parameters.
The interactions of the HNLs with the $W$- and $Z$-bosons, after the electroweak symmetry breaking, are described by the following Lagrangian:
\begin{eqnarray}
		{\cal L} &=& \frac{g}{\sqrt{2}}\ \sum_{\alpha}
		V_{\alpha N}\ \bar \ell_\alpha \gamma^{\mu} P_L N W^-_{L \mu}
		+\nonumber \\
		&&\frac{g}{2 \cos\theta_W}\ \sum_{\alpha, i}V^{L}_{\alpha i} V_{\alpha N}^*
		\overline{N} \gamma^{\mu} P_L \nu_{i} Z_{\mu},
		\label{eqn:CC-NC}
\end{eqnarray}
where $i=1,2,3$, and $\ell_\alpha$ ($\alpha=e,\,\mu, \tau$) are the charged leptons of the SM. 
$V^L$ denotes the PMNS lepton-mixing matrix and $V$ labels the mixing between the active neutrinos and the HNLs.
Furthermore, $P_L=(1-\gamma^5)/2$ is the left-chiral projection operator, and $W_{L\mu}$ and $Z_\mu$ are the fields of the $W$- and $Z$-bosons, respectively.

In phenomenological studies, often a simple ``3+1'' scenario is taken as a benchmark where there is only one HNL and it mixes with only one flavor of the active neutrinos; moreover, the mixing angle $V_{\alpha N}$ and the HNL's mass are treated as independent parameters.
Here, we consider the Majorana HNL and study the case where the HNL mixes with only $\nu_\alpha$ with $\alpha=e$ or $\mu$ (but not both).
We focus on the $Z$-pole operation mode of the CEPC and FCC-ee experiments, and restrict ourselves to the HNL production mode $Z\to N \nu_\alpha$, of which the partial decay width is computed analytically as follows:
\begin{eqnarray}
		\Gamma(Z\rightarrow  N \, \nu_{\alpha})&=& 2\cdot \Gamma(Z\rightarrow \nu_\alpha \bar{\nu}_\alpha) \cdot |V_{\alpha N}|^2   \nonumber\\
		&&   \cdot \left( 1 - \frac{m_N^2}{m_Z^2} \right)^2 \left( 1+\frac{1}{2}\frac{m_N^2}{m_Z^2} \right).
  \label{eqn:Z2nuNdecaywidth}
\end{eqnarray}
The decays of the HNL are also mediated with the mixing parameter, and all the two-body and three-body decay widths at tree-level are calculated with formulas given in Ref.~\cite{Atre:2009rg}.
For an HNL with mass below $m_Z\sim 90$ GeV and sufficiently small mixing angles, the HNL becomes naturally long-lived.
We will present sensitivity results in the $|V_{\alpha N}|^2$ vs.~$m_N$ plane.

\subsection{R-parity-violating supersymmetry and the lightest neutralino}\label{subsec:modelRPV}

In the R-parity-violating supersymmetry~\cite{Barbier:2004ez,Dreiner:1998wm,Mohapatra:2015fua}, a GeV-scale lightest neutralino~\cite{Domingo:2022emr} is allowed by all astrophysical and cosmological constraints~\cite{Grifols:1988fw,Ellis:1988aa,Lau:1993vf,Dreiner:2003wh,Dreiner:2013tja,Profumo:2008yg,Dreiner:2011fp}, as long as it is dominantly bino-like~\cite{Gogoladze:2002xp,Dreiner:2009ic} and can decay, e.g.~via an RPV coupling so as to avoid overclosing the Universe~\cite{Bechtle:2015nua}, the GUT relation between the gauginos $M_1\approx 0.5 M_2$ is lifted~\cite{Choudhury:1995pj,Choudhury:1999tn}, and the dark matter is not comprised of the lightest neutralino~\cite{Belanger:2002nr,Hooper:2002nq,Bottino:2002ry,Belanger:2003wb,AlbornozVasquez:2010nkq,Calibbi:2013poa}.
Further, the lightest neutralino is assumed to be the lightest supersymmetric particle.

We write down the general RPV superpotential below,
\begin{eqnarray}
		W_{\text{RPV}}&=&\mu_i H_u \cdot L_i + \frac{1}{2}\lambda_{ijk}L_i \cdot L_j \bar{E}_k \nonumber\\
		&& + \lambda'_{ijk} L_i \cdot Q_j \bar{D}_k + \frac{1}{2}\lambda''_{ijk}\bar{U}_i \bar{D}_j \bar{D}_k,
  \label{eqn:RPV-superpotential}
\end{eqnarray}
where the chiral superfields are defined as usual, and $\mu_i$, $\lambda_{ijk}$, $\lambda'_{ijk}$, and $\lambda''_{ijk}$ are RPV couplings.
We consider a simple case where only one RPV operator, $\lambda'_{112}L_1\cdot Q_1 \bar{D}_2$, is non-vanishing, allowing for quick comparison with existing works which took the same operator.

For the production of the lightest neutralino, we focus on the $Z$-boson decays: $Z\to \tilde{\chi}^0_1 \tilde{\chi}^0_1$; note that even though the vertex inducing this decay at tree level couples only to the Higgsino components of the lightest neutralino, we assume these components are sufficiently small so that the mass bounds on the Higgsinos are circumvented~\cite{ATLAS:2017vat} and the upper bounds on the Higgs invisible decay width~\cite{CMS:2023sdw,ATLAS:2022yvh} are also satisfied.
We note that associated production channels such as $Z\to \tilde{\chi}^0_1\,\tilde{\chi}^0_2$ can be relevant in other neutralino-mixing scenarios and may feature a larger rate; our sensitivity curves can be straightforwardly reinterpreted by rescaling the assumed production branching ratio together with the visible branching fraction and detector efficiency in Eq.~\eqref{eqn:NS-LLP}.
For numerical studies, we will fix the $Z$-boson's decay branching ratio Br$(Z\to \tilde{\chi}^0_1 \tilde{\chi}^0_1)$ at $10^{-3}$ for $m_{\tilde{\chi}^0_1}\ll m_Z/2$ so that the strong constraints on the $Z$-boson invisible decay width~\cite{ParticleDataGroup:2022pth,ParticleDataGroup:2024cfk} is also satisfied; note that this choice of the Br$(Z\to \tilde{\chi}^0_1 \tilde{\chi}^0_1)$)is also allowed by the Higgsino mass bounds~\cite{Helo:2018qej}

The decay of the lightest neutralino is then induced by the $\lambda'_{112}$ coupling via an off-shell sfermion.
For the calculation of the decay widths of the lightest neutralino, we follow the treatment of Ref.~\cite{Wang:2019orr}, where it is found that for $m_{\tilde{\chi}^0_1} \lesssim 3.5$ GeV the lightest neutralino decays dominantly into a kaon and an SM lepton, and above this threshold three-body decays into a lepton and two jets are dominant; the corresponding formulas for estimating the two-body and three-body decay widths are given in Ref.~\cite{deVries:2015mfw} and Ref.~\cite{Wang:2019orr}, respectively.
In particular, for $m_{\tilde{\chi}^0_1}\lesssim m_Z/2$ and sufficiently small $\lambda'_{112}$ coupling, the lightest neutralino travel a certain distance before decaying.
The sensitivity results will be presented in the $(m_{\tilde{\chi}^0_1}, \frac{\lambda'_{112}}{m^2_{\tilde{f}}})$ plane, where $m_{\tilde{f}}$ is the sfermion mass which we assume to be degenerate among all the sfermions.
We note that the present upper bound on $\lambda'_{112}/m^2_{\tilde{f}}$ is~\cite{Barger:1989rk,Allanach:1999ic}
\begin{eqnarray}
   \lambda'_{112}/m^2_{\tilde{f}} < \frac{2.1\times 10^{-4}}{\text{GeV} \cdot m_{\tilde{f}}}.
   \label{eqn:rpv_bound}
\end{eqnarray}

\subsection{Axion-like particles}\label{subsec:modelALPs}

Axions and axion-like particles are pseudoscalar bosons proposed for solving 
the QCD strong CP problem~\cite{Peccei:1977ur,Peccei:2006as,ParticleDataGroup:2024cfk} and can, additionally, serve as dark matter candidates~\cite{Dine:1982ah,Abbott:1982af,Preskill:1982cy,Marsh:2015xka,Lambiase:2018lhs,Auriol:2018ovo,Houston:2018vrf}.
They arise as the pseudo-Goldstone bosons when the global symmetry of an appended $U(1)_{\text{PQ}}$ gauge group breaks down at a high-energy scale~\cite{Peccei:1977ur,Peccei:2006as}.
Different from the axions, the ALPs do not fix the relation between the ALP mass and the high symmetry-breaking scale, imply a richer phenomenology at colliders such as LHC~\cite{Bauer:2017ris,Alonso-Alvarez:2023wni,Ren:2021prq,Ghebretinsaea:2022djg,Baldenegro:2019whq,Florez:2021zoo}.
In principle, the ALP can couple to not only the SM gauge bosons, but also the Higgs boson and the matter fields.
Here, we focus on a scenario where the ALP is only coupled to the SM electroweak gauge bosons, and we write down the corresponding general effective Lagrangian~\cite{Georgi:1986df}:
	\begin{eqnarray}
		{\cal L}_{\rm eff}
		&\supset & \, \frac12 \left( \partial_\mu a\right)\!\left( \partial^\mu a\right)
		- \frac{m_{a}^2}{2}\,a^2
		+ g^2\,C_{WW}\,\frac{a}{\Lambda}\,W_{\mu\nu}^A\,\tilde W^{\mu\nu,A} \nonumber \\
		&+& g^{\prime\,2}\,C_{BB}\,\frac{a}{\Lambda}\,B_{\mu\nu}\,\tilde B^{\mu\nu} \,,
  \label{eqn:ALP_Lag_b4SB}
	\end{eqnarray}
 where $m_a$ labels the mass of the ALP $a$, $\Lambda$ is the global-symmetry breaking scale, and $C_{WW}$ and $C_{BB}$ are coupling constants.
 Further $W^A_{\mu\nu}$ and $B_{\mu\nu}$ are the field strength tensors of the $SU(2)_L$ and $U(1)_Y$ gauge groups, respectively, and $g$ and $g'$ are the corresponding gauge-group coupling constants.

After electroweak symmetry breaking, the effective Lagrangian including the interactions of the ALP with the SM photon and the $Z$-boson is expressed as~\cite{Bauer:2018uxu}
	\begin{equation}
		\begin{aligned}
			{\cal L}_{\rm eff}
			\supset&\, e^2\,C_{\gamma\gamma}\,\frac{a}{\Lambda}\,F_{\mu\nu}\,\tilde F^{\mu\nu}
			+ \frac{2e^2}{s_w c_w}\,C_{\gamma Z}\,\frac{a}{\Lambda}\,F_{\mu\nu}\,\tilde Z^{\mu\nu}  \\
			&+ \frac{e^2}{s_w^2 c_w^2}\,C_{ZZ}\,\frac{a}{\Lambda}\,Z_{\mu\nu}\,\tilde Z^{\mu\nu} \,,
		\end{aligned}
  \label{eqn:ALP_Lag_afterSB}
	\end{equation}
 where $F_{\mu\nu}$ and $Z_{\mu\nu}$ are the field strength tensors of the photon and the $Z$-boson, respectively, $s^2_w=\sin^2{\theta_w}\approx 0.23$ with $\theta_w$ being the weak mixing angle and $c^2_w=\cos^2{\theta_w}$, and $e= g\, s_w$.

The Wilson coefficients in Eq.~\eqref{eqn:ALP_Lag_afterSB} are related those in Eq.~\eqref{eqn:ALP_Lag_b4SB} with~\cite{Bauer:2018uxu}
 \begin{eqnarray}
		C_{\gamma\gamma} &=& C_{WW}+C_{BB}, \notag \\
		C_{\gamma Z} &=& c^2_w\,C_{WW}-s^2_w\,C_{BB}, \notag \\
		C_{ZZ} &=& c^4_w\,C_{WW}+ s^4_w\,C_{BB}\, .
	\end{eqnarray}

\begin{figure}[t]
	\centering
	\includegraphics[height=4cm, width=6cm]{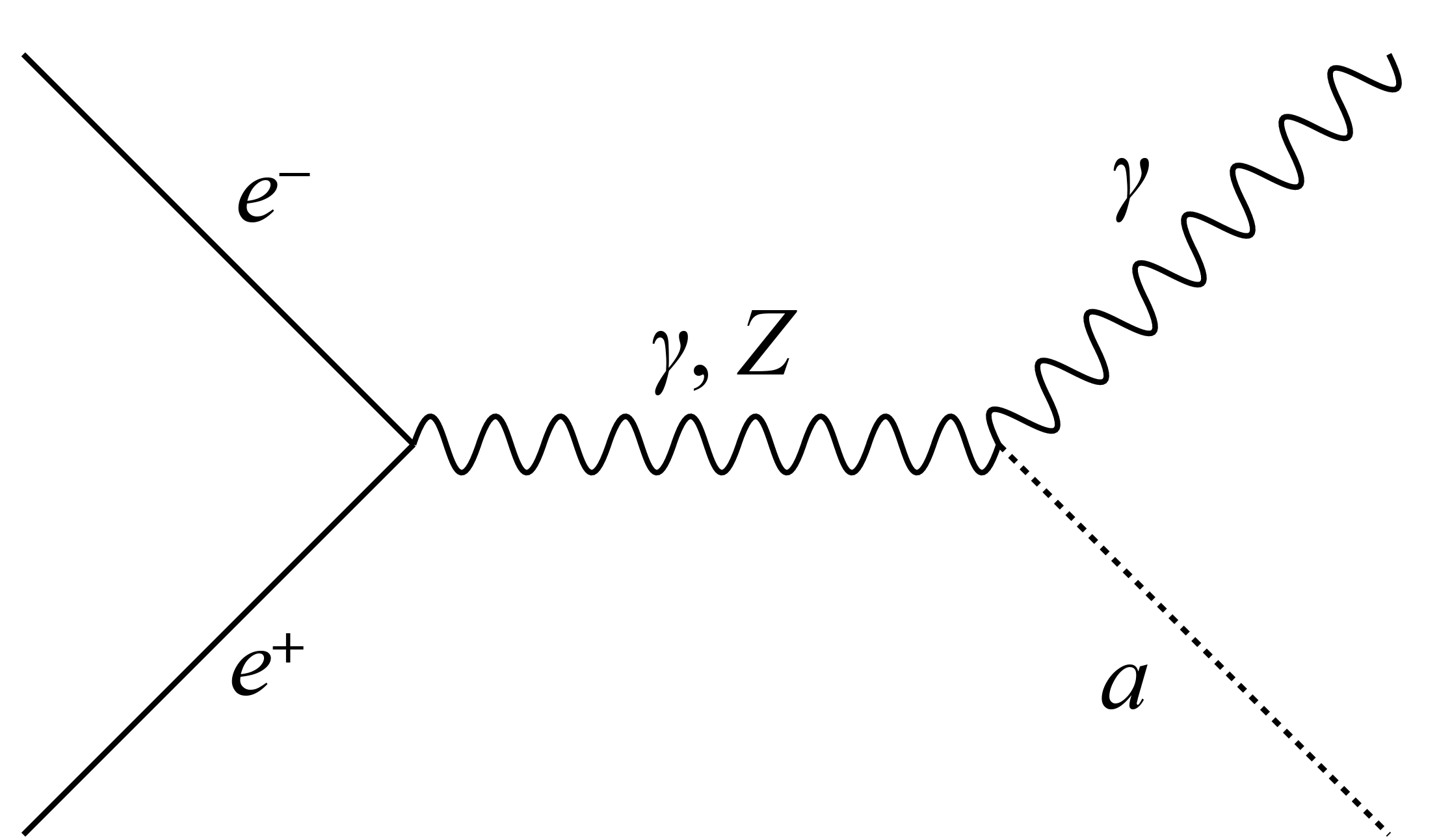}
	\caption{The ALP production process, $e^- e^+ \to   \gamma \, a$, at electron-positron colliders~\cite{Tian:2022rsi}.}
	\label{fig:eegga}
\end{figure}

As in Ref.~\cite{Tian:2022rsi} where the authors estimated the sensitivity reach of a series of potential far detectors at the CEPC or FCC-ee to this ALP scenario, we focus on the ALP production associated with a photon in the process, $e^- e^+\to \gamma \, a$~\footnote{The other two possible production processes, $e^- e^+ \to Z \, a$ and $Z\to \gamma \, a$, are not considered here.}, shown in Fig.~\ref{fig:eegga}.
The corresponding differential cross section is reproduced from Ref.~\cite{Bauer:2018uxu} as
\begin{align}
		\frac{d \sigma(e^- e^+\to\gamma \, a)}{d\Omega} =&
		\, 2 \pi \alpha \alpha^2(s) \frac{s^2}{\Lambda^2} \left( 1 - \frac{m_a^2}{s} \right)^3 \left(1 + \cos^2 \theta \right) \notag \\
		& \cdot \left( |V_\gamma(s)|^2 + |A_\gamma(s)|^2 \right),
		\label{eqn:ALP_prod_diff_XS}
	\end{align}
	where $s$ is the CM energy squared and
	\begin{align}
		V_\gamma(s) &= \frac{C_{\gamma \gamma}}{s} + \frac{g_V}{2 c_w^2 s_w^2}\frac{C_{\gamma Z}}{s - m_Z^2+ i \, m_Z \, \Gamma_Z}\, , \notag\\ 
		A_\gamma(s) &= \frac{g_A}{2 c_w^2 s_w^2}\frac{C_{\gamma Z}}{s - m_Z^2+ i \,m_Z \, \Gamma_Z}\, , 
	\end{align}
	with $g_V = 2 s_w^2 - 1/2$ and $g_A=-1/2$.

For $m_a\lesssim m_Z$, the ALP dominantly decays to a pair of photons with the partial decay width~\cite{Bauer:2017ris,Bauer:2018uxu}
 \begin{equation}
		\Gamma(a\to\gamma\gamma)  = 4\pi\alpha^2 m_a^3 \, \left| \frac{C_{\gamma\gamma} }{\Lambda} \right|^2 \, .
		\label{eqn:ALP2gammagamma-width}
\end{equation}	

Assuming the ALP total decay width $\Gamma_a$ is saturated by Eq.~\eqref{eqn:ALP2gammagamma-width}, we can compute the boosted decay length of the ALP with the following expression~\cite{Tian:2022rsi},
	\begin{equation}
		\lambda_a^{\text{lab}} \approx 15\,\mathrm{m} 
		\left( \frac{s-m_a^2}{m_a \sqrt{s}} \right) 
		\left( \frac{\GeV}{m_a} \right)^3
		\left( \frac{\Lambda}{\TeV} \right)^2 \left( \frac{10^{-4}}{C_{\gamma\gamma}} \right)^2 \, ,
		\label{eqn:lambda}
	\end{equation}
as the $Z$-bosons at the CM energy $\sqrt{s}=91.2$ GeV are produced at rest.

We emphasize that in this study, the production of the ALP is induced by both $C_{\gamma\gamma}$ and $C_{\gamma Z}$ couplings and its decay proceeds via the $C_{\gamma\gamma}$ coupling only.
We therefore end up with three model parameters: $m_a$, $C_{\gamma\gamma}/\Lambda$, and $C_{\gamma Z}/\Lambda$.

\section{Monte-Carlo simulation and numerical computation}\label{sec:simulation}

In this section, we elaborate on the computation of the signal-event numbers in the considered benchmark theoretical scenarios, including the MC simulation method we apply.
We start with explaining how we estimate the total production numbers of the LLPs, $N^{\text{prod}}_{\text{LLP}}$, in each scenario.
For the cases of the light scalar, the HNL, and the light neutralino, we make use of the following formula,
\begin{eqnarray}
    N_{\text{LLP}}^{\text{prod}} = N_M\cdot  \text{Br}(M\rightarrow n_{\text{LLP}}\,\text{LLP}+Y)\cdot n_{\text{LLP}},
\end{eqnarray}
where $N_M$ is the total number of the mother particle $M$ of the LLP with $M$ denoting either the SM Higgs boson or the $Z$-boson.
$n_{\text{LLP}}=1,2,\ldots$ is the number of the LLP(s) produced in each signal decay of $M$.
$Y$ represents any additional particles associated with the LLP production.
For the scenarios of the light scalar and the light neutralino, we treat Br($M\rightarrow n_{\text{LLP}} \,\text{LLP(s)}+Y)=$Br$(h\to XX)$ and Br$(Z\to \tilde{\chi}^0_1\tilde{\chi}^0_1)$, respectively, as independent parameters.
For the HNL study, we compute Br$(Z\to N \, \nu_\alpha )$ with the model parameters $m_N$ and $|V_{\alpha N}|^2$ analytically using Eq.~\eqref{eqn:Z2nuNdecaywidth}.

For the scenario of the ALP coupled to the electroweak gauge bosons, we numerically calculate the scattering cross sections of $e^- e^+\to \gamma \, a$ for various values of $m_a$, $C_{\gamma\gamma}/\Lambda$ and $C_{\gamma Z}/\Lambda$, with MadGraph 5~\cite{Alwall:2011uj,Alwall:2014hca} and the UFO (Universal FeynRules Output)~\cite{Degrande:2011ua,Darme:2023jdn} model file ``ALP linear Lagrangian''~\cite{Brivio:2017ije}, and parameterize these cross sections with
\begin{eqnarray}
    	\sigma(e^- e^+ \to \gamma\,a)     	\approx 16 \text{ fb} \cdot \left( \frac{\TeV}{\Lambda} \right)^2  \left( 1 - \frac{m_a^2}{s} \right)^3 \nonumber \\
    	\left(  \left| C_{\gamma\gamma} \right|^2 + 2680\left| C_{\gamma Z} \right|^2 - 0.082\left| C_{\gamma\gamma} C_{\gamma Z} \right|   \right).
    	\label{eqn:ALP_prod_XS}
\end{eqnarray}
The dependence of Eq.~\eqref{eqn:ALP_prod_XS} on the model parameters is verified to be consistent with that of Eq.~\eqref{eqn:ALP_prod_diff_XS}.
We thus compute the total production numbers of the ALP with,
\begin{eqnarray}
    N_{\rm{ALP}}^{\rm{prod}} = \sigma(e^- e^+ \rightarrow \gamma \, a)  \cdot  \mathcal{L}_Z,
\end{eqnarray}
where $\mathcal{L}_Z$ is the integrated luminosity at the $Z$-pole.

For the MC generation of the signal events for the light scalar, HNL, and light neutralino, we utilize the PYTHIA8 program with version 8.309~\cite{Sjostrand:2006za,Sjostrand:2014zea} which also performs showering, and follow the same simulation procedures as described in Ref.~\cite{Wang:2019xvx}.
The CM energies are set to be 240 GeV and 91.2 GeV for the light-scalar case, and the HNL and light neutralino cases, respectively.
For the ALPs, we generate the signal scattering events with MadGraph5 at the CM energy $\sqrt{s}=91.2$ GeV as described above, and the output LHE~\cite{Alwall:2006yp} file is interfaced with PYTHIA8 for showering.
For all LLP cases, PYTHIA8 is interfaced to the HepMC2 with version 2.06.09~\cite{Dobbs:2001ck} and provides the truth-level information of the kinematics of the simulated LLPs in the standard event record format.

The average decay probability of the LLPs in the detector's fiducial volume (f.v.), $\langle P[\text{LLP}\text{ in f.v.}]\rangle$, is estimated with the following formula:
\begin{eqnarray}
\langle P[\text{LLP}\text{ in f.v.}]\rangle=\frac{1}{N^{\text{MC}}_{\text{LLP}}}\sum_{i=1}^{N^{\text{MC}}_
    		{\text{LLP}}}P[(\text{LLP})_i\text{ in f.v.}]\,,
    	\label{eqn:average_decay_prob}
\end{eqnarray}
where $N^{\text{MC}}_{\text{LLP}}$ is the total number of MC-simulated LLPs.
$P[(\text{LLP})_i\text{ in f.v.}]$ is the individual decay probability of each generated LLP in the simulation, computed with 
\begin{equation}
P[(\text{LLP})_i\text{ in f.v.}] = e^{\left( -{ D_i^\text{l.m.} }/{ \lambda_i } \right)} - e^{\left( -{D_i^\text{a.c.}}/{\lambda_i} \right)}\,.
		\label{eqn:individual_decayProb}
\end{equation}
Here, $D_i^\text{l.m.}$ and $D_i^{\text{a.c.}}$ are the distances from the IP for the $i$-th simulated LLP to travel to leave the main detector and to arrive at the cavern surface, respectively, if it has not decayed beforehand.
Furthermore, $P[(\text{LLP})_i\text{ in f.v.}]=0$ if the $i$-th simulated LLP does not travel in a direction inside the coverage of the fiducial volume, e.g.~it travels towards the cavern's floor.
Here, $\lambda_i$ is the lab-frame decay length of the LLP and is computed with $\lambda_i=\beta_i \gamma_i c\tau$, where $\beta_i$ and $\gamma_i$ are the speed and boost factor of the $i$-th generated LLP, respectively, and $c\tau$ is the proper decay length of the LLP.

Finally, the total signal-event number is calculated with
 	\begin{eqnarray}
		N_{\text{LLP}}^{\text{signal}}=N_{\text{LLP}}^{\text{prod}} \cdot \left\langle P[\text{LLP in f.v.}]\right\rangle \cdot \text{Br(LLP}\rightarrow\text{visible}),\nonumber\\
		\label{eqn:NS-LLP}
	\end{eqnarray}
where $\text{Br(LLP}\rightarrow\text{visible})$ is the LLP's decay branching ratio into visible final states for the detector.
We note that for a realistic estimate, one should take into account accurate values of the visible decay branching ratio and the detector efficiency depending on the type and kinematics of the final-state particles. 
In this work, for simplicity, we assume a flat 100\% value for both of them.
A realistic consideration of such factors will weaken the sensitivities accordingly.

As discussed in Sec.~\ref{sec:detector-setup}, given the absence of shielding between the main detector and the proposed cavern-surface detector, we do not simply assume vanishing background in this work.
Instead, we will show sensitivity results for both 3 and 20 signal events for comparison purpose in the following section.
These two choices correspond to exclusion bounds at 95\% confidence level (C.L.) for 0 and 100 background events, respectively.
In addition, for simplicity we will write $N_{\text{LLP}}^{\text{signal}}$ as $N_S$ in the rest of the paper.

\section{Sensitivity results}
\label{sec:results}

We present our numerical results in this section, for each benchmark scenario separately.
Not only will we show the sensitivity reach in each model, but also the average decay probabilities, since the latter are important for understanding the final results.
For the proposed detector, we will plot sensitivity curves of both 3 and 20 signal events, to illustrate the effect of potential background events, cf.~Sec.~\ref{sec:detector-setup}.
In addition, we make comparisons with studies on other lepton-collider far detectors, CEPC/FCC-ee's main detectors, as well as LHC's far detectors.

For the Higgs-factory mode at the CM energy $\sqrt{s}=240$ GeV, we choose to confine ourselves to the case of integrated luminosity $\mathcal{L}_{h} = 5.6$ ab$^{-1}$ for the future high-energy electron-positron colliders, delivering $1.14\times 10^6$ Higgs bosons~\cite{CEPCStudyGroup:2018ghi}.
We note that, for example, CEPC with potential upgrades of the collider technologies may deliver up to 20 ab$^{-1}$ integrated luminosity at the Higgs-factory mode over the whole period of the experiment's operation time~\cite{CEPCPhysicsStudyGroup:2022uwl}, which can lead to stronger discovery sensitivities. 
At the $Z$-pole, we study two benchmark integrated luminosities of 16 ab$^{-1}$ and 150 ab$^{-1}$ corresponding to total production of $7\times 10^{11}$ and $5\times 10^{12}$ $Z$-bosons.

\subsection{Exotic Higgs decays}\label{subec:exotichiggsdecays}

\begin{figure}[t]
	\centering
	\includegraphics[width=\columnwidth]{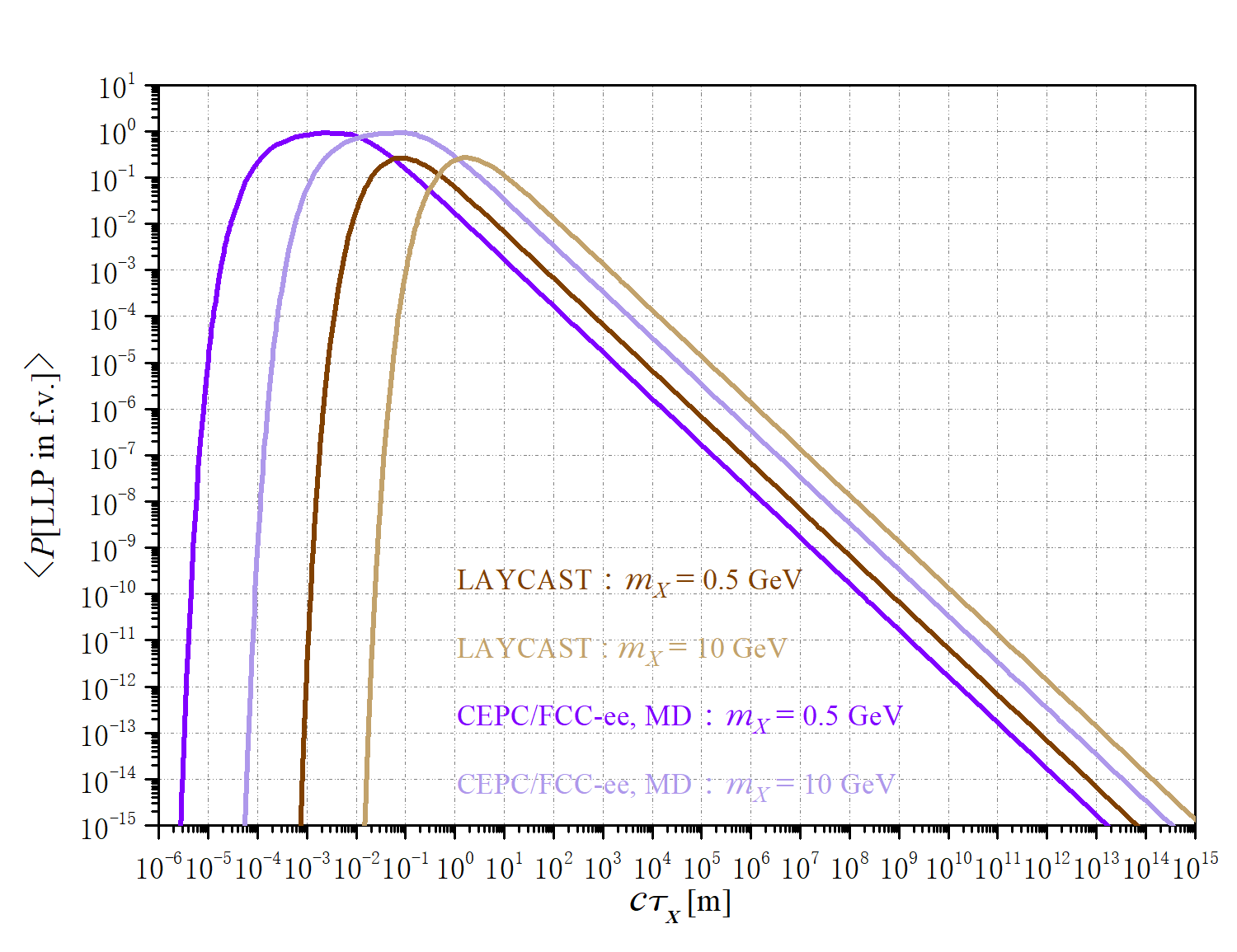}
	\caption{The average decay probability of the light scalar $X$ as functions of $c\tau_X$, for LAYCAST and the CEPC/FCC-ee's main detector and for $m_X=0.5$ GeV and 10 GeV.}
	\label{fig:H2XX-Avg}
\end{figure}

Fig.~\ref{fig:H2XX-Avg} is a plot of $\langle P[\text{LLP}\text{ in f.v.}]\rangle$ as functions of $c\tau_X$, for both the CEPC/FCC-ee's main detector and the LAYCAST experiment, with $m_X=0.5$ GeV and 10 GeV.
The maximal average decay probability that can be attained for LAYCAST is about 0.26, at $c\tau_X=0.08$ m and 1.65 m with $m_X=0.5$ GeV and $m_X=10$ GeV, respectively.
Because of their closer distance to the IP, the main detectors are more sensitive to smaller $c\tau_X$ values, and can reach up to 0.93 for the average decay probability at $c\tau_X=0.003$ m and 0.057 m for the same pair of $X$ masses.

For the two mass choices, the two $c\tau_X$ values corresponding to the maximal average decay probability differ by a factor $\sim 20$ which is exactly the ratio between the two chosen mass values of the light scalar.
This is because for the maximal average decay probability, there is a corresponding value of the boosted decay length $\beta_X \gamma_X c \tau_X$ approximately determined by the experiment's geometry and distance to the IP, where $\beta_X \gamma_X=|\vec{p_X}|/m_X$, with $|\vec{p_X}|$ being the 3-momentum magnitude of the light scalar $X$.
Roughly speaking, then, varying $m_X$ by a factor 20 leads to a general shift of $\beta_X \gamma_X$ by 20, and thus $c\tau_X$ should change by $1/20$ so as to keep the boosted decay length unchanged, leading to the observed horizontal shift of the curves.

In the prompt regime (small $c\tau_X$), the exponential functions for the decay-position distributions change rapidly with $c\tau_X$ (see Eq.~\eqref{eqn:individual_decayProb}), leading to the large slope of the displayed curves in Fig.~\ref{fig:H2XX-Avg}.
On the other hand, for large $c\tau_X$ values, the exponential function can be expanded and as a result, $\langle P[\text{LLP}\text{ in f.v.}]\rangle$ becomes linearly proportional to $1/c\tau_X$ to first order, as shown in Fig.~\ref{fig:H2XX-Avg}.
In the remaining benchmark scenarios, the same trend of the average decay probability curves will also be observed.

\begin{figure}[t]
	\centering
	\includegraphics[width=\columnwidth]{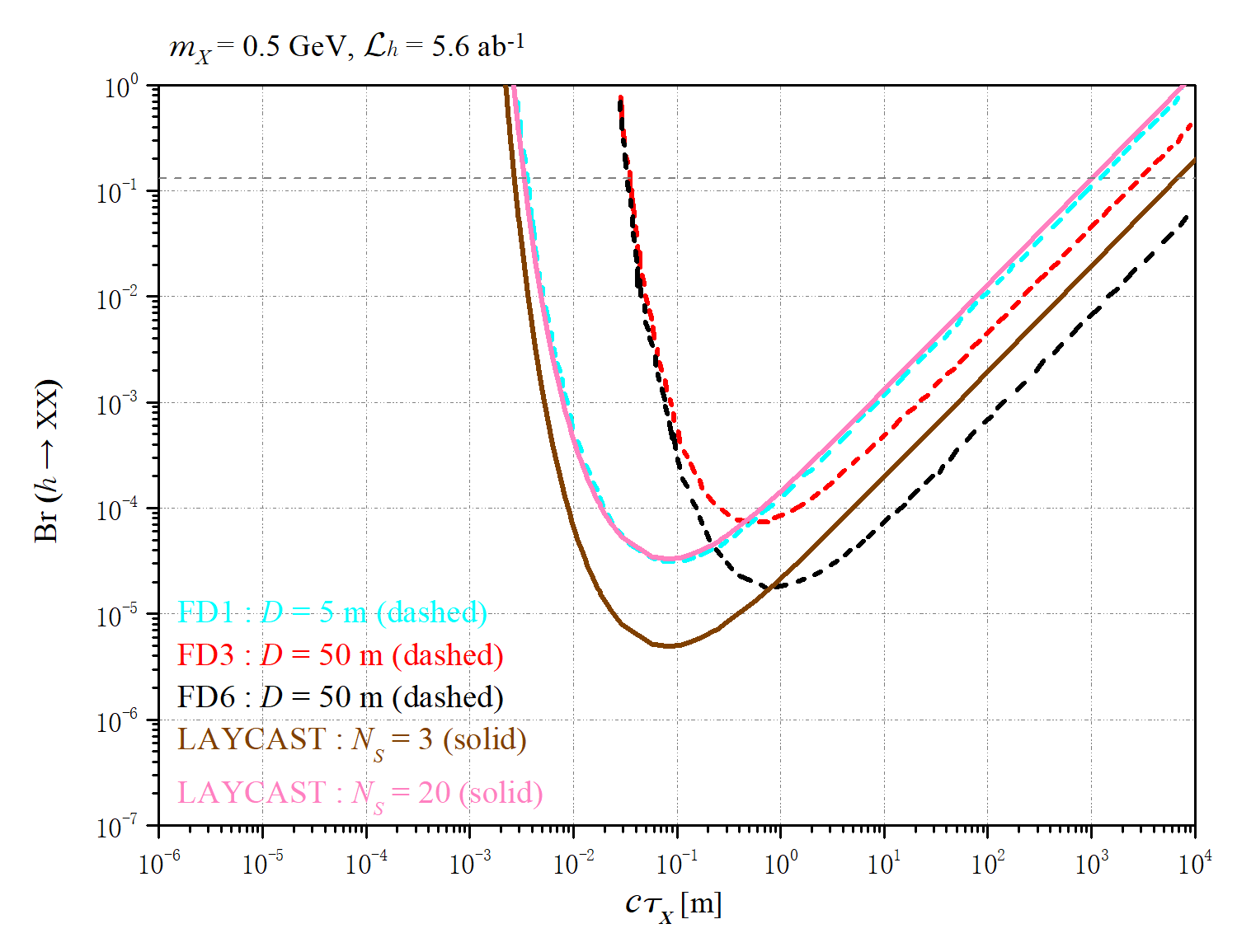}	
	\includegraphics[width=\columnwidth]{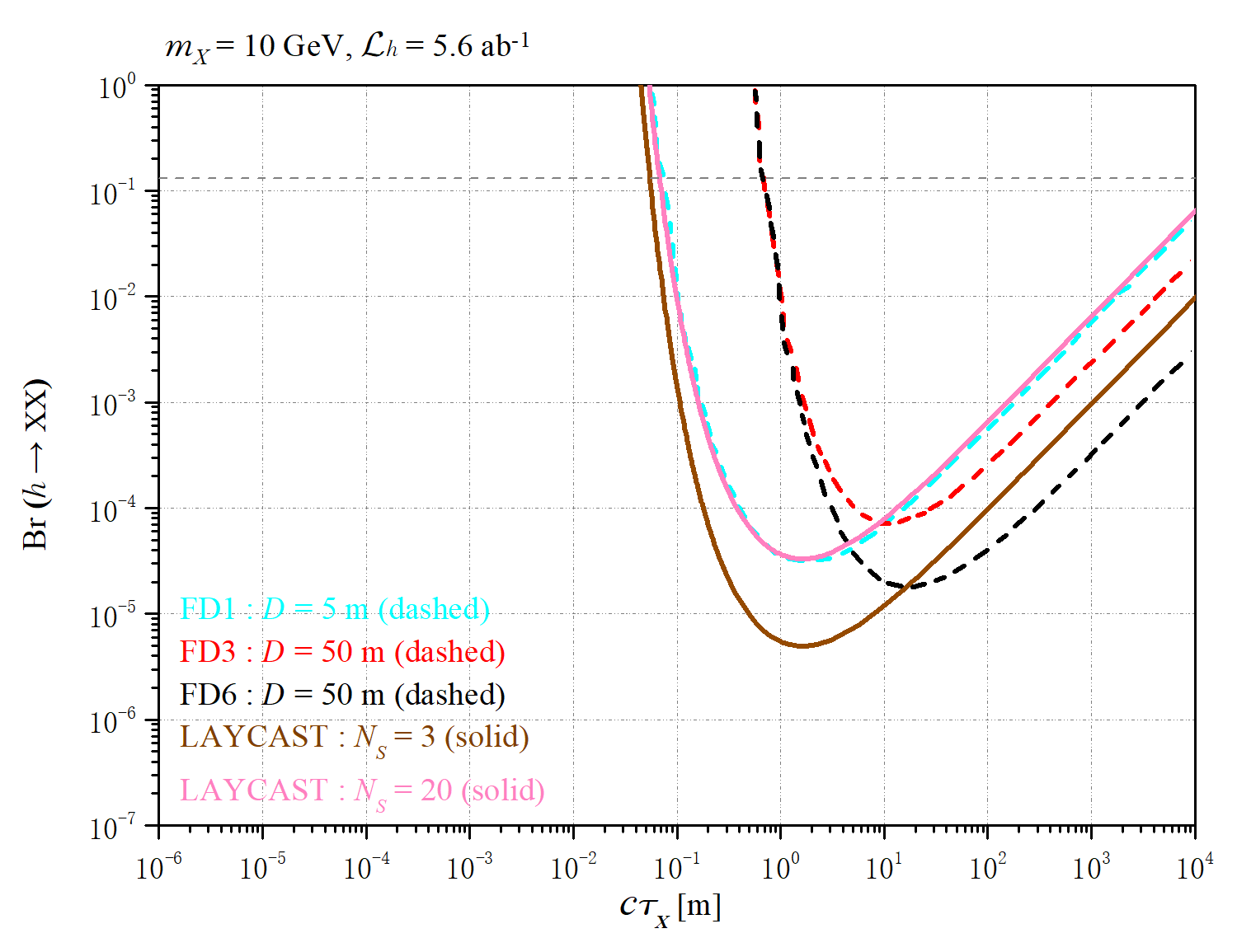}
	\caption{Sensitivity reaches of LAYCAST compared to those of FD1, FD3, and FD6 investigated in Ref.~\cite{Wang:2019xvx}, shown in the Br($h \rightarrow XX$) vs.~$c\tau_X$ plane for $m_X = 0.5$ GeV (\textit{upper panel}) and 10 GeV (\textit{lower panel}).
Together shown is also the current limit on the Higgs invisible decay branching ratio at 13\%~\cite{ATLAS:2022vkf}, as a horizontal gray dashed line.}
	\label{fig:H2XX}
\end{figure}

In Fig.~\ref{fig:H2XX}, we compare sensitivity reaches of the proposed experiment shown in the $(c\tau_X,\text{ Br}(h\to X X))$ plane to those of FD1, FD3, and FD6 studied in Ref.~\cite{Wang:2019xvx}.
Two plots are for $m_X=0.5$ GeV and 10 GeV, respectively; their main difference is a horizontal shift, as discussed above.
For LAYCAST, we show both $N_S=3$ and $N_S=20$ sensitivity curves~\footnote{Here, the obtained sensitivities are for an integrated luminosity of $\mathcal{L}_{h} =$ 5.6 ab$^{-1}$.
For a higher integrated luminosity of 20 ab$^{-1}$, the discovery sensitivities to Br$(h\to XX)$ can be enhanced by a factor of $5.6/20$ for each $c\tau_X$ value.}.

We find that it is more sensitive in the low and medium $c\tau_X$ regions ($c\tau_X \lesssim 1~(10)$ m) for $m_X=0.5$ GeV ($m_X=20$ GeV).
For longer decay lengths, FD3 and especially FD6 are expected to perform better.
We note that FD1 is estimated to be similarly sensitive to the long-lived light scalar as LAYCAST with $N_S=20$.
Among these experiments, only the LAYCAST experiment with zero background can reach Br$(h\to XX)$ down to the order of $10^{-6}$ while all the other setups can probe the branching ratio in the order of $10^{-5}$ at the best.
These are all more than 3 order of magnitude lower than the current upper bound on the Higgs invisible decay branching ratio $13\%$~\cite{ATLAS:2022vkf} shown as the horizontal gray dashed lines in the plots.

\begin{figure}[t]
	\centering
	\includegraphics[width=\columnwidth]{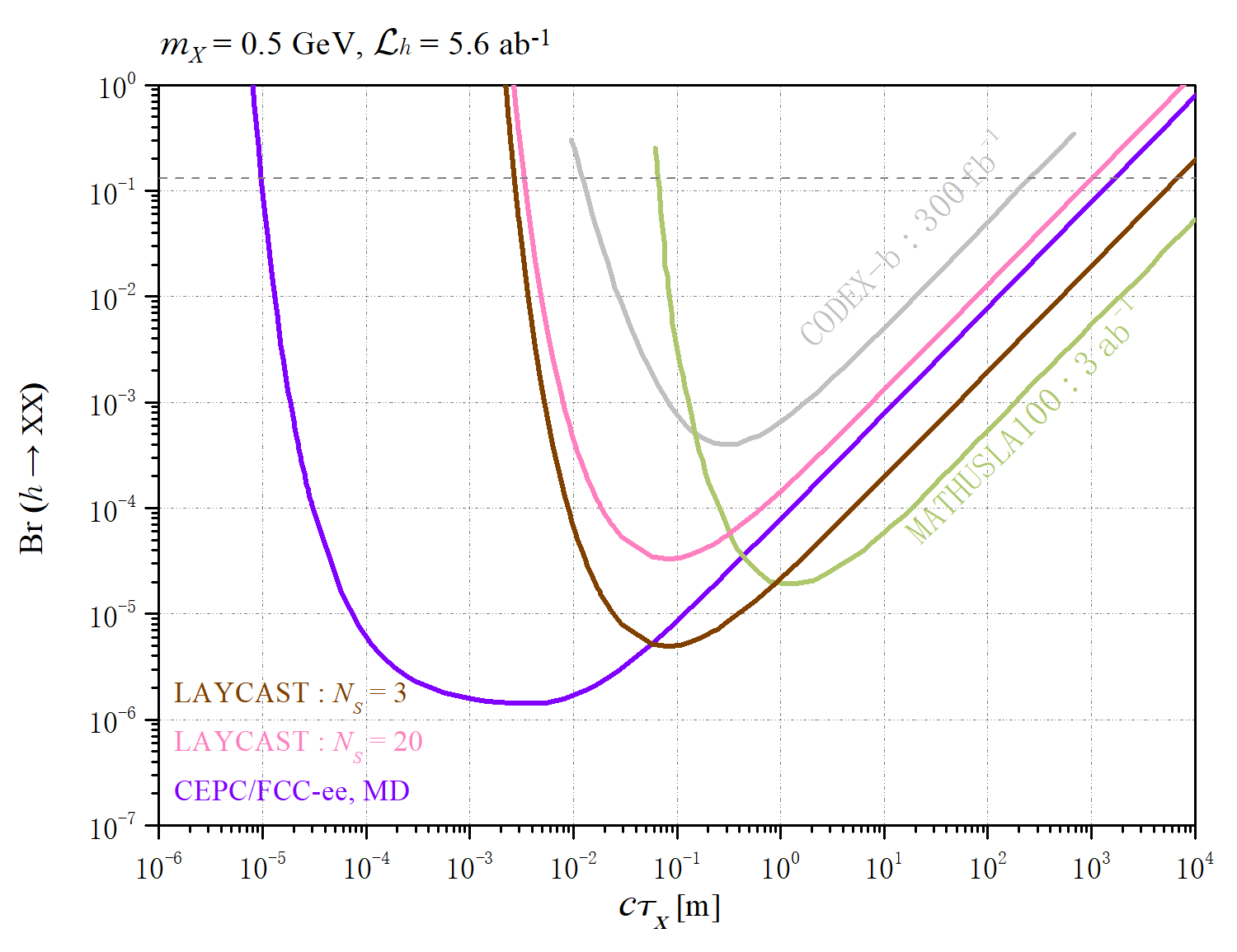}
	\includegraphics[width=\columnwidth]{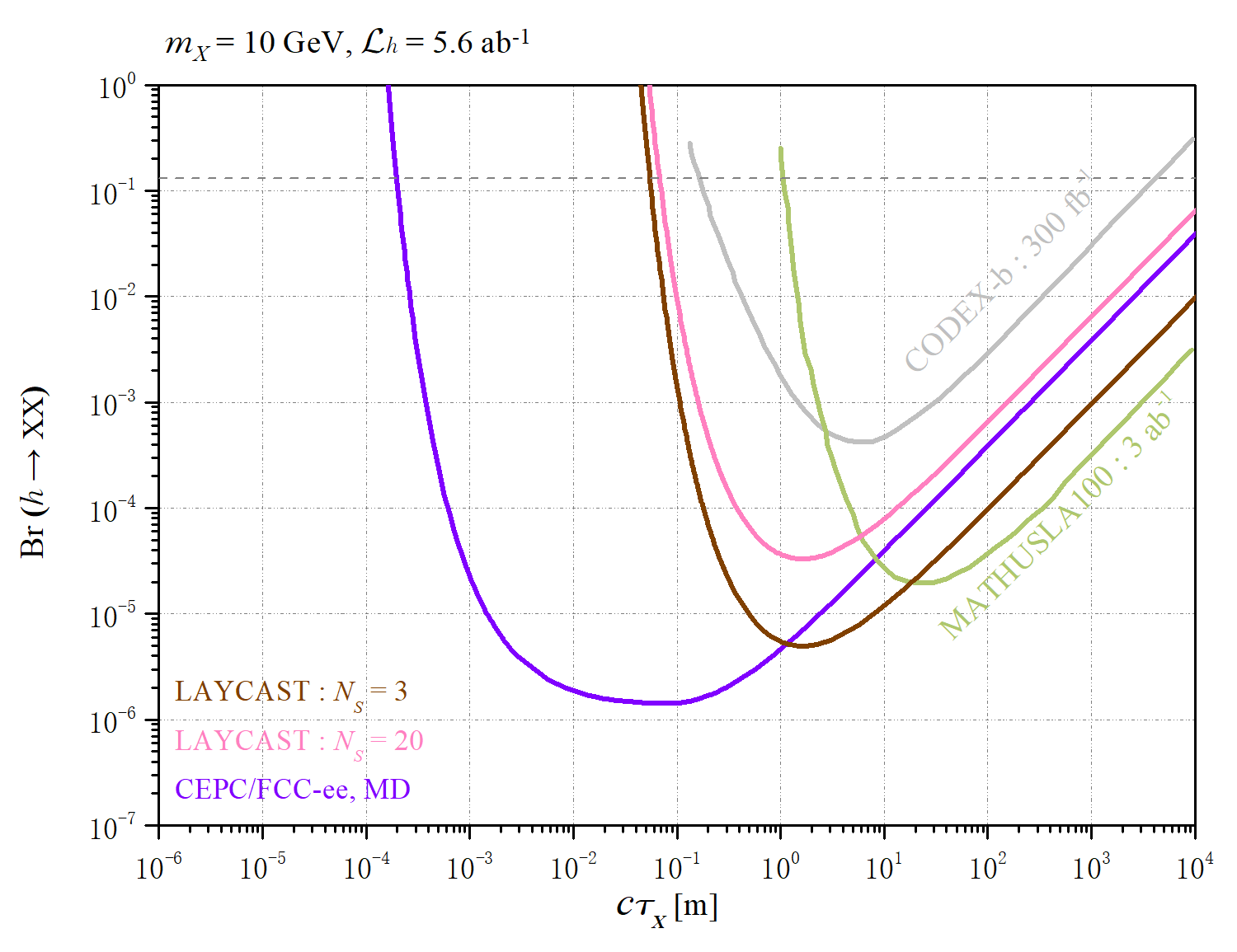}
	\caption{Sensitivity reaches of LAYCAST compared to the CEPC/FCC-ee's main detector and the LHC's far detectors CODEX-b and MATHUSLA~\cite{Wang:2019xvx} for $m_X = 0.5$ GeV (\textit{upper panel}) and 10 GeV (\textit{lower panel}).
 The present bound on the SM Higgs invisible decay branching ratio 13\%~\cite{ATLAS:2022vkf} is plotted as a horizontal gray dashed curve.  }
	\label{fig:H2XX-other-exp}
\end{figure}

In Fig.~\ref{fig:H2XX-other-exp} we further overlap the sensitivity reach of LAYCAST with that of the CEPC/FCC-ee's main detector and those of the proposed LHC's far detectors CODEX-b~\cite{Gligorov:2017nwh} and MATHUSLA~\cite{MATHUSLA:2018bqv}.
The results of the MD are calculated with the method described in this article, and those of CODEX-b and MATHUSLA are reproduced from Ref.~\cite{Wang:2019xvx}.
As expected, the main detector shows dominant sensitivity in the short-$c\tau_X$ regime, while medium and long decay-length regimes LAYCAST can compete with the CODEX-b and MATHUSLA, and show promising sensitivities at $c\tau_X$ values between the most sensitive positions of the main detector and MATHUSLA.

\subsection{Heavy neutral leptons}\label{subsec:HNL}

\begin{figure}[t]
	\centering
	\includegraphics[width=\columnwidth]{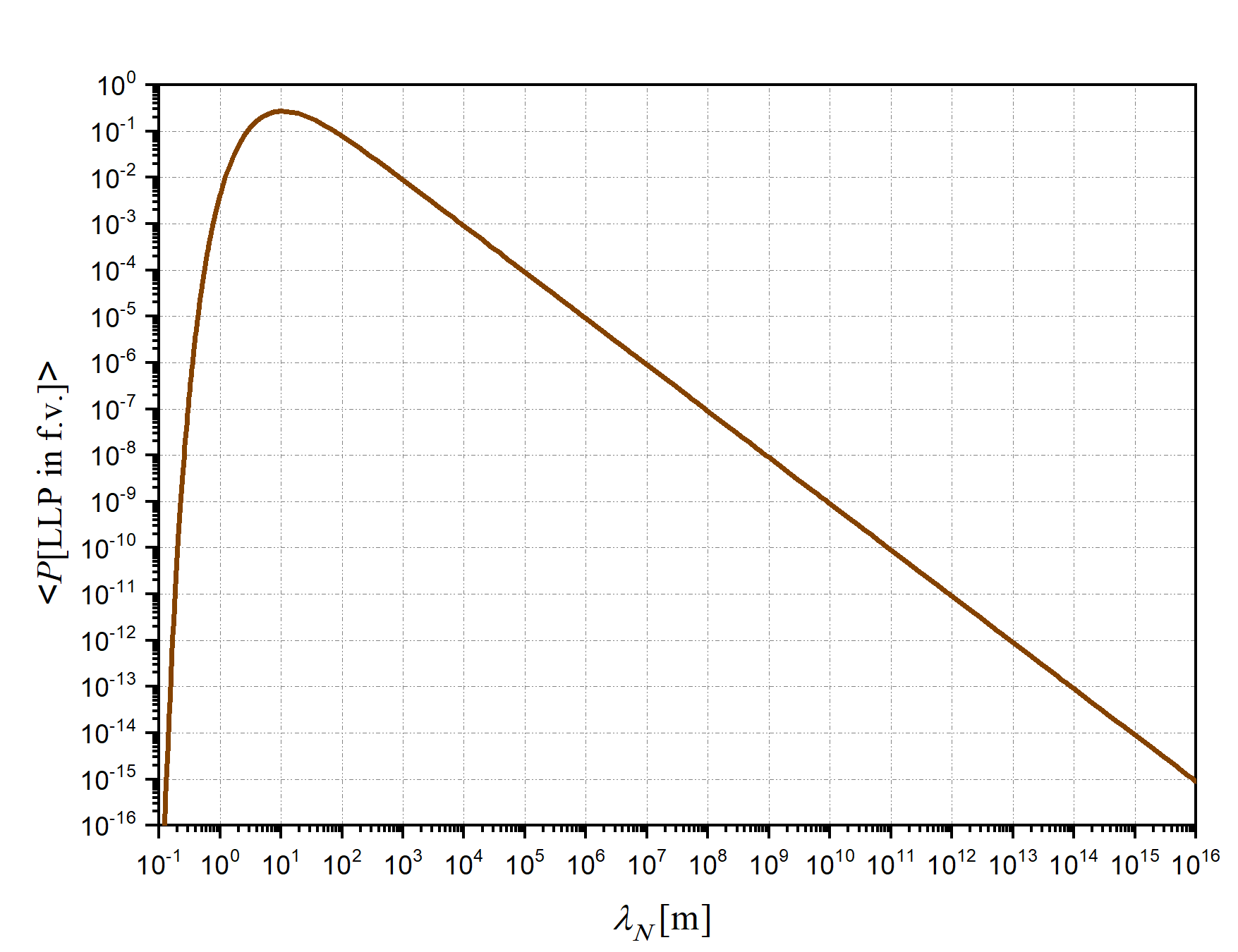}
	\caption{The average decay probability of the HNLs in the fiducial volume of the LAYCAST experiment as a function of the boosted decay length of the HNL.}
	\label{fig:Z-NV-Avg}
\end{figure}

For the HNL study where the HNL is produced in $Z\to \nu_\alpha N$, we also start with showing a plot of $\langle P[\text{LLP}\text{ in f.v.}]\rangle$, now as a function of the boosted decay length $\lambda_N$ of $N$, for the proposed experiment; see Fig.~\ref{fig:Z-NV-Avg}.
We find the maximal average decay probability 0.27 is achieved at $\lambda_N\sim 10$ m, in general alignment with the cavern size.
The result essentially does not depend on the mass of the HNL.

\begin{figure}[t]
\centering
\includegraphics[width=\columnwidth]{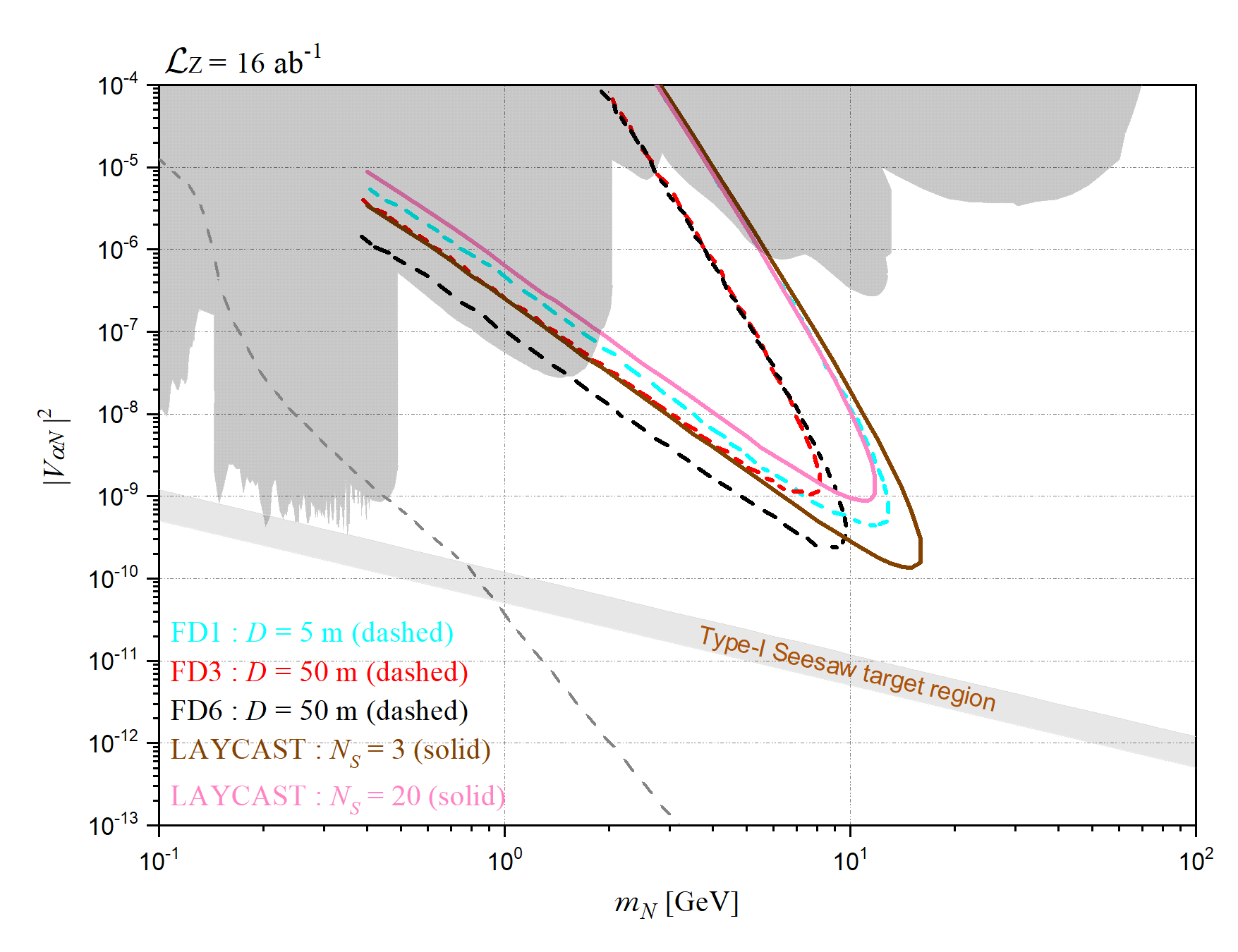}
\includegraphics[width=\columnwidth]{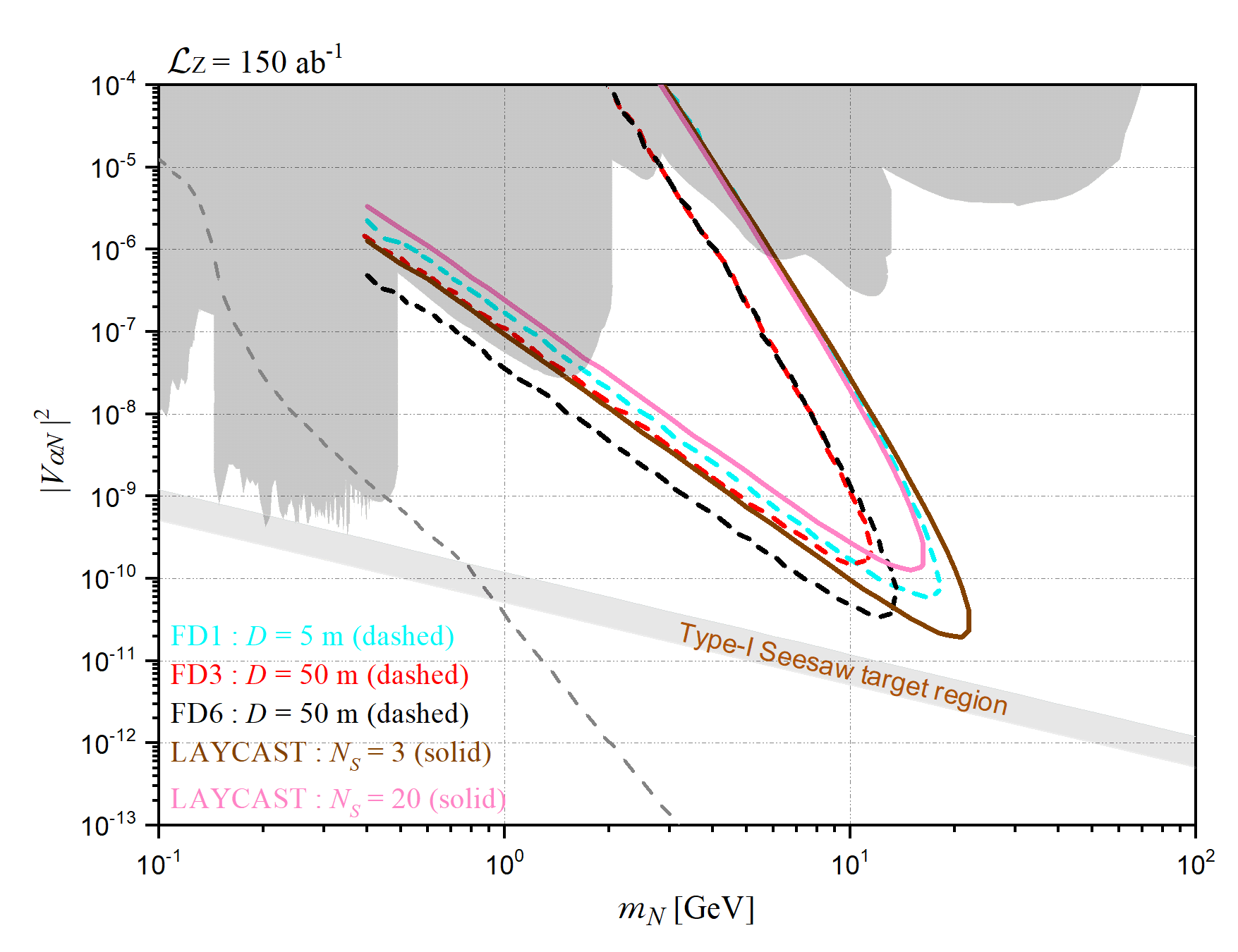}
\caption{Sensitivity reaches of LAYCAST, and FD1, FD3, and FD6 from Ref.~\cite{Wang:2019xvx}, displayed in the $|V_{\alpha N}|^2$ vs.~$m_N$ plane with $\mathcal{L}_Z$ = 16 ab$^{-1}$ (\textit{upper panel}) and 150 ab$^{-1}$ (\textit{lower panel}).
The gray area in the background is the parameter region currently excluded by experiments~\cite{PIENU:2017wbj,Bryman:2019bjg,NA62:2020mcv,T2K:2019jwa,Barouki:2022bkt,CMS:2022fut,CMS:2024ake,CMS:2024xdq,ATLAS:2022atq,L3:2001zfe} on $|V_{eN}|^2$.
The gray band labeled with ``Type-I Seesaw target region'' corresponds to the parameter space explaining the active neutrino mass between 0.05 eV~\cite{Canetti:2010aw} and 0.12 eV~\cite{Planck:2018vyg} with the type-I seesaw relation $|V_{\alpha N}|^2=m_{\nu_\alpha}/m_N$.
The gray dashed curve in the lower left part is the lower bound on the mixing angle, derived from consideration of Big Bang Nucleosynthesis~\cite{Sabti:2020yrt,Boyarsky:2020dzc}.
}
\label{fig:HNL}
\end{figure}

Again, now in Fig.~\ref{fig:HNL}, we compare our proposal with FD1, FD3, and FD6 from Ref.~\cite{Wang:2019xvx}, shown in the plane $|V_{\alpha N}|^2$ vs.~$m_N$ with $\alpha=e,\mu$.
The upper (lower) plot is for an integrated luminosity of $\mathcal{L}=16$ ab$^{-1}$ ($\mathcal{L}=150$ ab$^{-1}$).

In both plots, we find that LAYCAST has the largest reach in $m_N$, touching 16 GeV and 22 GeV for $\mathcal{L}=16$ ab$^{-1}$ and 150 ab$^{-1}$, respectively, if no background is present.
Even for the case of 100 background events, it can still probe $m_N$ up to about 12 and 16 GeV for the two integrated luminosities.
FD1 has a slightly better sensitivity reach than our proposal with $N_S=20$ has, and FD3 and FD6 can probe slightly smaller $m_N$ and $|V_{\alpha N}|^2$ regimes.
We conclude that the advantage of LAYCAST in this work is mainly the sensitivity to slightly larger masses of the HNL.

In the background of these plots, we show as gray area the current bounds on $|V_{eN}|^2$ as a function of $m_N$ obtained from PIENU~\cite{PIENU:2017wbj}, KENU~\cite{Bryman:2019bjg}, NA62~\cite{NA62:2020mcv}, T2K~\cite{T2K:2019jwa}, BEBC~\cite{Barouki:2022bkt}, CMS~\cite{CMS:2022fut,CMS:2024ake,CMS:2024xdq}, ATLAS~\cite{ATLAS:2022atq}, and L3~\cite{L3:2001zfe}.
We show also a gray band covering the parameter region that could explain the small but non-vanishing active-neutrino mass with the type-I seesaw relation $|V_{\alpha N}|^2= m_{\nu_\alpha}/m_N$, for $m_{\nu_\alpha}$ between 0.05 eV and 0.12 eV obtained from neutrino-oscillation experiments~\cite{Canetti:2010aw} and cosmology observations~\cite{Planck:2018vyg}, respectively.
In addition, the lower-left gray dashed curve is the lower bound on the parameter space, obtained from Big Bang Nucleosynthesis~\cite{Sabti:2020yrt,Boyarsky:2020dzc}.

\begin{figure}[t]
\centering
\includegraphics[width=\columnwidth]{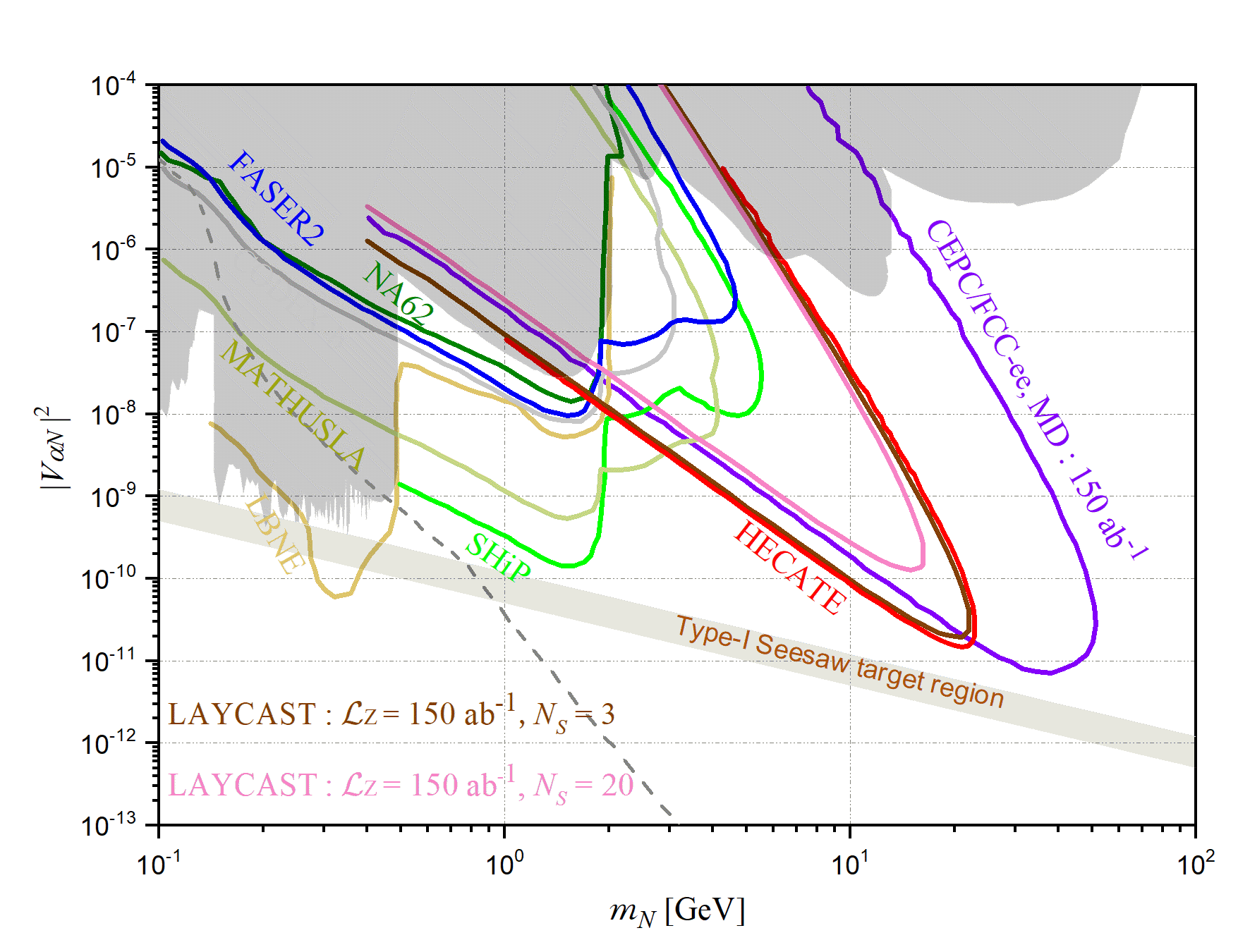}
\includegraphics[width=\columnwidth]{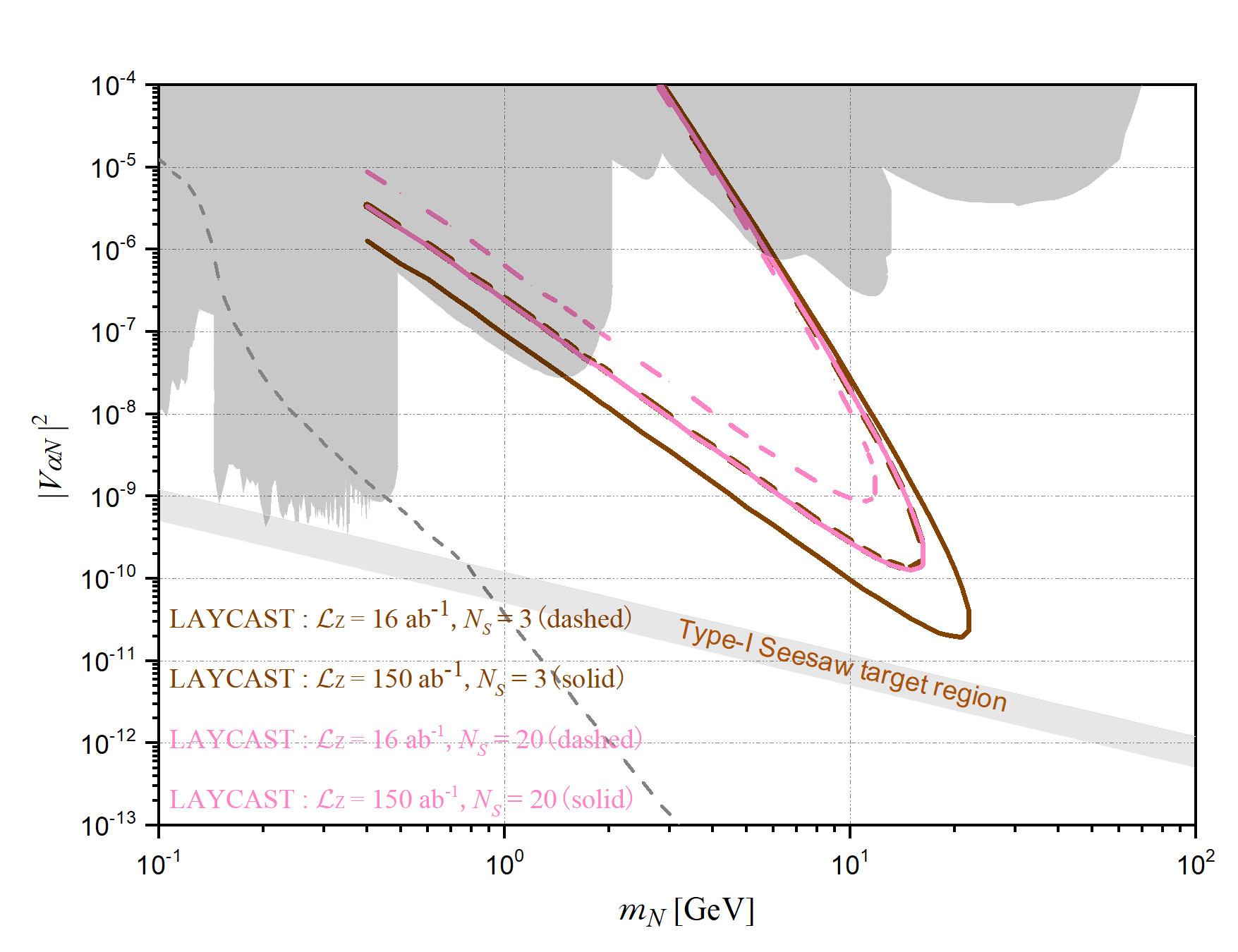}
\caption{\textit{Upper panel}: comparison of LAYCAST with other experiments extracted from Refs.~\cite{Helo:2018qej,Wang:2019xvx,Bondarenko:2018ptm,Drewes:2018gkc,LBNE:2013dhi,Chrzaszcz:2020emg}.
The HECATE result~\cite{Chrzaszcz:2020emg} corresponds to nine signal events and has been adjusted to account for the difference in the considered integrated luminosities.
         \textit{Lower panel}: sensitivity reaches of LAYCAST with different luminosities.}
\label{fig:HNL-other-exp}
\end{figure}

In the upper plot of Fig.~\ref{fig:HNL-other-exp}, we place the sensitivity reach of LAYCAST for $\mathcal{L}=150$ ab$^{-1}$ together with that of other future experiments including the LHC far detectors~\cite{Helo:2018qej} (CODEX-b, FASER2, MATHUSLA), the CEPC/FCC-ee's main detectors with 150 ab$^{-1}$~\cite{Wang:2019xvx}, and beam-dump experiments such as SHiP~\cite{Bondarenko:2018ptm}, NA62~\cite{Drewes:2018gkc}, and LBNE~\cite{LBNE:2013dhi}.
In addition, the HECATE curve is reproduced here from the red solid curve (with $l_0 = 4$ m, $l_1 = 15$ m) in Fig.~1 of Ref.~\cite{Chrzaszcz:2020emg}, with a correction factor applied in order to align the integrated luminosities considered~\footnote{Ref.~\cite{Chrzaszcz:2020emg} assumes an integrated luminosity half that we assume. This leads to a reduction factor 2 on the signal-event number. We extract the HECATE curve for 150 ab$^{-1}$ from Ref.~\cite{Chrzaszcz:2020emg} and apply a factor $1/\sqrt{2}$ to $|V_{\alpha N}|^2$ in the lower part of the curve while keeping the upper part unchanged, and present it in the upper plot of Fig.~\ref{fig:HNL-other-exp}.}.
We find that compared to the LHC's far detectors and the beam-dump experiments, LAYCAST clearly shows a large advantage in probing the mixing parameters for $m_N \gtrsim 3$ GeV.
Our proposal can also extend the sensitivity reach of the main detector to some extent for $m_N\lesssim 20$ GeV, in the case that both experiments suffer from no background events.
Our results with $N_S=3$ is almost identical to those of HECATE with 9 signal events, mainly as a result of the more realistic setup of LAYCAST despite its bigger fiducial volume.

The lower plot of Fig.~\ref{fig:HNL-other-exp} shows the sensitivity reach of the LAYCAST experiment for both integrated luminosities $\mathcal{L}_Z=16$ ab$^{-1}$ and 150 ab$^{-1}$, and both $N_S=3$ and $N_S=20$.
This plot shows the change in the sensitivity when $\mathcal{L}_Z$ or the background level varies.

\subsection{Light neutralinos from $Z-$boson decays}\label{subsec:lightneutralinos}

Here, we refrain from showing a plot of the average decay probability of the lightest neutralino at the proposed experiment, since it would look similar to Fig.~\ref{fig:Z-NV-Avg}.

\begin{figure}[t]
\centering
\includegraphics[width=\columnwidth]{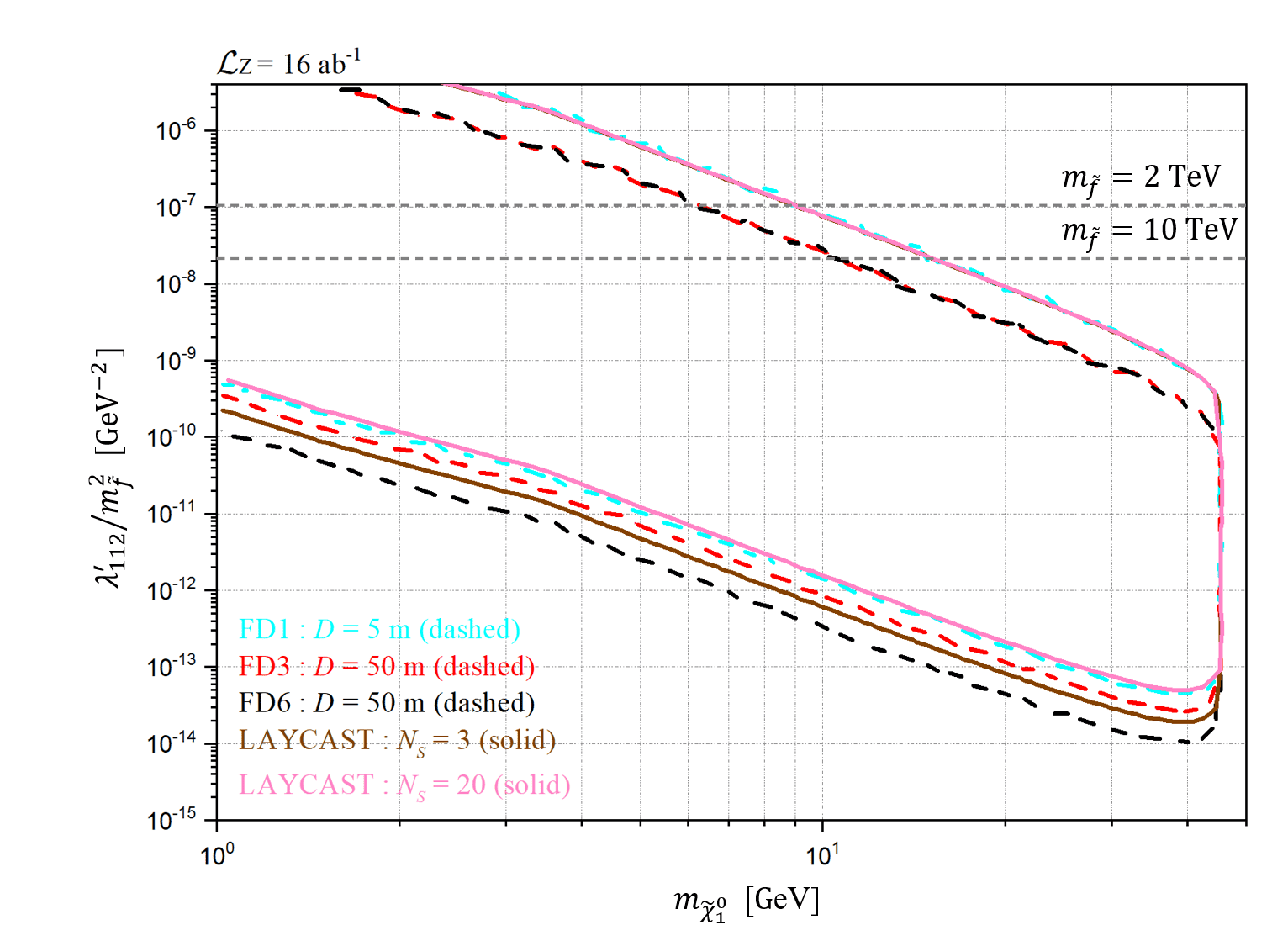}
\includegraphics[width=\columnwidth]{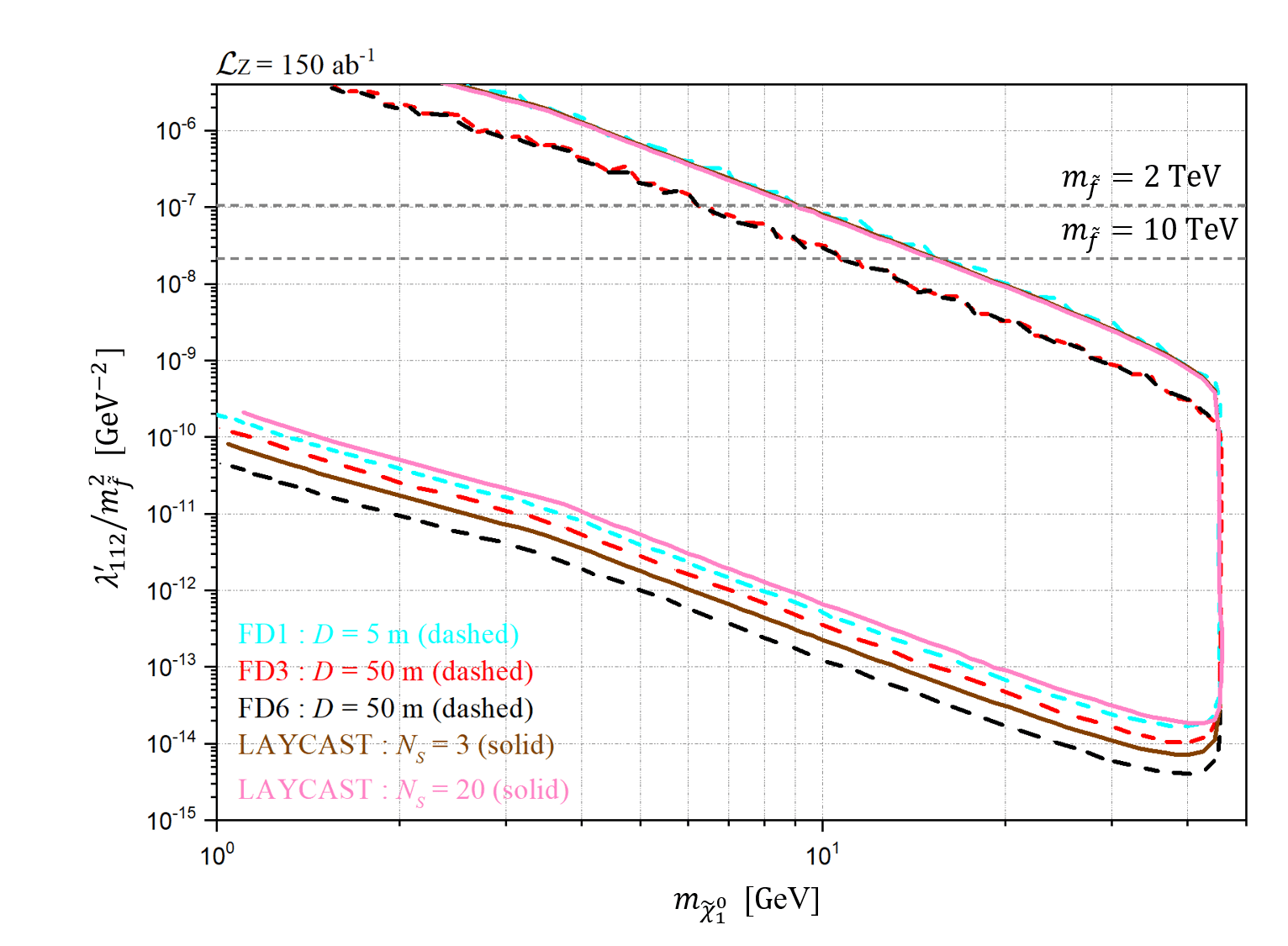}
\caption{Sensitivity reach of LAYCAST in comparison with that of FD1, FD3, and FD3 from Ref.~\cite{Wang:2019xvx}, shown in the $\lambda'_{112}/m^2_{\tilde{f}}$ vs.~$m_{\tilde{\chi}_1^0}$ plane.
We have assumed Br($Z\rightarrow\tilde{\chi}_1^0\tilde{\chi}_1^0$) = $10^{-3}$ for $m_{\tilde{\chi}^0_1} \ll m_{Z}/2$.
The two plots are for $\mathcal{L}_Z$ = 16 ab$^{-1}$ (\textit{upper panel}) and 150 ab$^{-1}$ (\textit{lower panel}), respectively.
The two horizontal dashed curves correspond to the current bounds on the $\lambda'_{112}/m^2_{\tilde{f}}$ for $m_{\tilde{f}}=2$ TeV and 10 TeV.}
\label{fig:Z2n1n1:BrLim1}
\end{figure}

We present in Fig.~\ref{fig:Z2n1n1:BrLim1} the sensitivity reach of LAYCAST to the lightest neutralino in the RPV SUSY, shown in the $(m_{\tilde{\chi}^0_1}, \lambda'_{112}/m^2_{\tilde{f}})$ plane, together with the sensitivity reach of FD1, FD3, and FD6 from Ref.~\cite{Wang:2019xvx}.
The two plots are for $\mathcal{L}_Z=16$ ab$^{-1}$ and 150 ab$^{-1}$, respectively.
The current bounds on the coupling $\lambda'_{112}/m^2_{\tilde{f}}$, cf.~Eq.~\eqref{eqn:rpv_bound}, for $m_{\tilde{f}}=2$ TeV and 10 TeV are shown as horizontal gray dashed curves in the plots.
We find that these considered experiments can reach $\lambda'_{112}/m^2_{\tilde{f}}$ down to $10^{-15}-10^{-14}$ GeV$^{-2}$, at $m_{\tilde{\chi}^0_1}\sim 40$ GeV, orders of magnitude below the current bounds.
In the absence of background events, LAYCAST shows slightly stronger sensitivity reach than FD3, and for 100 background events, it is as sensitive as FD1.
FD6 is the strongest in both plots, mainly owing to its much larger fiducial volume.

\begin{figure}[t]
\centering
\includegraphics[width=\columnwidth]{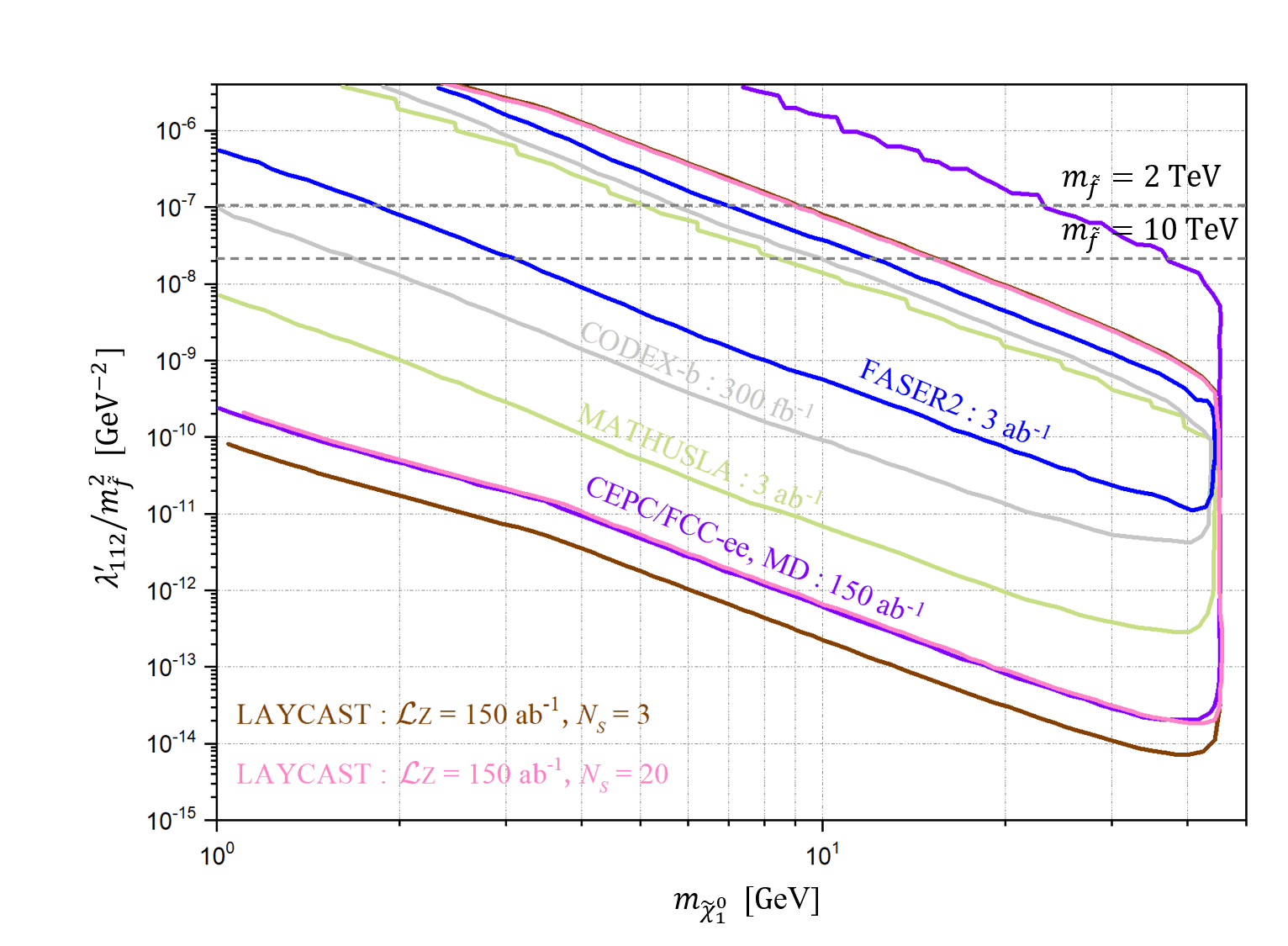}
\includegraphics[width=\columnwidth]{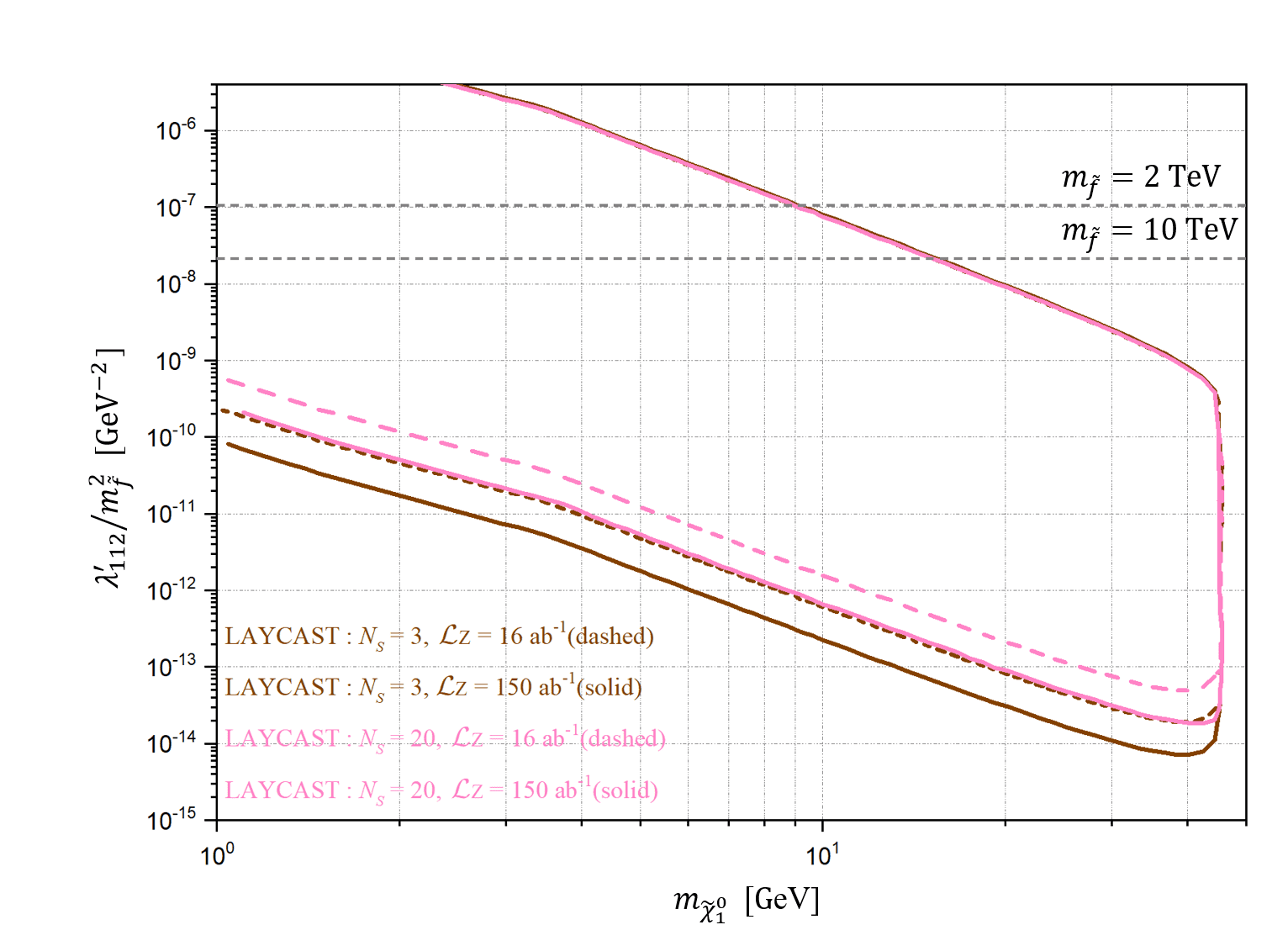}
\caption{\textit{Upper panel}: sensitivity reach of LAYCAST compared to those of the main detector~\cite{Tian:2022rsi} and the LHC’s far detectors~\cite{Helo:2018qej}.
\textit{Lower panel}: sensitivity reach of LAYCAST for different integrated luminosities.}
\label{fig:Z2n1n1:BrLim2}
\end{figure}

We further compare LAYCAST with some LHC's far detectors~\cite{Helo:2018qej} and the CEPC/FCC-ee's main detector~\cite{Wang:2019xvx}, in the upper plot of Fig.~\ref{fig:Z2n1n1:BrLim2}.
For an integrated luminosity of 150 ab$^{-1}$, the proposed experiment for $N_S=20$ can probe similarly small values of $\lambda'_{112}/m^^2_{\tilde{f}}$ as the main detector, while the latter is much more sensitive in the large couplings' limit.
None of FASER2, CODEX-b, and MATHUSLA can compete with the lepton-collider experiments shown here.

In the lower plot of Fig.~\ref{fig:Z2n1n1:BrLim2},  we plot the sensitivity reach of LAYCAST for $N_S=3$ and 20, and for $\mathcal{L}_Z=16$ ab$^{-1}$ (dashed) and 150 ab$^{-1}$ (solid).
We observe similar hierarchies between these curves as we find in the lower plot of Fig.~\ref{fig:HNL-other-exp}.

\subsection{Axion-like particles}
\label{subsec:axion-like particle}

\begin{figure}[t]
\centering
\includegraphics[width=\columnwidth]{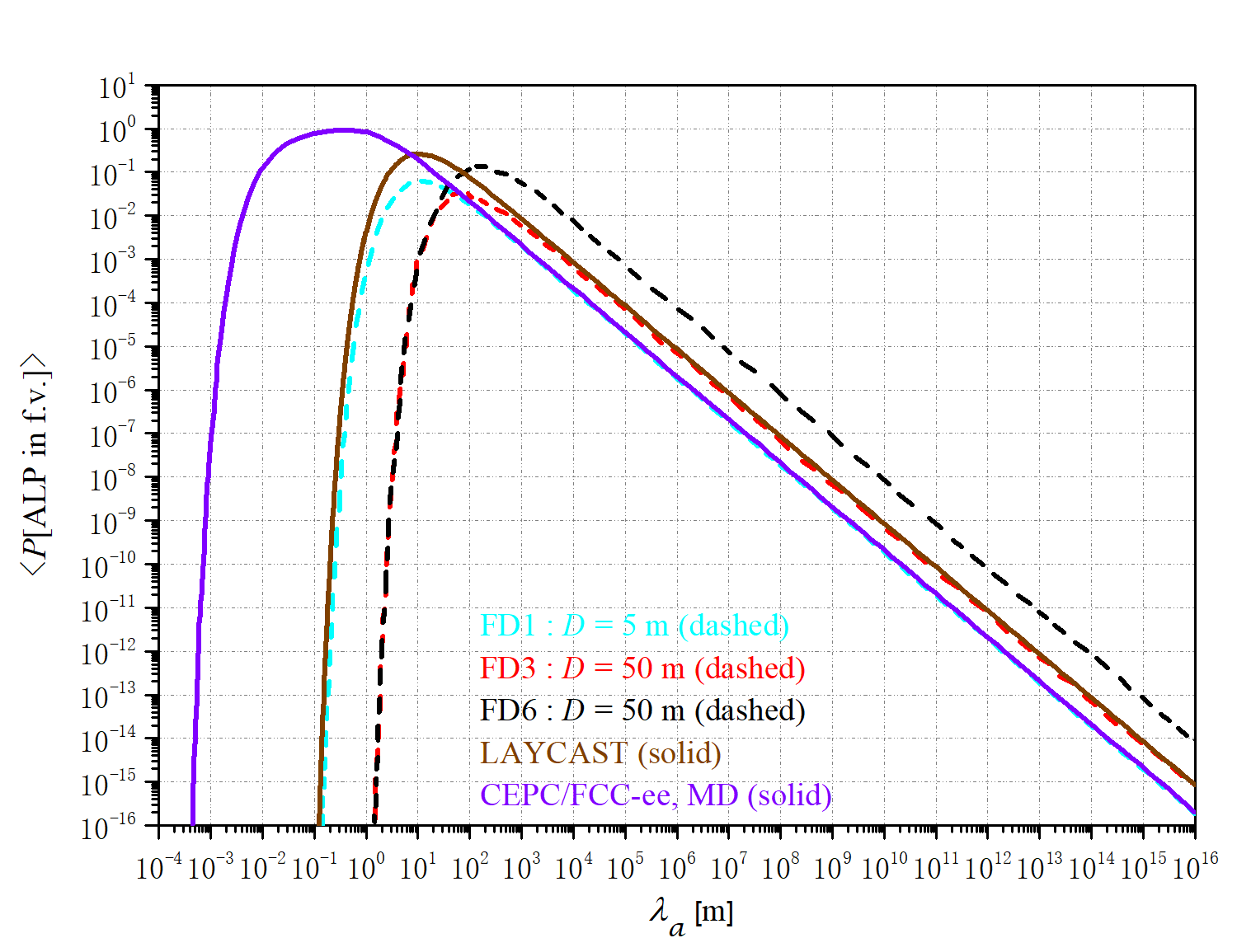}
\caption{The average decay probability of the ALP at the proposed experiment, the FCC-ee/CEPC's main detector, and FD1, FD3, and FD6~\cite{Tian:2022rsi}, as functions of the ALP's boosted decay length $\lambda_a$.}
\label{fig:ALP_avg}
\end{figure}

In Fig.~\ref{fig:ALP_avg} we plot $\langle P[\text{LLP}\text{ in f.v.}]\rangle$ as functions of the ALP's boosted decay length, $\lambda_a$, for not only LAYCAST, but also the FCC-ee/CEPC's main detector, and the proposed far-detector setups from Ref.~\cite{Tian:2022rsi}.
The results of the MD are calculated with the method described in this article, and those of  FD1, FD3, and FD6 are extracted from Ref.~\cite{Tian:2022rsi}.
We find the maximal reach in $\langle P[\text{LLP}\text{ in f.v.}]\rangle$ for LAYCAST is about 0.26 at $\lambda_a\sim 10$ m, roughly the same as in Fig.~\ref{fig:H2XX-Avg} and Fig.~\ref{fig:Z-NV-Avg}.
The main detector is more sensitive to smaller $\lambda_a$ values, and can reach the maximal average decay probability of 0.93 at $\lambda_a\sim 0.33$ m.
We observe that the average decay probability of the ALPs at LAYCAST is slightly better than that at FD1, especially for $\lambda_a \sim 10$ m, while FD6 is more sensitive to larger $\lambda_a$ values.
We note that as in the previous scenarios, these results are insensitive to the LLP's mass.

We present numerical results for the ALP model in three cases:
(1) $C_{\gamma Z}=0, C_{\gamma\gamma}\neq 0$, (2) $C_{\gamma Z}=C_{\gamma\gamma}\neq 0$, and (3) independent and non-vanishing $C_{\gamma Z}$ and $C_{\gamma\gamma}$ parameters.
For the first case where $C_{\gamma Z}=0$, we take into account the existing bounds on the coupling $C_{\gamma\gamma}/\Lambda$ extracted from the summary in Ref.~\cite{Antel:2023hkf}; for the other two cases with non-zero $C_{\gamma Z}$, these existing bounds, in principle, should only be loosely related as they are obtained under the assumption of the ALP coupled to the photons only.

\subsubsection{$C_{\gamma Z} =0$}
\label{subsubsec:cgz0}

\begin{figure}[t]
\centering
\includegraphics[width=\columnwidth]{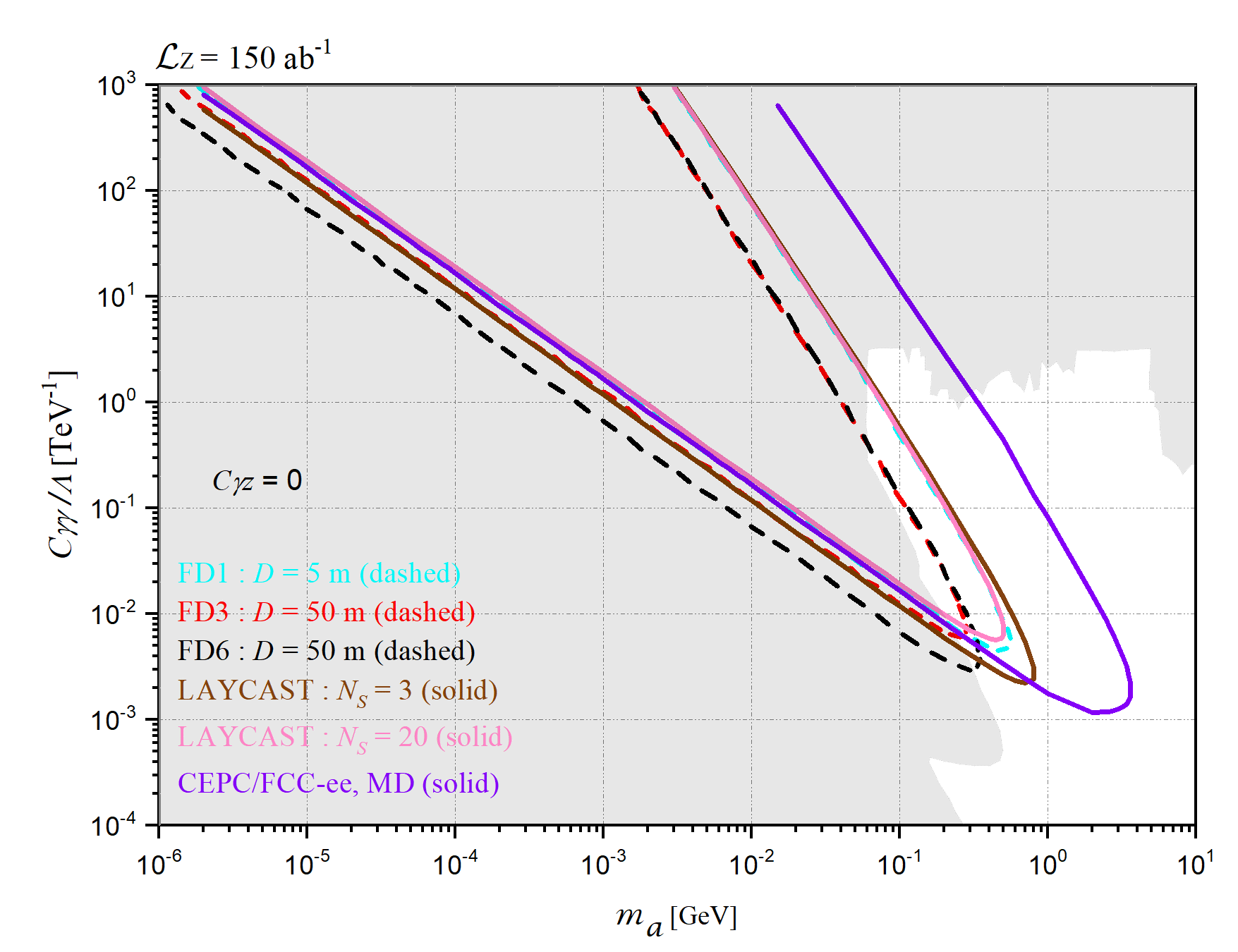}
\includegraphics[width=\columnwidth]{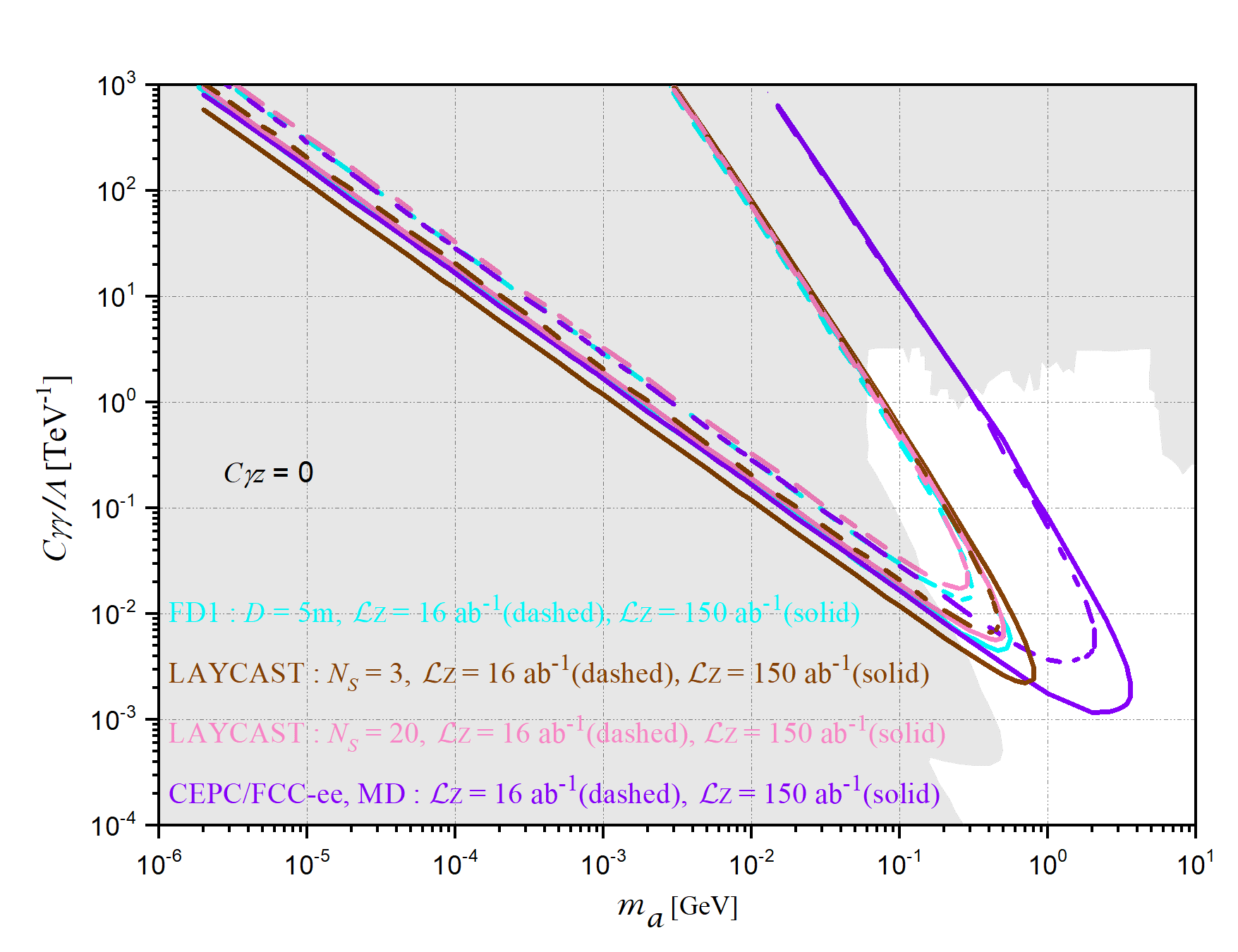}
\caption{\textit{Upper panel}: sensitivity reach of LAYCAST compared to that of the CEPC/FCC-ee's main detector, as well as those of FD1, FD3, and FD6 from Ref.~\cite{Tian:2022rsi}, for $\mathcal{L}_Z$ = 150 ab$^{-1}$ and $C_{\gamma Z}=0$, presented in the $C_{\gamma\gamma}/\Lambda$ vs.~$m_a$ plane.
\textit{Lower panel}: sensitivity reach of LAYCAST, FD1, as well as the main detector, for $\mathcal{L}_Z=16$ ab$^{-1}$ and 150 ab$^{-1}$.
In both plots, the gray area in the background is the presently excluded region of the parameter space $(m_a, C_{\gamma\gamma}/\Lambda)$, reproduced from Ref.~\cite{Antel:2023hkf} (see also the references therein).}
\label{fig:ALP1}
\end{figure}

Here, $C_{\gamma Z}$ assumed to be vanishing, and we have only two parameters: $C_{\gamma \gamma}/\Lambda$ and $m_a$.
Therefore, we present the sensitivity results in the plane $C_{\gamma\gamma}/\Lambda$ vs.~$m_a$ in Fig.~\ref{fig:ALP1}.
The gray area in both plots corresponds to the currently excluded region in the plane spanned by $m_a$ and $C_{\gamma\gamma}/\Lambda$; see Ref.~\cite{Antel:2023hkf} and the references therein~\footnote{In Ref.~\cite{Antel:2023hkf} the ALP-photon interaction Lagrangian has the term $-\frac{g_{a\gamma}}{4}a F_{\mu\nu}\tilde{F}^{\mu\nu}$ in its Eq.~(24), in comparison to our $e^2 C_{\gamma\gamma}\frac{a}{\Lambda}F_{\mu\nu}\tilde{F}^{\mu\nu}$ in Eq.~\eqref{eqn:ALP_Lag_afterSB}. This relation allows to convert the bounds on $g_{a\gamma}$ given in Ref.~\cite{Antel:2023hkf} to those displayed in Fig.~\ref{fig:ALP1}.}.

The upper plot of Fig.~\ref{fig:ALP1} shows the sensitivity reach of LAYCAST, the CEPC/FCC-ee's main detector, as well as FD1, FD3, and FD6 from Ref.~\cite{Wang:2019xvx}, for $\mathcal{L}_Z=150$ ab$^{-1}$.
We observe that the CEPC/FCC-ee's main detector can reach the largest ALP mass, while LAYCAST with $N_S=3$ and FD6 can probe smaller values of $C_{\gamma\gamma}/\Lambda$ for $m_a \lesssim \mathcal{O}(1)$ GeV.
FD1 and the new proposal with $N_S=20$ are found again to have almost identical discovery potentials.
These experiments all could probe $C_{\gamma\gamma}/\Lambda$ down to the order of $10^{-3}$ TeV$^{-1}$.
Taking into account the existing constraints shown in the background, we find that mainly the experiments located closer to the IP can probe relatively larger currently allowed regions where the ALPs are not so long lived (with large $m_a$ and $C_{\gamma\gamma}/\Lambda$).

In the lower plot, we illustrate the effect of the integrated luminosity on the sensitivity results of experiments including LAYCAST, the CEPC/FCC-ee's main detector, and FD1~\cite{Tian:2022rsi}.
The CEPC and FCC-ee benchmark integrated luminosities, $\mathcal{L}_Z=16$ ab$^{-1}$ (dashed) and 150 ab$^{-1}$ (solid) are considered.
We observe that even with the minor integrated luminosity of 16 ab$^{-1}$ large unexcluded parameter space can be probed by these experiments.

\subsubsection{$C_{\gamma Z} = C_{\gamma \gamma}$}\label{subsubsec:cgzEqcgg}

\begin{figure}[t]
\centering
\includegraphics[width=\columnwidth]{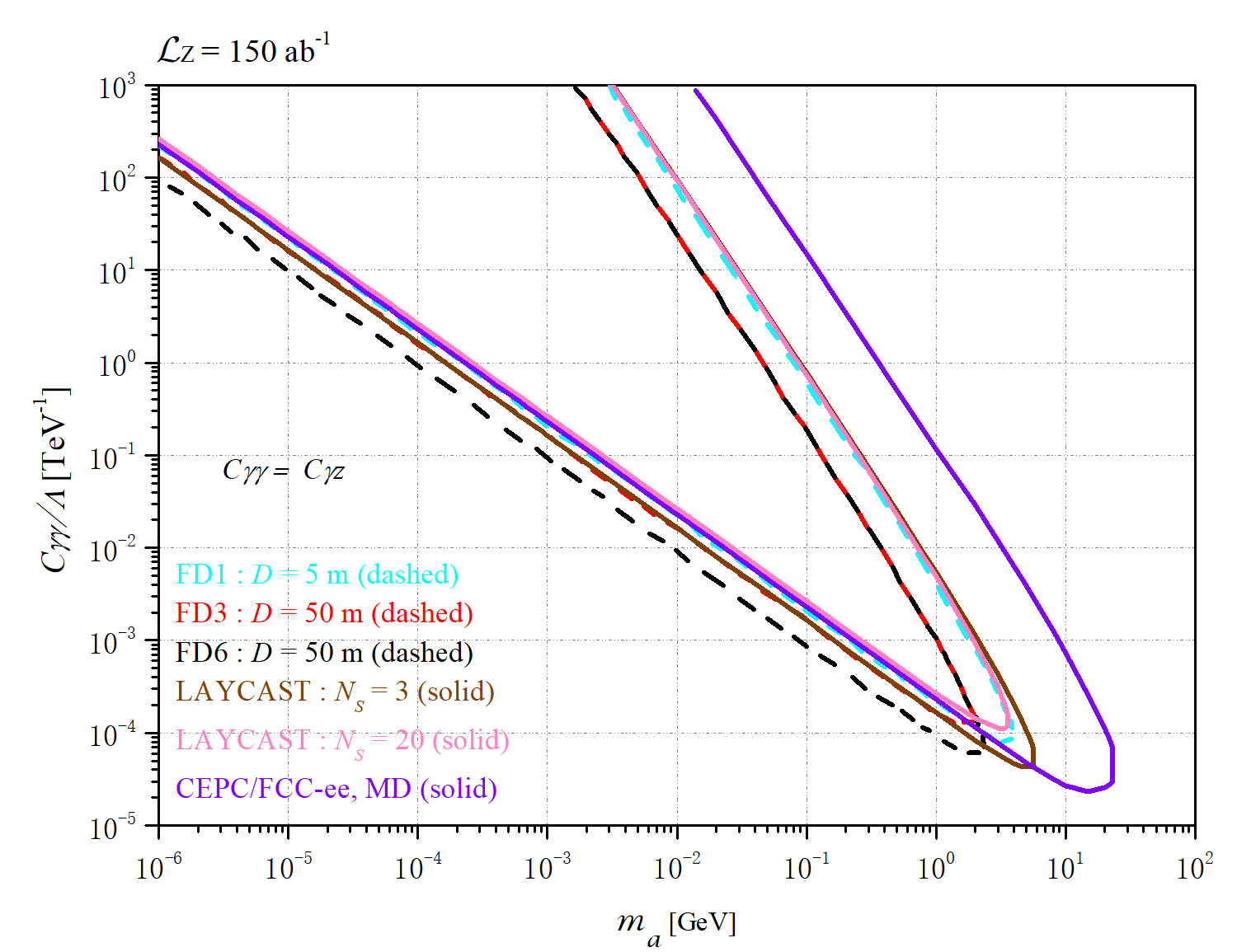}
\includegraphics[width=\columnwidth]{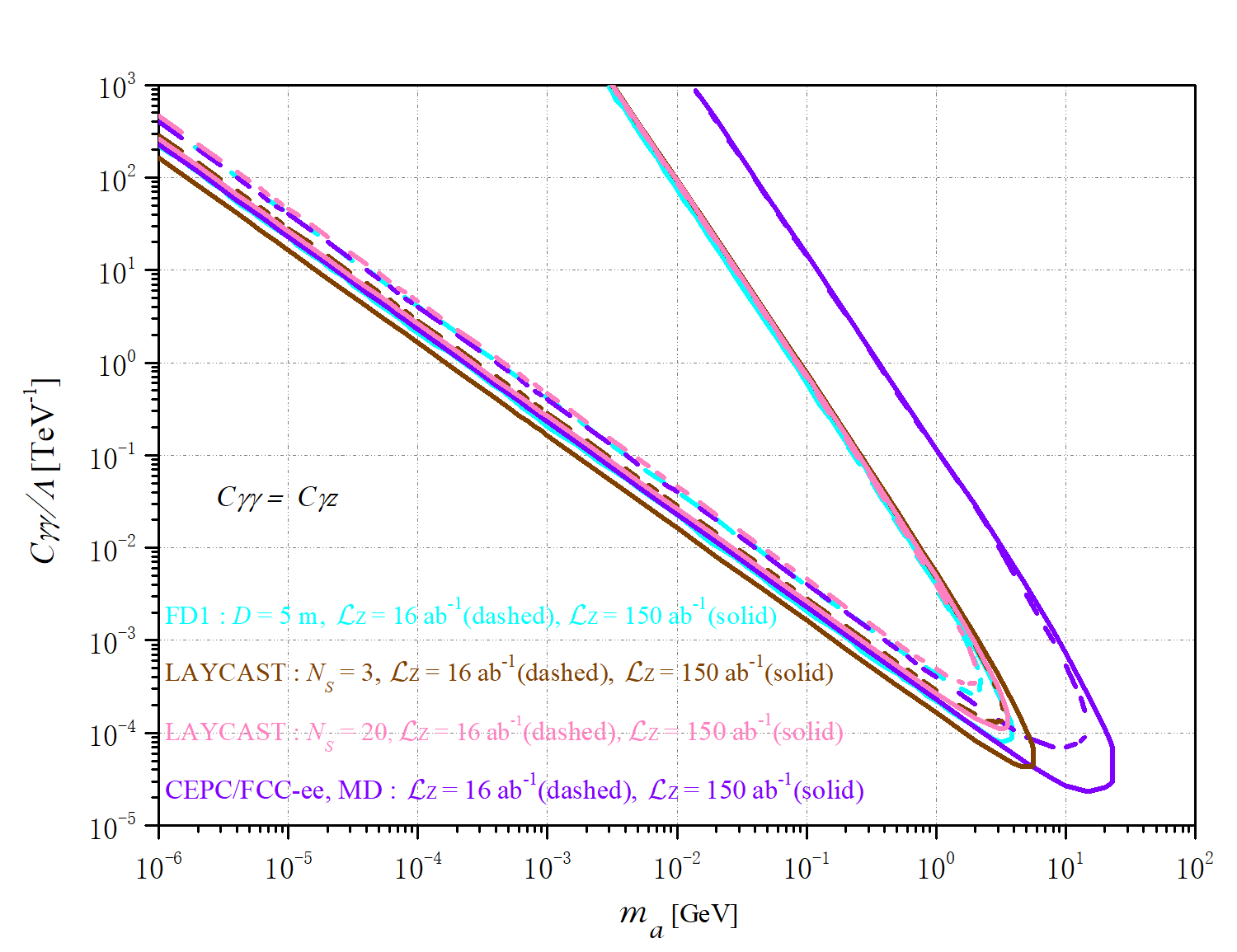}
\caption{The same plots as those in Fig.~\ref{fig:ALP1}, but for $C_{\gamma Z} = C_{\gamma \gamma}$.}
\label{fig:ALP2}
\end{figure}

In this scenario, the model parameters $C_{\gamma Z}$ and $C_{\gamma\gamma}$ are assumed to be equal.
We present the sensitivity results in Fig.~\ref{fig:ALP2}, in the $C_{\gamma\gamma}/\Lambda$ vs.~$m_a$ plane.

Fig.~\ref{fig:ALP2} is the same as Fig.~\ref{fig:ALP1}, except that now we switch on $C_{\gamma Z}$ and assume it is equal to $C_{\gamma\gamma}$.
The $C_{\gamma Z}$ coupling does not induce the ALP's decays, but it enhances the production rates of the ALP, cf.~Eq.~\eqref{eqn:ALP_prod_XS}.
We thus find generally stronger sensitivity results compared to those given in Fig.~\ref{fig:ALP1}, while the comparisons between the various experiments investigated remain unchanged.
Now these experiments can probe $C_{\gamma\gamma}/\Lambda$ down to the orders of magnitude $10^{-5}-10^{-4}$ TeV$^{-1}$ depending on the integrated luminosity.

\subsubsection{Independent and non-vanishing $C_{\gamma Z}$ and $C_{\gamma \gamma}$ parameters}
\label{subsubsec:cgzcgg}

In this case, we assume that $C_{\gamma\gamma}$ and $C_{\gamma Z}$ are decoupled and vary independently.
Given the three free parameters now, we will show sensitivity results in the $(C_{\gamma Z}/\Lambda, C_{\gamma\gamma}/\Lambda )$ plane for fixed ALP masses.

\begin{figure}[t]
\centering
\includegraphics[width=\columnwidth]{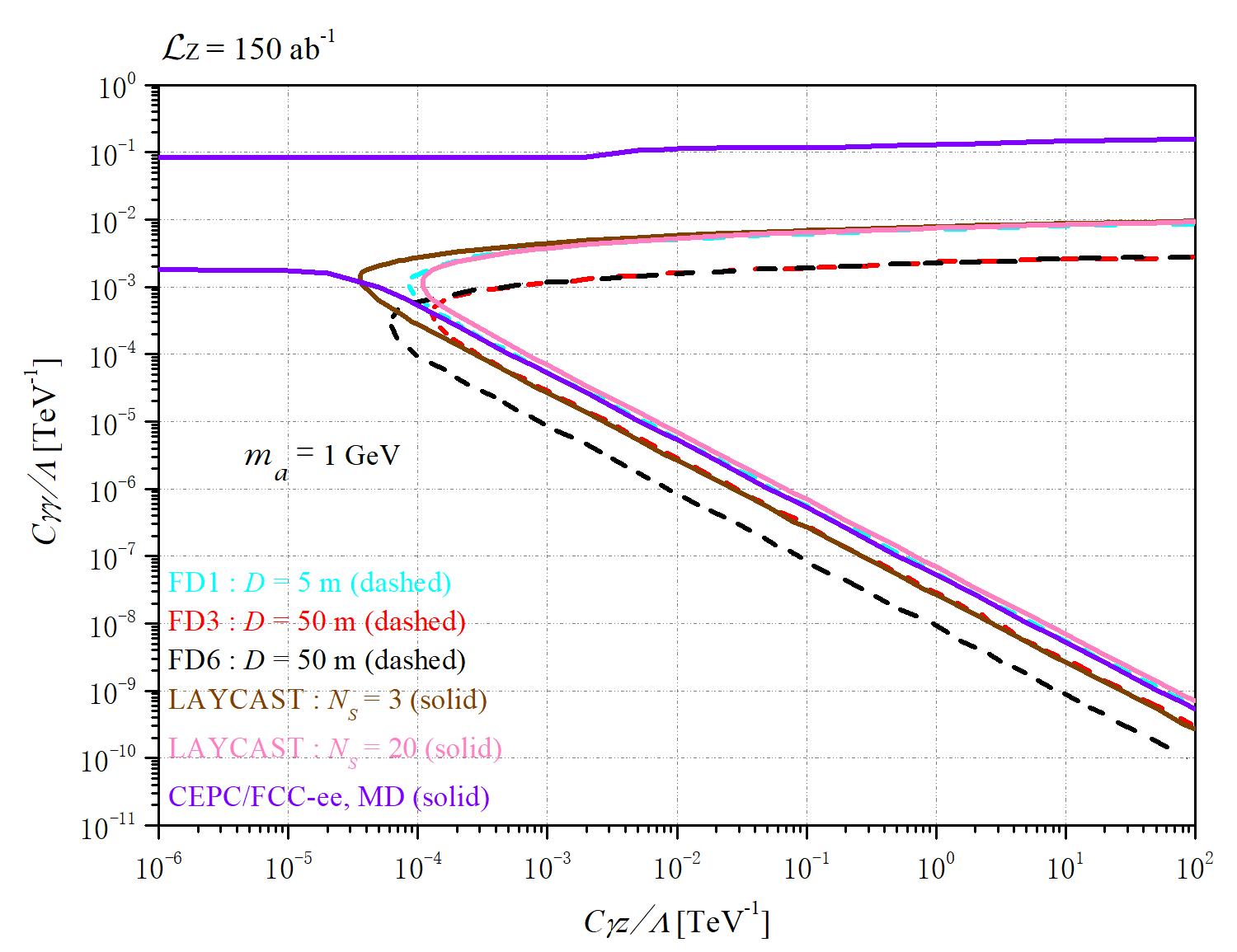}	 
\includegraphics[width=\columnwidth]{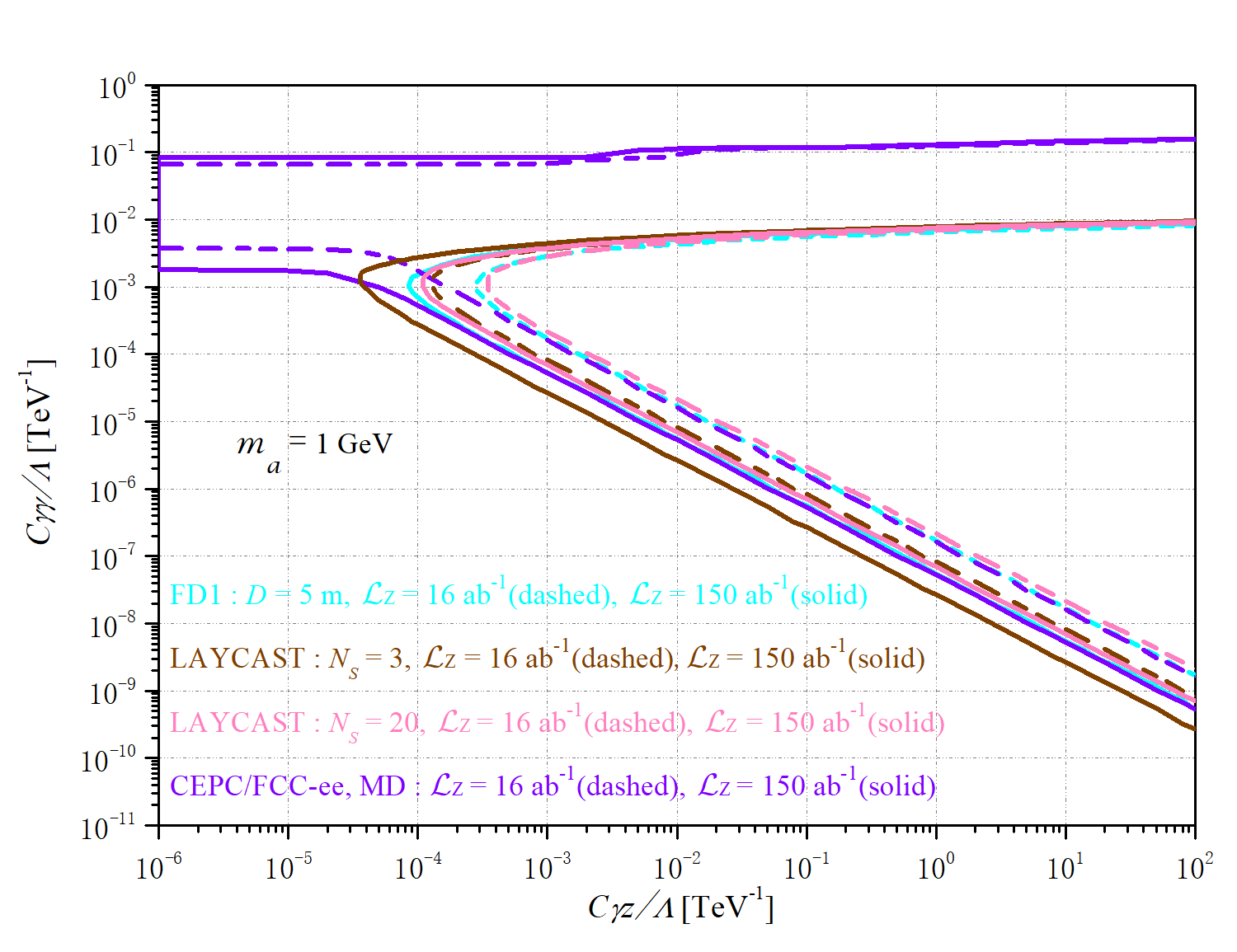}
\caption{\textit{Upper panel}: sensitivity reach of LAYCAST, the CEPC/FCC-ee's main detector, as well as FD1, FD3, and FD6 from Ref.~\cite{Tian:2022rsi} for $m_a=1$ GeV and $\mathcal{L}_Z=150$ ab$^{-1}$, shown in the plane $C_{\gamma\gamma}/\Lambda$ vs.~$C_{\gamma Z}/\Lambda$.
\textit{Lower panel}: sensitivity reach of LAYCAST, FD1, as well as the CEPC/FCC-ee's main detector with $m_a=1$ GeV, for $\mathcal{L}_Z=16$ ab$^{-1}$ and 150 ab$^{-1}$.}
\label{fig:ALP3}
\end{figure}

The upper plot of Fig.~\ref{fig:ALP3} shows the sensitivities of LAYCAST compared to those of the CEPC/FCC-ee's main detector, and FD1, FD3, and FD6 from Ref.~\cite{Tian:2022rsi}.
We fix $m_a$ at 1 GeV, and take $\mathcal{L}_Z=150$ ab$^{-1}$ as the benchmark integrated luminosity.
Among the far detectors including LAYCAST, FD6 can probe the smallest values of $C_{\gamma\gamma}/\Lambda$ for $C_{\gamma Z}/\Lambda$ down to about $7\times 10^{-5}$ TeV$^{-1}$.
LAYCAST with $N_S=20$ shows similar sensitivity reach as FD1, and for $N_S=3$ it can probe $C_{\gamma\gamma}/\Lambda$ for the smallest values of $C_{\gamma Z}/\Lambda$.
The CEPC/FCC-ee's main detector shows sensitivities to $C_{\gamma\gamma}/\Lambda$ for $C_{\gamma Z}/\Lambda$ below $5\times 10^{-5}$ TeV$^{-1}$ where all the far-detector experiments have no sensitivity; in fact, even for vanishing $C_{\gamma Z}$, the main detector is sensitive to $C_{\gamma\gamma}/\Lambda$ between $2\times 10^{-3}$ TeV$^{-1}$ and $9\times 10^{-2}$ TeV$^{-1}$, in good agreement with the upper plot of Fig.~\ref{fig:ALP1}.

The lower plot shows the change in the sensitivities for LAYCAST, the CEPC/FCC-ee's main detector, as well as FD1~\cite{Tian:2022rsi}, resulting from differences in the integrated luminosity $\mathcal{L}_Z=16$ ab$^{-1}$ and 150 ab$^{-1}$.

\begin{figure}[t]
\centering
\includegraphics[width=\columnwidth]{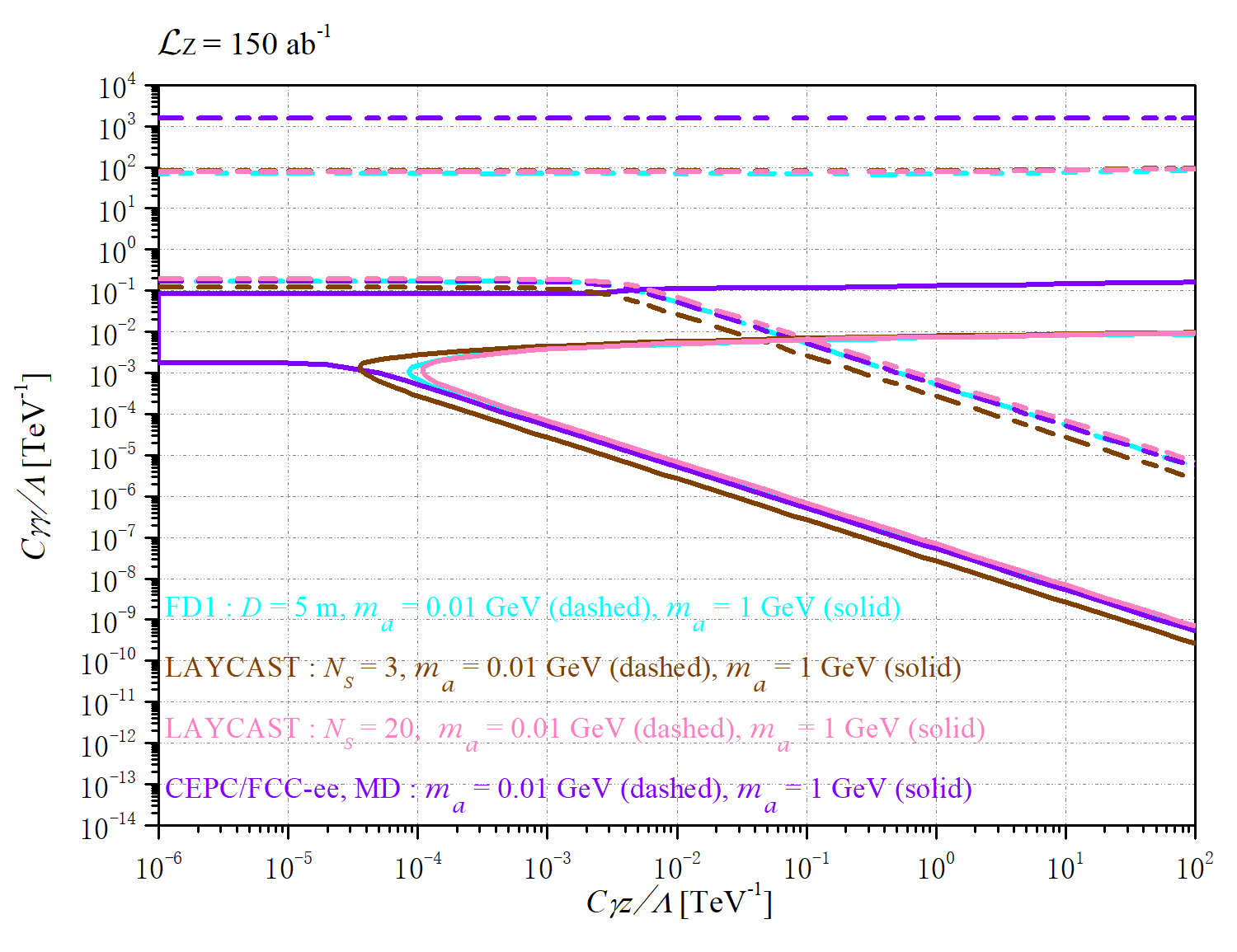}
\includegraphics[width=\columnwidth]{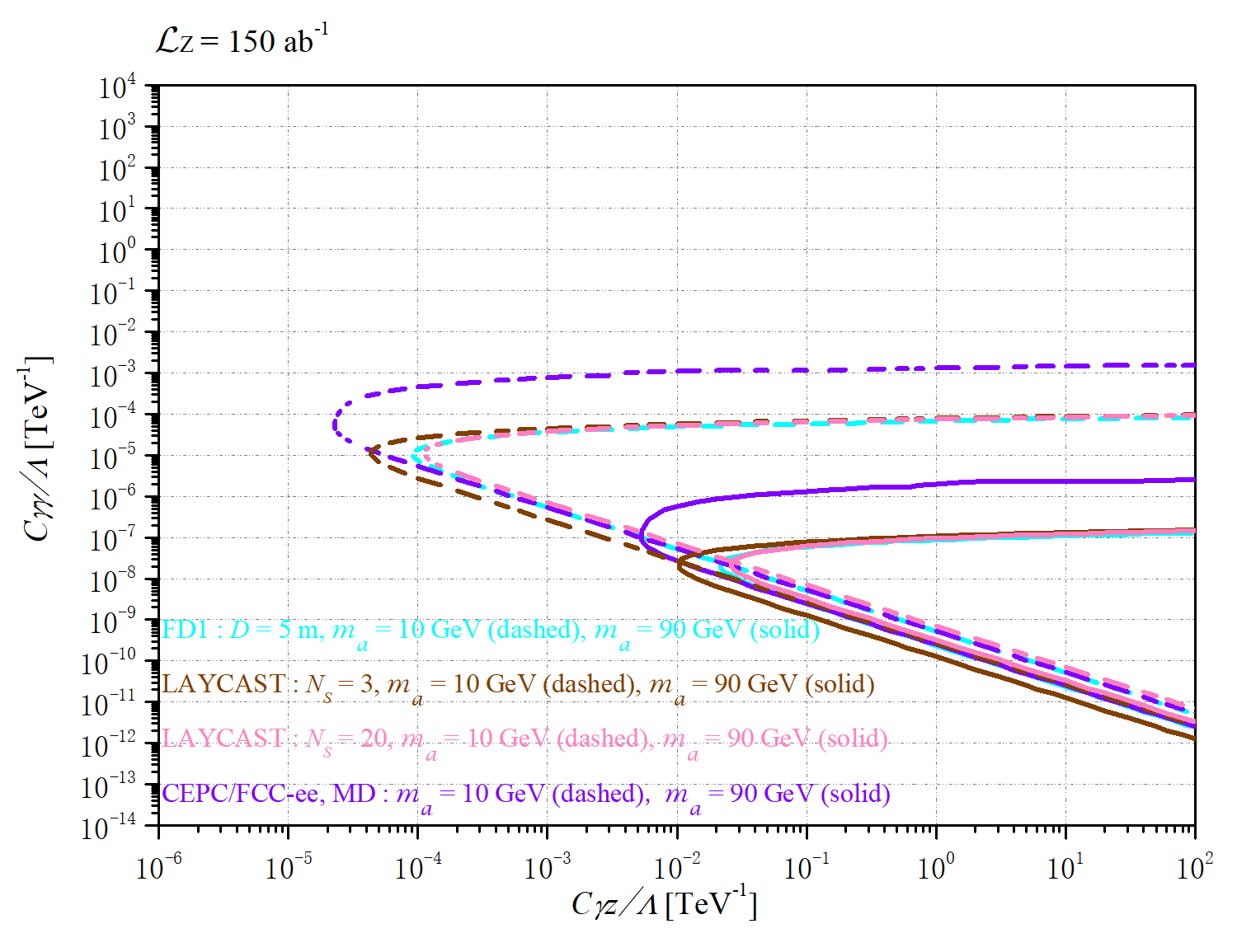}
\caption{ \textit{Upper panel}: sensitivity reaches of representative detectors and integrated luminosity of $\mathcal{L}_Z$ = 150 ab$^{-1}$ in the $C_{\gamma\gamma}/\Lambda$ vs $C_{\gamma Z}/\Lambda$ plane for $m_a=0.01$ GeV and 1 GeV.
\textit{Lower panel}: the same format for $m_a=10$ GeV and 90 GeV.}
\label{fig:ALP4}
\end{figure}

Finally, to understand how the sensitivities change with varying ALP mass, in Fig.~\ref{fig:ALP4}, we show two sensitivity plots in the plane $C_{\gamma\gamma}/\Lambda$ vs $C_{\gamma Z}/\Lambda$, for $m_a=0.01$ GeV and 1 GeV (\textit{upper panel}) and $m_a=10$ GeV and 90 GeV (\textit{lower panel}).
The integrated luminosity is fixed at 150 ab$^{-1}$ in these plots.
We take into account LAYCAST, FD1 (for which the results are extracted from Ref.~\cite{Tian:2022rsi}), and the CEPC/FCC-ee's main detector.
In each plot, for the lower (higher) mass value the sensitivity curves are plotted in the dashed (solid) style.

In the upper plot, we find that for $m_a=0.01$ GeV now all of the considered detectors show a funnel behavior at the small values of $C_{\gamma Z}/\Lambda$, and this can be explained with the same argument we provide above for the sensitivity curves of the CEPC/FCC-ee's main detector shown in Fig.~\ref{fig:ALP3}.
For $m_a=10$ GeV and 90 GeV, no such funnel is observed for the same reasons.

In general, these two plots show that for increasing ALP mass, the sensitive regions in general shift in the lower-right direction, requiring larger values of $C_{\gamma Z}/\Lambda$ and smaller values of $C_{\gamma\gamma}/\Lambda$.
This can be understood as follows.
For a heavier ALP, its decay width increases thus making the ALP more prompt and less likely to decay inside the fiducial volume.
To restore the decay probabilities in the fiducial volume, a smaller decay coupling, $C_{\gamma\gamma}/\Lambda$ is required; however, this would also decrease the production rate, which can be offset, in turn, by a larger value of $C_{\gamma Z}/\Lambda$.

\section{Conclusions}
\label{sec:conclusions}

In this work, we have proposed a dedicated LLP detector, LAYCAST, for future electron-positron colliders such as CEPC and FCC-ee, based on and intended to emulate the proposal from Ref.~\cite{Chrzaszcz:2020emg}.
Using the main detector of the circular lepton colliders as an effective veto for (most) SM background events, the idea of this work is to regard the space between the main detector and the surface of the main experimental cavern of CEPC and FCC-ee, as the fiducial volume where an LLP may decay with the decay products detected by the layered scintillators or RPCs to be installed on the inner surface of the cavern.
Besides, for load bearing and other reasons, we assume that LAYCAST would not be mounted on the floor of the experiment hall.

To good approximation we treat the fiducial volume as a cuboid, excluding the internal space occupied by the cylindrical main detector.
We assess the sensitivity reach of LAYCAST at CEPC and FCC-ee to a series of benchmark models with LLPs: a light scalar $X$ from exotic Higgs decays $h\to X X$ that arise in a minimal extension of the SM Higgs sector with an SM-singlet scalar field, an HNL $N$ from $Z$-boson decays $Z\to \nu_\alpha N$, the lightest neutralino in the RPV SUSY from $Z$-boson decays $Z\to \tilde{\chi}^0_1\tilde{\chi}^0_1$, and an ALP $a$ coupled to the SM photon and the $Z$-boson produced in direct scattering processes $e^- e^+ \to \gamma \, a$ via an $s$-channel photon or $Z$-boson.
We check these sensitivity results against existing bounds, and also compare them with those obtained for the CEPC/FCC-ee's main detector, LHC's far detectors, as well as proposed LLP detectors at future $e^- e^+$ colliders~\cite{Chrzaszcz:2020emg,Wang:2019xvx}.
Moreover, since the new proposed detector is capable of accepting LLPs from most directions, it is not easy to add additional shielding for the detector, and it may suffer from non-vanishing background events. 
We, thus, show the sensitivity reach at 95\% C.L.~for both 0 and 100 background events, corresponding to 3 and 20 signal events, respectively.

Compared to the FD1, FD3, and FD6 setups initially studied in Ref.~\cite{Wang:2019xvx} where zero background is (legitimately) assumed, we find that the sensitivity results of LAYCAST with $N_S=20$ are similar to those of FD1, while FD3 and FD6 are sensitive to relatively larger proper decay lengths $c\tau$ (corresponding to smaller mass or decay coupling) of the LLPs.
The latter two's stronger sensitivities are mainly due to their much larger volumes.

The main detectors at CEPC and FCC-ee, owing to their proximity to the IP, are most sensitive to LLPs with small $c\tau$, and can also cover the intermediate-$c\tau$ regime thanks to their large acceptance.
LAYCAST is instead most competitive at larger $c\tau$, and in the zero-background limit it can probe regions beyond the main detectors.
As a concrete SM background case, Appendix~\ref{app:KL_background} presents a dedicated estimate of the $K_L$ contribution from hadronic $Z$ decays (with $K_L$ decaying in the cavern volume before the first LAYCAST layer), showing that the MD$\times$FD requirements suppress it to a negligible level, especially for the Big layout.
If $\mathcal{O}(100)$ background events were nevertheless present for LAYCAST, it would not outperform a background-free main-detector search; conversely, if the main detector also suffers from sizable backgrounds, LAYCAST can again access to parameter space that becomes inaccessible to the main detector (Appendix~\ref{appendix:NS_main_detector}).

Compared to the far detectors at the LHC, LAYCAST is more sensitive for smaller $c\tau$ values.
In the long-lifetime regime, we find it can probe larger LLP masses or smaller decay couplings.
In addition, HECATE~\cite{Chrzaszcz:2020emg} focused on an HNL mixed with the muon neutrino.
We observe that the sensitivity reach of HECATE with 9 signal events shown in the plane $(m_N, |V_{\alpha N}|^2)$ is almost the same as that of LAYCAST studied in this work with 3 signal events.
This is mainly due to the more realistic setup of the new detector.

Furthermore, in Appendix~\ref{appendix:cavern_geometries}, we study numerically different sensitivity reaches between the geometries of the cavern that we use in this work~\cite{CEPC_cavern_old_slides} and the slightly larger, updated configuration~\cite{CEPCStudyGroup:2023quu}, and find that the effect is essentially small for such a variation in the cavern geometries.

Finally, we comment that we have fixed the distance between the IP and the cavern floor to be 5 m corresponding to a support holding the main detector, and have excluded the cavern floor from the locations where the scintillators can be installed.
Via numerical checks, we find that these two factors do not play an important role in the sensitivity-reach estimates.
These findings are explained in detail in Appendix~\ref{appendix:IP} and Appendix~\ref{appendix:underside}.

\appendix



\section{Effects of varying the cavern geometries}
\label{appendix:cavern_geometries}

As mentioned in Footnote~\ref{footnote:cavern_geometry}, the latest design of the CEPC cavern~\cite{CEPCStudyGroup:2023quu} employs dimensions (L$\times$W$\times$H) of $50\text{ m}\times 30\text{ m}\times 30\text{ m}$ with a support length of 12.15 m, slightly larger than the geometry setup used for numerical studies in this work ($40\text{ m}\times 20\text{ m}\times 30\text{ m}$ with a support of 5 m height~\cite{CEPC_cavern_old_slides}.)
In this appendix, we use the light scalar and the HNL scenarios to illustrate the effect on the average decay probability and the sensitivity reach at LAYCAST, arising from this difference in the cavern geometries.

\begin{figure}[H]
\centering
\includegraphics[width=\columnwidth]{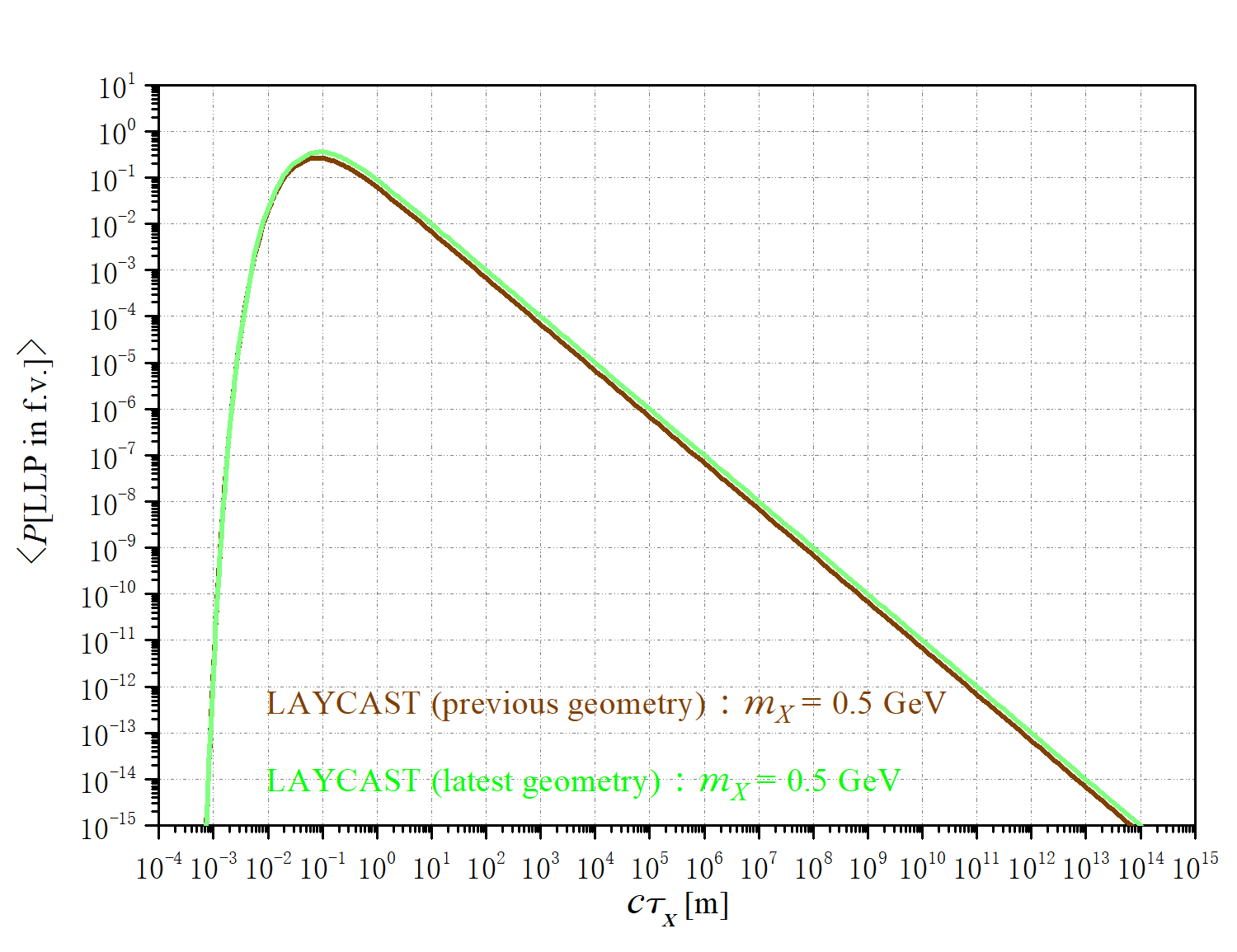}
\includegraphics[width=\columnwidth]{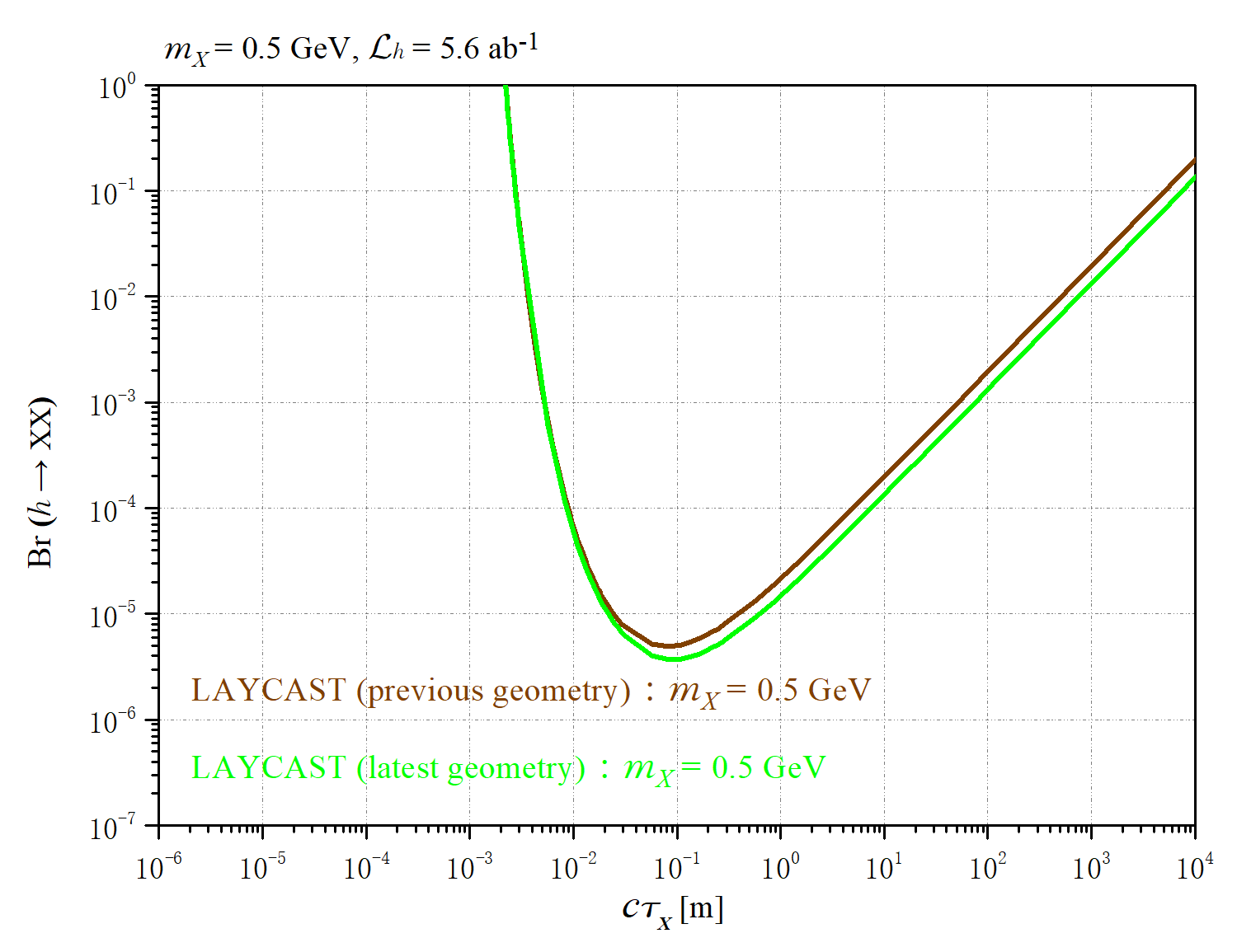}
\caption{\textit{Upper panel}: the average decay probability at LAYCAST as functions of the proper decay length $c\tau_X$, for our used CEPC cavern geometry (brown) and the latest geometry (green), in the light-scalar model with $m_X=0.5$ GeV.
\textit{Lower panel}: the sensitivity reach of LAYCAST in the $(c\tau_X, \text{Br}(h\to X X))$ plane for $m_X=0.5$ GeV, comparing the two geometrical configurations of the CEPC cavern.}
\label{fig:H-XX-cavern}
\end{figure}

In Fig.~\ref{fig:H-XX-cavern} we display two comparison plots shown in the plane $\langle P[\text{LLP}\text{ in f.v.}]\rangle$ vs.~$c\tau_X$ and Br$(h\to X X)$ vs.~$c\tau_X$, respectively, with $m_X=0.5$ GeV.
The brown curve corresponds to the geometrical setup of the CEPC experiment cavern we have used for the main results in the present work, and the green one to the latest design of the main cavern.
We find that with the latest design of the CEPC cavern, owing to its larger volume, LAYCAST is predicted to perform slightly better than with the setup we have been using.

\begin{figure}[H]
\centering
\includegraphics[width=\columnwidth]{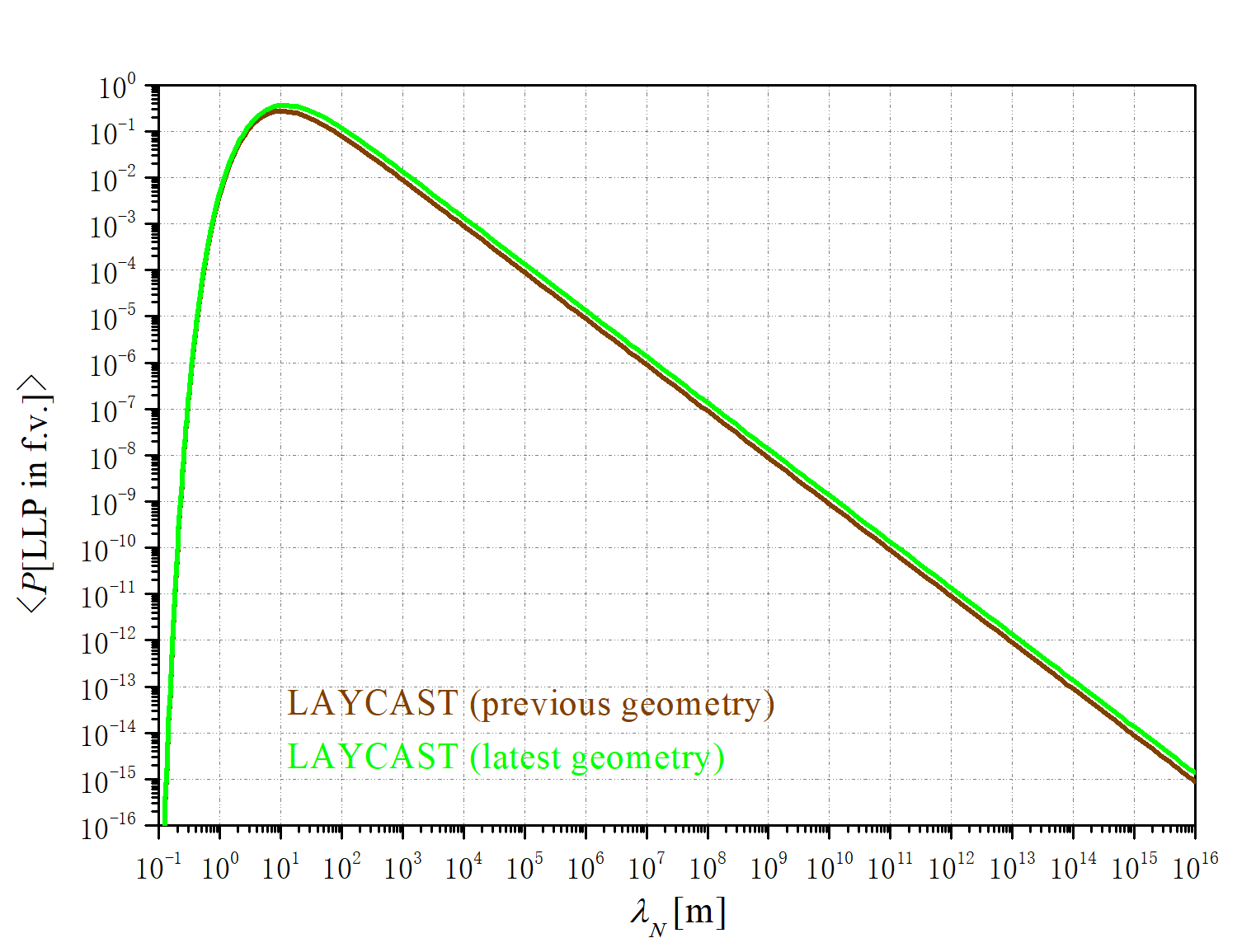}
\includegraphics[width=\columnwidth]{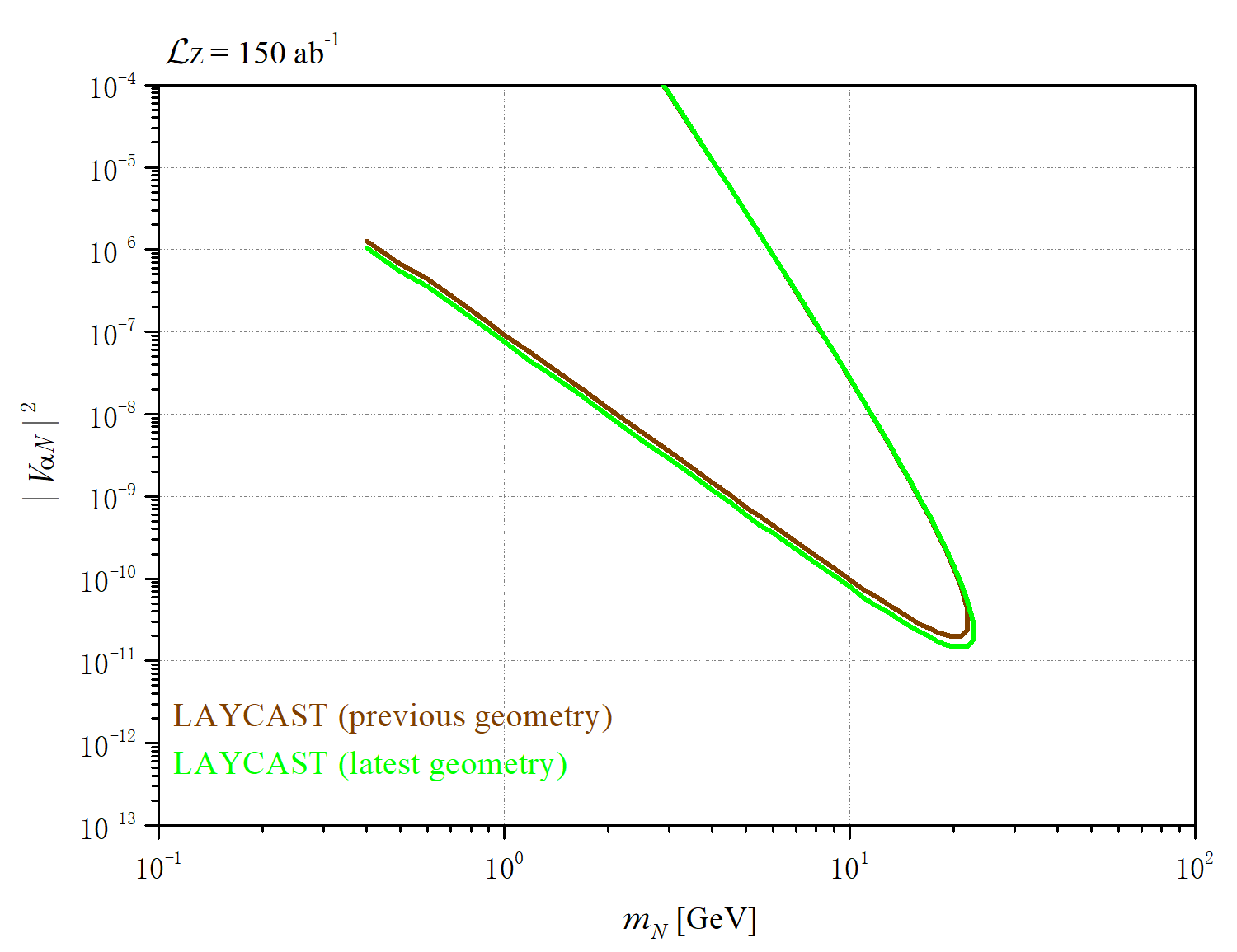}
\caption{\textit{Upper panel}: the average decay probability at LAYCAST as functions of the boosted decay length $\lambda$, for our used CEPC cavern geometry (brown) and the latest geometry (green), in the HNL model.
\textit{Lower panel}: the sensitivity reach of LAYCAST in the $(m_N, |V_{\alpha N}|^2)$ plane, comparing the two geometrical configurations of the CEPC cavern.}
\label{fig:Z-NV-cavern}
\end{figure}

Fig.~\ref{fig:Z-NV-cavern} makes the same comparison as Fig.~\ref{fig:H-XX-cavern}, except that it is now for the HNL model.
We draw the similar conclusion that with the two geometrical configurations of the CEPC cavern the proposed LAYCAST experiment is projected to have almost the same constraining power.

\section{Effects of varying the position of the collision point}
\label{appendix:IP}

Because of the load bearing and other factors in the main cavern, the beam collision point has been set at 5 m above the cavern's floor.
We take the ALP model as an example to showcase the effect of the relative position of the IP from the cavern floor, by numerically estimating the proportion of the long-lived ALPs traveling towards all six planes of the cuboid fiducial volume, for $l_{\text{support}}=5$ m and 15 m, where $l_{\text{support}}$ is the length of the vertical support.
For $l_{\text{support}}=15$ m, the IP is roughly in the middle of the experiment cavern.

\begin{table}[H]
\begin{tabular}{ccccccc}
	\hline
	\hline	
proportion & $x$ & $-x$	& $y$ & $-y$ & $z$ & $-z$\\
        \hline
$l_{\text{support}}=5$ m & 0.2148 & 0.2218 & 0.0617 & 0.3448 & 0.0805 & 0.0763\\
	\hline
$l_{\text{support}}=15$ m & 0.2577 & 0.2661 & 0.1290 & 0.1383 & 0.1043 & 0.1046\\
	\hline
	\hline    
\end{tabular}
\caption{The proportion of LLP events detected on each surface of the fiducial volume for the IP located at 5 m and 15 m above the floor of the experiment cavern.
$(-)x, (-)y,$ and $(-)z$ represent the surfaces lying in the corresponding side and perpendicular to the corresponding axis.}
\label{tab:CutEffiHLLHC}
\end{table}

The results are listed in Table~\ref{tab:CutEffiHLLHC}.
We observe that for the case with a smaller value of $l_{\text{support}}$, the proportion of the ALPs leaving the cavern through the floor (the $-y$ column) is larger and correspondingly fewer ALPs travel towards the other five surfaces.
This is mainly due to the larger solid-angle coverage between the IP and the cavern floor.
Therefore, the conclusion is for a longer vertical support, the cavern floor is less important to some extent.
We note also that these results are almost independent of the ALP mass.

\section{Effects of adding the cavern floor}
\label{appendix:underside}

\begin{figure}[h]
\centering
\includegraphics[width=\columnwidth]{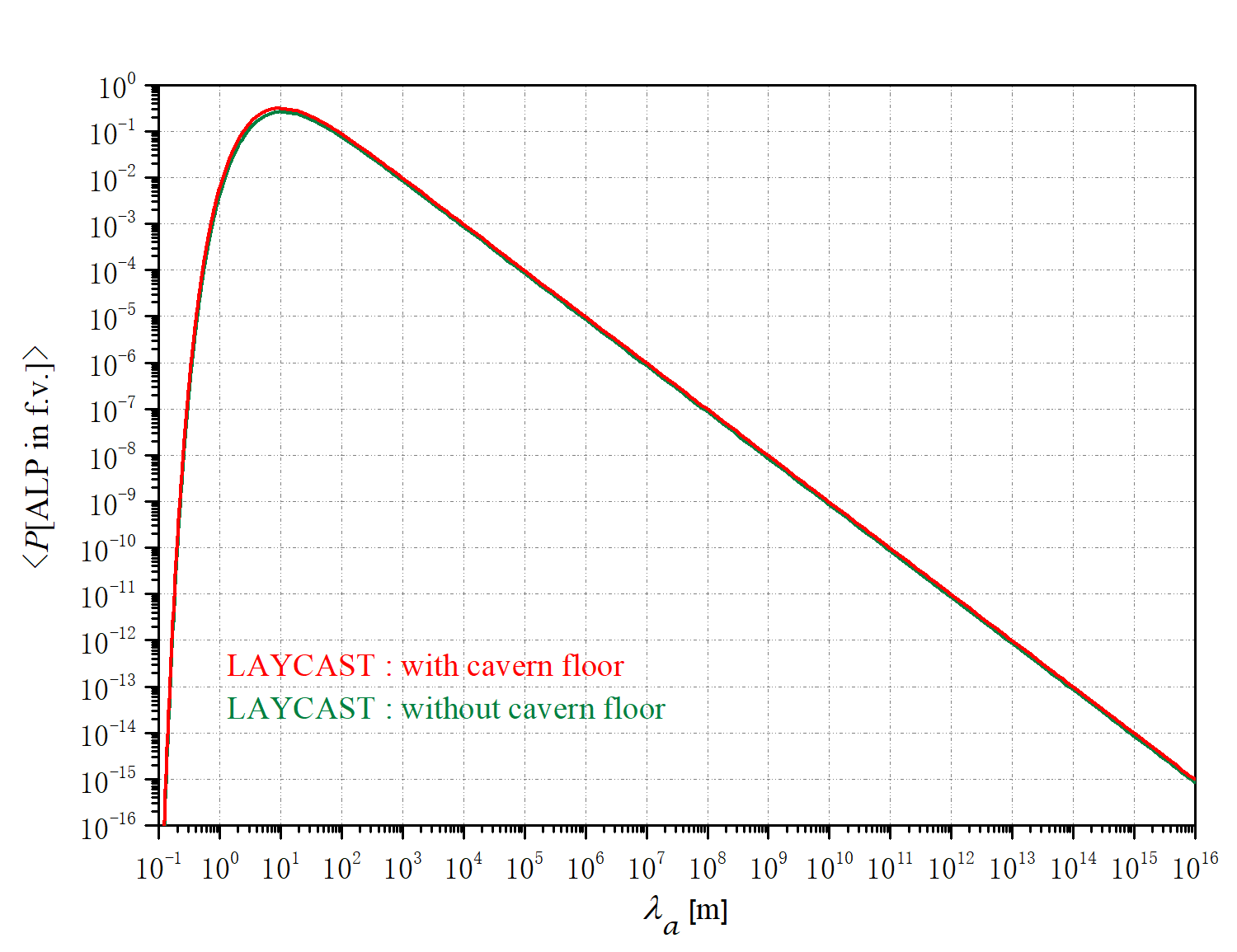}
\caption{The average decay probability of the ALP at LAYCAST as a function of the boosted decay length of the ALP $\lambda$, for the two cases of including and excluding the bottom surface of the cavern as part of the LAYCAST's fiducial volume.}
\label{fig:ALP10}
\end{figure}

In the main text, we have assumed that it is impossible to install detector materials on the cavern floor for the purpose of detecting LLPs.
In order to analyze the effect of this exclusion on the final sensitivity results, we check the average decay probabilities of the long-lived ALP for both cases of including and excluding the cavern floor as part of the fiducial volume of LAYCAST, and present the results in Fig.~\ref{fig:ALP10}, showing two curves for the two cases, respectively, where we have fixed the support length to be 5 m.
The two curves shown in the figure almost completely overlap with each other.
More concretely, the maximal average decay probability that can be achieved for $\lambda$ just below 10 m in both cases, is 0.32 and 0.26, for including and excluding the cavern floor, respectively.
We thus conclude that the cavern floor has only a small effect on the final sensitivities.
In addition, we remind that as discussed in Sec.~\ref{sec:results}, the mass of the LLP is unimportant in the considered benchmark scenarios for the plots of the average decay probability as functions of the lab-frame decay length of the LLP.

\section{$K_L$ background estimate}
\label{app:KL_background}

Neutral long-lived kaons from hadronic $Z$ decays can constitute a non-negligible SM background for LAYCAST.
The $K_L$ mean proper decay length is $c\tau\simeq 15.34~\mathrm{m}$~\cite{ParticleDataGroup:2024cfk}, so a sizeable fraction may traverse the main detector and decay in the cavern volume between the main-detector outer surface and the first LAYCAST layer.
Photons from such decays (directly or via $\pi^0$ production) may reach LAYCAST and mimic displaced, signal-like activity.
This appendix summarizes our estimate of the $K_L$ background for the two main-detector layouts.
As a concrete benchmark, we use $e^+e^-\to Z\to a\gamma$ with an ultra-long-lived ALP $a$, and illustrate how the combined main-detector $\times$ far-detector (MD$\times$FD) requirements suppress the $K_L$ contribution.

\begin{figure}[htbp]
\centering
\includegraphics[width=\columnwidth]{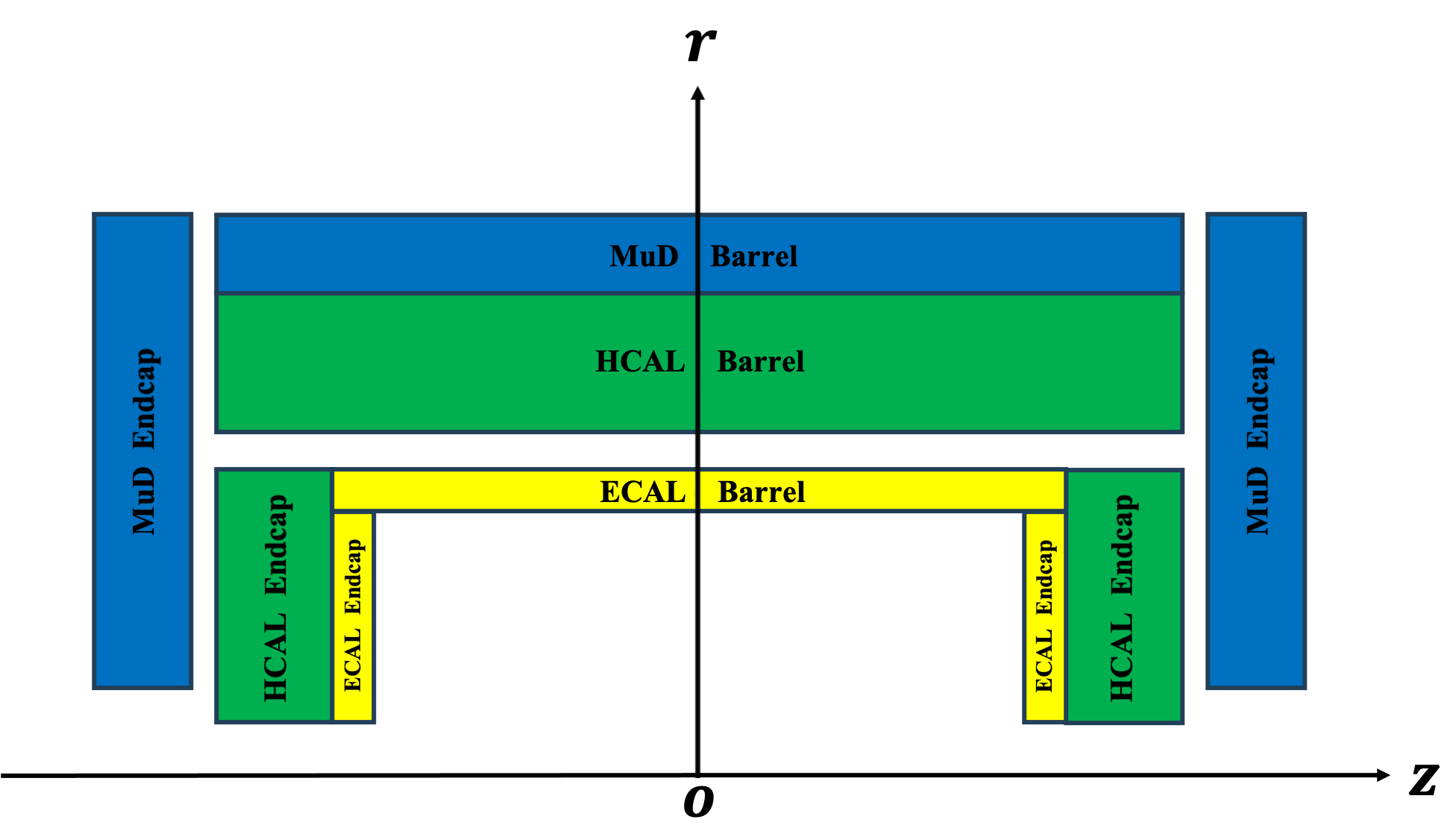}
\caption{The side cross-section views of the main detector.}
\label{fig:mainD}
\end{figure}

\paragraph*{Geometry of the main detector.}
Fig.~\ref{fig:mainD} shows a side cross section of the simplified main detector used in this study.
The electromagnetic calorimeter (ECAL), hadronic calorimeter (HCAL), and muon detector (MuD) are approximated as coaxial hollow cylinders.
Table~\ref{tab:mainD} summarizes the corresponding geometry inputs (barrel and endcaps) for the ``Small'' and ``Big'' detector designs; all dimensions are in mm.

\begin{table*}[htbp]
\squeezetable
\begin{ruledtabular}
\begin{tabular}{c c| c c| c c}
\multirow{2}{*}{Detector} & \multirow{2}{*}{Component} & \multicolumn{2}{c|}{Small design}  & \multicolumn{2}{c}{Big design} \\ 
\cline{3-6}
 &  & $Z$ range & Radius range & $Z$ range & Radius range \\ 
\hline
\multirow{3}{*}{ECAL} & Barrel       & $(-3370,\,3370)$   & $(2000, \,2330)$ & $(-2350, \,2350)$ & $(1810, \,2140)$  \\
 & Endcap Left  & $(-3370,\,-3040)$  & $(400, \,2000)$  & $(-2680, \,-2350)$ & $(329, \,2140)$  \\
 & Endcap Right & $(3040,\,3370)$    & $(400, \,2000)$  & $(2350, \,2680)$ & $(329, \,2140)$  \\
 \hline
\multirow{3}{*}{HCAL} & Barrel       & $(-4445,\,4445)$   & $(2585, \,3660)$ & $(-2680, \,2680)$ & $(2140, \,3380)$  \\
 & Endcap Left  & $(-4445,\,-3370)$  & $(400, \,2330)$  & $(-4143, \,-2680)$ & $(329, \,3380)$  \\
 & Endcap Right & $(3370,\,4445)$    & $(400, \,2330)$  & $(2680, \,4143)$ & $(329, \,3380)$  \\
 \hline
\multirow{3}{*}{MuD} & Barrel       & $(-4445,\,4445)$   & $(3660, \,4260)$ & $(-4143, \,4143)$ & $(4400, \,6080)$  \\
 & Endcap Left  & $(-5560,\,-4660)$  & $(650, \,4260)$  & $(-5863, \,-4143)$ & $(329, \,6080)$  \\
 & Endcap Right & $(4660,\,5560)$    & $(650, \,4260)$  & $(4143, \,5863)$ & $(329, \,6080)$  \\
\end{tabular}
\end{ruledtabular}
\caption{Subdetector geometry summary, and the unit is mm.}
\label{tab:mainD}
\end{table*}

\paragraph*{$K_L$ production at the $Z$ pole.}
The signal process $e^+e^-\!\to Z\!\to a\gamma$ features a single isolated photon in the MD and no additional visible activity, while the ultra-long-lived ALP $a$ traverses the MD and produces displaced photon activity in the LAYCAST layers after decaying in the cavern volume.

The relevant $K_L$ background is therefore not inclusive $Z\to$ hadrons production, but the rare topology in which
(i) a $K_L$ is produced at the IP and exits the MD, and
(ii) the accompanying hadronic system is sufficiently quiet in the MD such that the event passes the ``single-photon + veto'' requirement (i.e.\ the MD reconstructs only one isolated photon and no additional hadronic activity).
For this purpose we model the background with two parton-level channels:
(a)~$e^+e^-\to s\bar s$, where the recoil hemisphere fakes a photon candidate (merged $\pi^0\!\to\gamma\gamma$ or jet$\to\gamma$), and
(b)~$e^+e^-\to s\bar s\gamma$, where the photon arises from QED radiation and the hadronic activity is suppressed by the MD veto.
In both cases, any additional hadrons near the $K_L$ flight direction would fail the MD veto and are not part of the targeted background configuration.

As shown in Sec.~\ref{sec:results}, at the $Z$ pole we consider $\mathcal{L}_Z=16~\mathrm{ab}^{-1}$ and $150~\mathrm{ab}^{-1}$, corresponding to
$N_Z=7\times10^{11}$ and $5\times10^{12}$ on-shell $Z$ bosons, respectively.
To estimate the number of potentially relevant $K_L$ mesons we do \emph{not} normalize to the inclusive neutral-kaon multiplicities measured at LEP \cite{ParticleDataGroup:2024cfk},
since those correspond to generic hadronic $Z$ events with substantial hadronization activity, which are efficiently rejected by the MD veto.
Instead we introduce an effective probability $P_{\rm iso}(s/\bar s\to K_L)$ for an $s$ or $\bar s$ quark to yield an ``isolated'' $K_L$ configuration compatible with the MD single-photon requirement, and we conservatively take
$P_{\rm iso}=1/4$.
Using the PDG down-type average fraction $\mathrm{Br}(Z\to dd+ss+bb)/3\simeq 0.156$ as a proxy for $\mathrm{Br}(Z\to s\bar s)$~\cite{ParticleDataGroup:2024cfk}, we obtain
\begin{align}
N_{K}
&\approx 2\,N_Z\cdot \mathrm{Br}(Z\to s\bar s)\cdot P_{\rm iso}(s/\bar s\to K_L)
\;\approx\;0.078\,N_Z
\nonumber\\
&\approx
\left\{
\begin{array}{ll}
5.5\times10^{10}, & (\mathcal{L}_Z=16~\mathrm{ab}^{-1}),\\[2pt]
3.9\times10^{11}, & (\mathcal{L}_Z=150~\mathrm{ab}^{-1}).
\end{array}
\right.
\end{align}

For the polar angle $\theta$ (defined with respect to the beam axis), we adopt the standard even component
\begin{equation}
\frac{1}{N_K}\frac{\mathrm{d} N_K}{\mathrm{d} \cos\theta}=\frac{3}{8}\bigl(1+\cos^2\theta\bigr),
\label{eq:KL_ang}
\end{equation}
which corresponds to the familiar $(1+\cos^2\theta)$ dependence for $e^+e^-\to f\bar f$ in the ultrarelativistic limit~\cite{ParticleDataGroup:2024cfk}.
At the $Z$ pole a forward--backward term proportional to $\cos\theta$ can appear, but it largely cancels in symmetric acceptances and is neglected here for a baseline estimate.

Based on Eq.~\eqref{eqn:ALP_prod_diff_XS} \cite{Bauer:2018uxu},
the benchmark signal process $e^+e^-\!\to Z\!\to a\gamma$ follows the same $\theta$ weighting as Eq.~\eqref{eq:KL_ang}, 
so that central vs.~forward regions receive identical relative weights in signal and background integrals.

\subsection{Combined main/far-detector strategy for suppressing the $K_L$ background}
\label{app:KL_MDxFD}

Here, we summarize the principles and working formulae underlying the combined MD$\times$FD strategy
for suppressing the long-lived neutral-kaon background at a Tera-$Z$ $e^+e^-$ collider ($\sqrt{s}=91.2$~GeV).
In this appendix, the far detector (FD) refers to the LAYCAST layers installed on the cavern surface.
The signal is the two-body process $e^+e^-\!\to a\gamma$, where the axion-like particle $a$ is ultra-long-lived, traverses the main detector, and decays in the cavern volume, yielding displaced photon activity in LAYCAST.
In contrast, Standard-Model $K_L$ mesons are abundantly produced in hadronic $Z$ decays and can occasionally escape the main detector and decay in the same cavern volume,
producing photons predominantly via neutral-hadronic decays involving $\pi^0\!\to\gamma\gamma$ and, subdominantly, radiative modes~\cite{ParticleDataGroup:2024cfk}.

The MD$\times$FD selection combines 
(i) two-body photon kinematics and tight photon identification in the main detector with
(ii) a displaced two-photon vertex and energy--momentum consistency in the far detector, together with an event-level matching between the two subsystems
(e.g.\ pointing and timing consistency).

For the two-body signal at the $Z$ pole, the photon energy is sharply peaked,
\begin{equation}
E_\gamma^{\rm sig}=\frac{s-m_a^2}{2\sqrt{s}} \,.
\end{equation}
The main-detector signature is therefore a single isolated high-energy photon plus large missing energy with minimal additional activity.
Tight electromagnetic-cluster identification (e.g.\ shower-shape, conversion veto, pointing) and isolation, together with narrow windows in $E_\gamma$,
strongly suppress generic $Z\!\to$ hadrons events in which a $K_L$ is produced alongside sizeable hadronic activity.

The expected number of $K_L$ background events passing the full MD$\times$FD analysis can be factorized as
\begin{equation}
N_B \simeq N_{K} \,\cdot\,
P_{\rm sel}\,\cdot\,
P_{\rm det}\,\cdot\,
\mathrm{Br}\!\left(K_L \to \gamma\gamma + X\right),
\label{eq:KL_NB_factor}
\end{equation}
where $P_{\rm sel}$ denotes the efficiency of the combined MD and FD analysis cuts (including the MD--FD matching),
and $P_{\rm det}$ encodes the probability for a $K_L$ to (i) escape the main detector without being absorbed and (ii) decay inside the fiducial volume of the far detector.

We note that, once the di-photon $\gamma\gamma$ is detected by the far detector, the diphoton invariant mass $m_{\gamma\gamma}$ can be reconstructed.
For the signal, $m_{\gamma\gamma}$ peaks at $m_a$ (up to detector resolution and photon-energy scale).
For $K_L$-induced photon pairs, however, $m_{\gamma\gamma}$ is in general not correlated with $m_a$: photons typically originate from neutral-hadronic decays involving $\pi^0\!\to\gamma\gamma$ (leading to structures near $m_{\pi^0}$) or from broader multi-body final states, while the rare two-body mode $K_L\!\to\!\gamma\gamma$ yields a narrow peak near $m_{K}$~\cite{ParticleDataGroup:2024cfk}.
Therefore, requiring $m_{\gamma\gamma}$ to lie within a narrow window around $m_a$ can provide an additional handle to further suppress the $K_L$ background (except in the special case where $m_a$ happens to be close to $m_{\pi^0}$ or $m_K$).
We do not impose this requirement in the baseline estimate; our quoted $K_L$ background should thus be regarded as conservative.
This effect can be incorporated by multiplying Eq.~\eqref{eq:KL_NB_factor} by an additional factor $P_{m_{\gamma\gamma}}$, which we conservatively set to unity in this estimate.

\subsection{Selection of final state and kinematics in the main detector}
\label{subsec:kinematics}

This subsection evaluates the selection factor $P_{\rm sel}$ entering Eq.~\eqref{eq:KL_NB_factor}.
It quantifies the probability for a hadronic $Z$ event containing a $K_L$ to pass the
main-detector (MD) preselection and to satisfy kinematic criteria similar to the two-body signal.
We parameterize it as
\begin{equation}
P_{\rm sel}
=
\epsilon^{\rm MD}_{\text{pre}}
\,\cdot\,
\epsilon^{\rm MD}_{\text{kin}}\,,
\label{eq:selection_prob}
\end{equation}
where
\(\epsilon^{\rm MD}_{\text{pre}}\) is the probability that the hadronic event produces exactly one isolated photon
and exactly one candidate \(K_L\) in the MD with no other significant hadronic activity,
and \(\epsilon^{\rm MD}_{\text{kin}}\) encodes the efficiency for the surviving events to have photon and \(K_L\)
kinematics consistent with the two-body signal hypothesis.

\paragraph*{Event generation and MD preselection.}
To estimate the above factors, we generate hadronic background samples \(e^+e^-\!\to s\bar s\) and \(e^+e^-\!\to s\bar s\gamma\)
at \(\sqrt{s}=91.2~\mathrm{GeV}\) using MadGraph5\_aMC@NLO interfaced with PYTHIA8 for parton showering and hadronization~\cite{Alwall:2014hca,Sjostrand:2014zea},
followed by a fast detector simulation with Delphes using a CEPC configuration~\cite{deFavereau:2013fsa,Chen:2017yel}.
The photon in the \(s\bar s\gamma\) sample arises from QED radiation.
Although the $s\bar{s}\gamma$ background is more signal-like in kinematics, its cross section is much smaller,
\begin{equation}
\frac{\sigma_{s\bar s\gamma}}{\sigma_{s\bar s}}\simeq 4.4\times10^{-3}.
\end{equation}
We therefore define the total effective selection efficiency as the cross-section--weighted efficiency of the two components under the selection criteria,
\begin{equation}
\epsilon^{\rm MD}
=
\frac{\sigma_{s\bar s}\,\epsilon_{s\bar s}+\sigma_{s\bar s\gamma}\,\epsilon_{s\bar s\gamma}}
{\sigma_{s\bar s}+\sigma_{s\bar s\gamma}} \, .
\end{equation}

We first impose two simple MD-level preselection criteria: (i) exactly one reconstructed isolated photon; (ii) exactly one reconstructed jet (taken as a proxy for the recoiling hadronic system),
with no other significant activity.
Under these requirements the overall preselection efficiency factor can be written as the product of the isolated-photon selection efficiency and the fraction of events that contain a suitable $K_L$ candidate,
\begin{equation}
\epsilon^{\rm MD}_{\text{pre}}
=
\epsilon^{\rm MD}_{1\gamma}\cdot
P^{\rm MD}_{K_L,\text{pre}}
\simeq
(3.2\times10^{-2})\cdot(2.9\times10^{-1})
\simeq 9.3\times10^{-3}\,.
\end{equation}

\begin{figure}[h]
\centering
\includegraphics[width=\columnwidth]{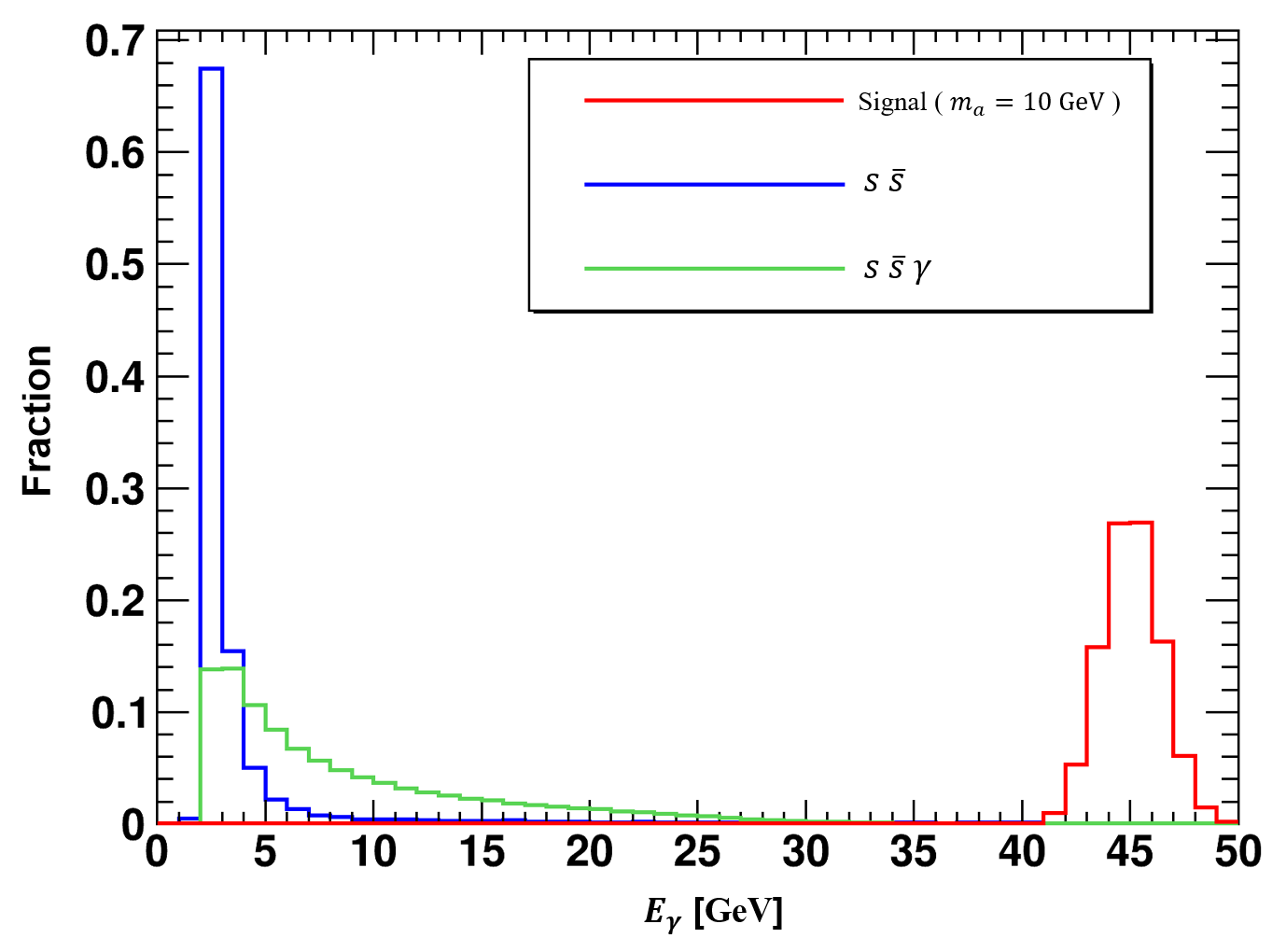}
\caption{
Photon-energy spectrum in the main detector for the signal $e^+e^-\to a\gamma$ with $m_a=10~\mathrm{GeV}$ and the background samples $e^+e^-\to s\bar s$ and $e^+e^-\to s\bar s\gamma$ at $\sqrt{s}=91.2~\mathrm{GeV}$. Distributions are shown after the MD preselection and normalized to unit area for shape comparison.
}
\label{fig:Egamma10}
\end{figure}

\begin{figure}[h]
\centering
\includegraphics[width=\columnwidth]{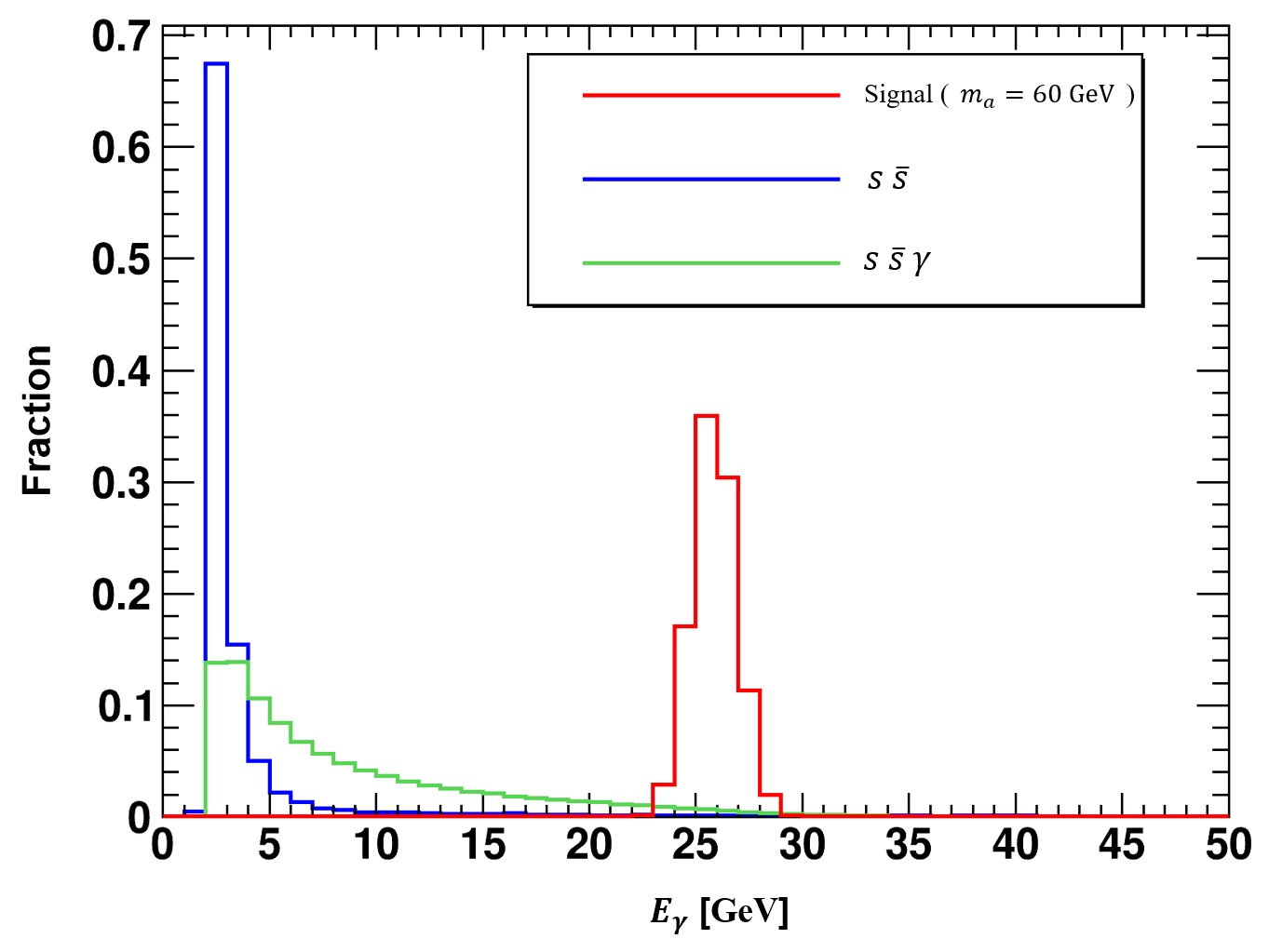}
\caption{
Same as Fig.~\ref{fig:Egamma10}, but for the signal mass $m_a=60~\mathrm{GeV}$.
}
\label{fig:Egamma60}
\end{figure}

\paragraph*{Kinematic consistency with the signal.}

Figures~\ref{fig:Egamma10} and \ref{fig:Egamma60} show the photon-energy spectra for the signal and the two background samples.
As expected from two-body kinematics, the signal peaks near $E_\gamma\simeq 45~\mathrm{GeV}$ ($m_a=10~\mathrm{GeV}$) and $25~\mathrm{GeV}$ ($m_a=60~\mathrm{GeV}$).
We define $\epsilon^{\rm MD}_{\text{kin}}$ via an $E_\gamma$ window around the signal peak,
\begin{equation}
\epsilon^{\rm MD}_{\text{kin}} \simeq
\begin{cases}
1.2\times10^{-2}, & 40<E_\gamma<50~\mathrm{GeV}\quad (m_a=10~\mathrm{GeV}),\\
3.3\times10^{-2}, & 20<E_\gamma<30~\mathrm{GeV}\quad (m_a=60~\mathrm{GeV}).
\end{cases}
\end{equation}
Combining the above factors gives
\begin{equation}
P_{\rm sel}\simeq
\begin{cases}
1.11\times10^{-4}, & m_a=10~\mathrm{GeV},\\
3.06\times10^{-4}, & m_a=60~\mathrm{GeV}.
\end{cases}
\end{equation}

\subsection{Detection in the FD}
\label{subsec:KL_FD_detection}

\paragraph*{Traversal of the main detector (survival).}
A $K_L$ produced at the IP must traverse the ECAL, HCAL, and muD before entering the cavern region. 
We model hadronic interactions in detector material with an exponential survival probability,
\begin{align}
P_{\rm surv}(\theta) &= \exp\!\big[-d_{\rm E}(\theta)-d_{\rm H}(\theta)-d_\mu(\theta)\big], \\
d_i(\theta) &= \frac{L_i(\theta)\, f_{\rm T}}{\lambda_i}\,, \qquad i\in\{{\rm E,H},\mu\},
\end{align}
where $L_i(\theta)$ are the path lengths through each subdetector along polar angle $\theta$, 
obtained piecewise from
Fig.~\ref{fig:mainD} and Table~\ref{tab:mainD}. 
They depend on the angle $\theta$ only, since the MD is azimuthally symmetric.
The factor $f_{\rm T}$ is an effective material fill factor (accounting for gaps and non-absorbing volumes),
and $\lambda_i$ are effective absorption lengths for neutral-kaon interactions, taken as
$\lambda_{\rm E}=9.6~\mathrm{cm}$, $\lambda_{\rm H}=17~\mathrm{cm}$, and $\lambda_{\mu}=16.8~\mathrm{cm}$
for tungsten, steel, and iron, respectively~\cite{CEPCStudyGroup:2018ghi}.
We normalize the optical depths at $\theta=\pi/2$ as $d_{\rm E}=0.875$ and $d_{\rm H}=4.7$ for both layouts, and use $d_\mu=2.39~(6.7)$ with $f_{\rm T}=0.67$ for the Small (Big) design.
Integrating $P_{\rm surv}(\theta)$ over the central polar range $0.17\le\theta\le\pi-0.17$ gives the fraction of kaons that exit the main detector.

\paragraph*{Decays in the cavern before reaching LAYCAST (geometry, lifetime, and azimuthal acceptance).}
Let $D_1(\theta)$ be the distance from the IP to the \emph{outer} surface of the main detector, and
$D_2(\theta,\varphi)$ the distance to the \emph{first LAYCAST layer} along direction $(\theta,\varphi)$.
Because LAYCAST is not azimuthally symmetric and does not provide full azimuthal coverage, $D_2$ depends on $\varphi$, and 
for directions without instrumentation---in particular, trajectories pointing toward the cavern floor where no LAYCAST layers are installed---we set the decay probability to zero.
At the $Z$ pole the $K_L$ decay length is
\begin{equation}
\lambda_K
=\beta_K\gamma_K c\tau_K 
=\frac{p_{K}}{m_K} c\tau_K 
\approx \frac{m_Z\,c\tau_K}{2m_K}
\approx 1.405~\mathrm{km}\,.
\end{equation}
The probability for a $K_L$ to decay in the cavern volume between the main detector and the first LAYCAST layer is
\begin{align}
P_{\rm decay}(\theta,\varphi)
&= e^{-D_1(\theta)/\lambda_K}-e^{-D_2(\theta,\varphi)/\lambda_K}
\; \\
&\approx\; \frac{D_2(\theta,\varphi)-D_1(\theta)}{\lambda_K}\,,
\end{align}
where the linearization holds when $\lambda_K\gg D_{1,2}$.

\begin{figure}[htbp]
\centering
\includegraphics[width=\columnwidth]{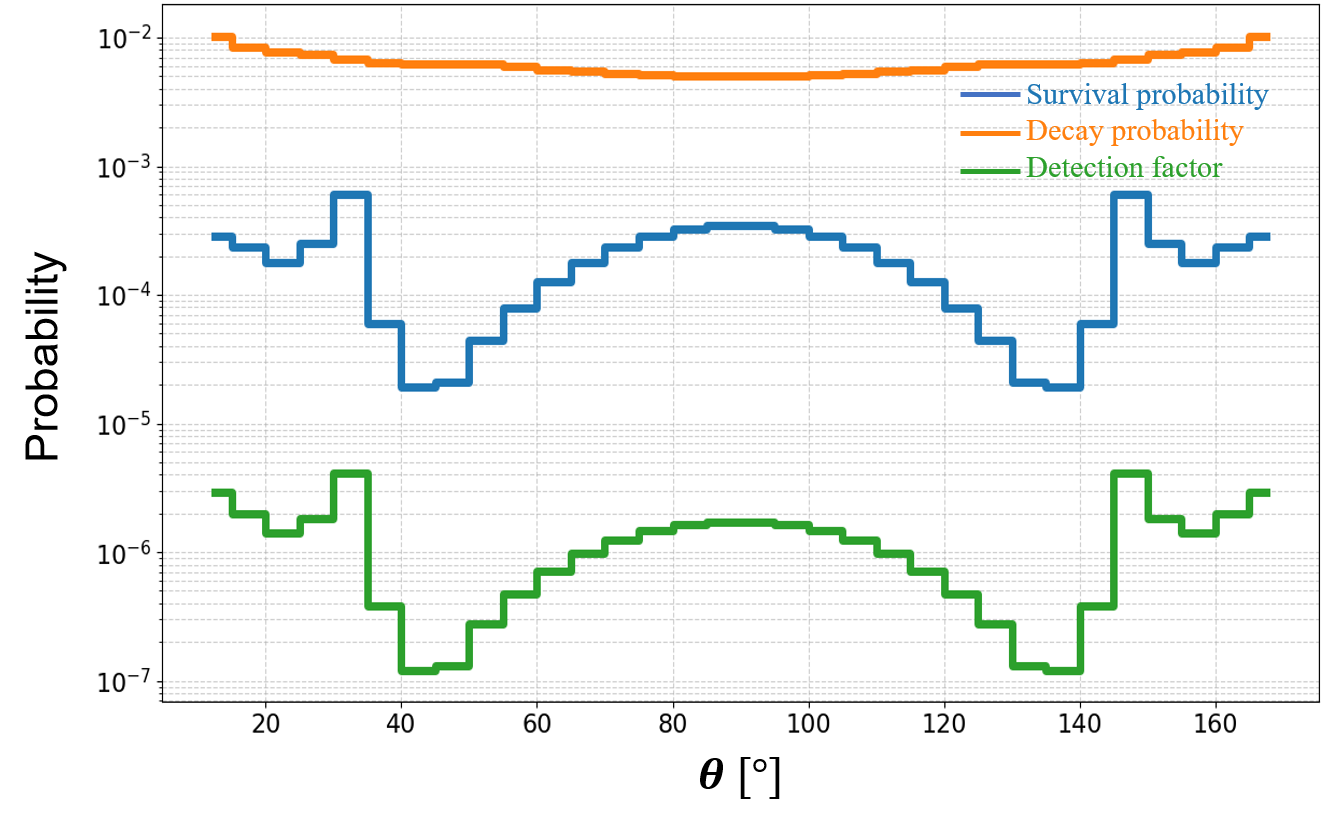}
\caption{Small main-detector layout: angular dependence of the $K_L$ survival probability $P_{\rm surv}$, cavern-decay probability $P_{\rm decay}$, and their product entering $P_{\rm det}$.}
\label{fig:KL_small}
\end{figure}

\begin{figure}[htbp]
\centering
\includegraphics[width=\columnwidth]{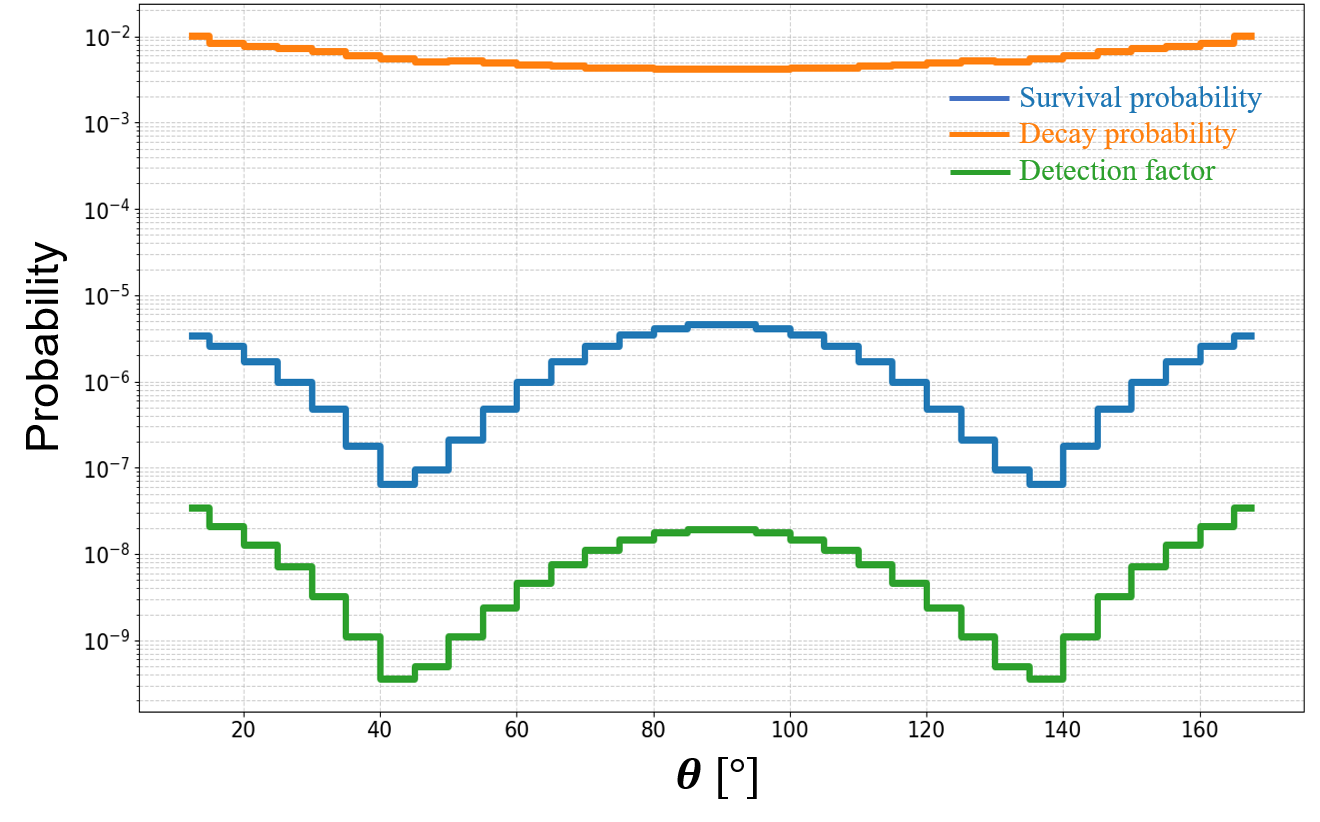}
\caption{Big main-detector layout: same as Fig.~\ref{fig:KL_small}.}
\label{fig:KL_big}
\end{figure}

The FD detection factor is obtained by combining the survival probability in the main detector with the probability to decay in the cavern region before the first LAYCAST layer, and integrating over the instrumented solid angle.
We evaluate this integral numerically via Monte Carlo integration: we generate $2\times10^{8}$ $K_L$ directions with the polar-angle distribution in Eq.~\eqref{eq:KL_ang} and a uniform azimuthal angle $\varphi\in[0,2\pi)$, apply the LAYCAST azimuthal acceptance, and average the product $P_{\rm surv}(\theta) \cdot P_{\rm decay}(\theta,\varphi)$.
The resulting $\theta$ dependences for the Small and Big layouts after integrating $\varphi$ are shown in Figs.~\ref{fig:KL_small} and~\ref{fig:KL_big}, respectively.

We integrate the above results to obtain
\begin{equation}
P_{\rm det} \simeq
\begin{cases}
1.21\times10^{-6}\quad (\text{Small MD}),\\
8.55\times10^{-9}\quad (\text{Big MD})\,.
\end{cases}
\end{equation}

Finally, to obtain a $\gamma\gamma$-like background we require at least one $\pi^0$ in the $K_L$ decay chain.
As a simple estimate we retain the dominant $\pi^0$-rich hadronic modes~\cite{ParticleDataGroup:2024cfk},
\begin{equation}
\mathrm{Br}(K_L\to \gamma\gamma +X) \approx
\mathrm{Br}(K_L\to \pi^0+X)
\approx\;0.32\,.
\end{equation}

\subsection{Final background estimate}
\label{subsec:KL_final}

We define the net per-$K_L$ probability to enter the $\gamma\gamma$ background category as
\begin{equation}
P_{\rm net} \;\equiv\; P_{\rm sel}\cdot P_{\rm det}\cdot \mathrm{Br}\!\left(K_L \to \gamma\gamma + X\right).
\label{eq:KL_Pnet}
\end{equation}
The expected number of background events then follows from Eq.~\eqref{eq:KL_NB_factor} as
$N_B = N_{K_L}\,\cdot P_{\rm net} \approx 0.078 \, N_Z \cdot P_{\rm net}$.
Table~\ref{tab:KL_final} summarizes the resulting $P_{\rm net}$ and $N_B$ for the two main detector layouts and benchmark ALP masses.

\begin{table}[htbp]
\centering
\begin{tabular}{cccccc}
\hline\hline
Layout & $m_a$ [GeV] & $P_{\rm net}$ & $\mathcal{L}_Z$ [ab$^{-1}$] & $N_B$ \\
\hline
\multirow{4}{*}{Small} & \multirow{2}{*}{10} & \multirow{2}{*}{$4.3 \times 10^{-11}$} & 16 & 2.4 \\
& & & 150 & 17 \\
 & \multirow{2}{*}{60} & \multirow{2}{*}{$1.18 \times 10^{-10}$} & 16 & 6.5 \\
& & & 150 & 46 \\
\hline
\multirow{4}{*}{Big} & \multirow{2}{*}{10} & \multirow{2}{*}{$3.04 \times 10^{-13}$} & 16 & 0.016 \\
& & & 150 & 0.12 \\
 & \multirow{2}{*}{60} & \multirow{2}{*}{$8.37 \times 10^{-13}$} & 16 & 0.046 \\
& & & 150 & 0.33 \\
\hline\hline
\end{tabular}
\caption{
Net per-$K_L$ probability $P$ and expected $K_L$ background yield $N_B$ after the full MD$\times$FD selection.
}
\label{tab:KL_final}
\end{table}

Overall, the MD$\times$FD strategy suppresses the $K_L$ background to ${\cal O}(1$--$10)$ events for the Small layout, and well below one event for the Big layout at the same integrated luminosities.
A full detector simulation will refine these numbers, but the conclusion is robust: the combined requirements of only one signal-like isolated photon in the main detector and FD-side displaced $\gamma\gamma$ activity, matched at the event level, render the $K_L$ contribution negligible, especially for the Big layout.

\paragraph*{Validation and dominant uncertainties.}
Most inputs to $N_B$ can be validated or constrained with data control samples and sidebands: the MD single-photon selection and kinematic windows can be calibrated in $Z\to q\bar q$ data using isolation and recoil-mass sidebands, while the MD--FD consistency can be benchmarked with geometry- and kinematics-based studies.
The $K_L$ rate relevant for this estimate is not the inclusive neutral-kaon multiplicity, but the rare ``quiet'' topology that survives the MD veto; it can be cross-checked with dedicated $Z\to$ hadrons control samples passing the same MD preselection.
The dominant uncertainties are expected to come from the effective absorption lengths (material modeling), the modeling of veto-surviving hadronization topologies, and the detailed cavern/acceptance geometry (notably $D_2(\theta,\varphi)$ and azimuthal holes); these will be refined with full simulation and in-situ calibrations.

\section{Sensitivity benchmarks for different signal-yield thresholds}
\label{appendix:NS_main_detector}

The numerical results given in Sec.~\ref{sec:results} show the already excellent sensitivity reaches of the CEPC/FCC-ee's main detectors; however, this is obtained under the assumption of zero background.
This may not be the case in reality and there can be multiple background events depending on the final-state signatures.
In order to illustrate the effect of non-vanishing background events on the sensitivity reach of the main detectors, we show here sensitivity curves of not only 3, but also 20 and 50 signal events corresponding to 95\% C.L.~exclusion bounds for 100 and 625 background events, respectively.
For illustrative purpose, we take the light scalar and the ALP models as examples.

\begin{figure}[t]
\centering
\includegraphics[width=\columnwidth]{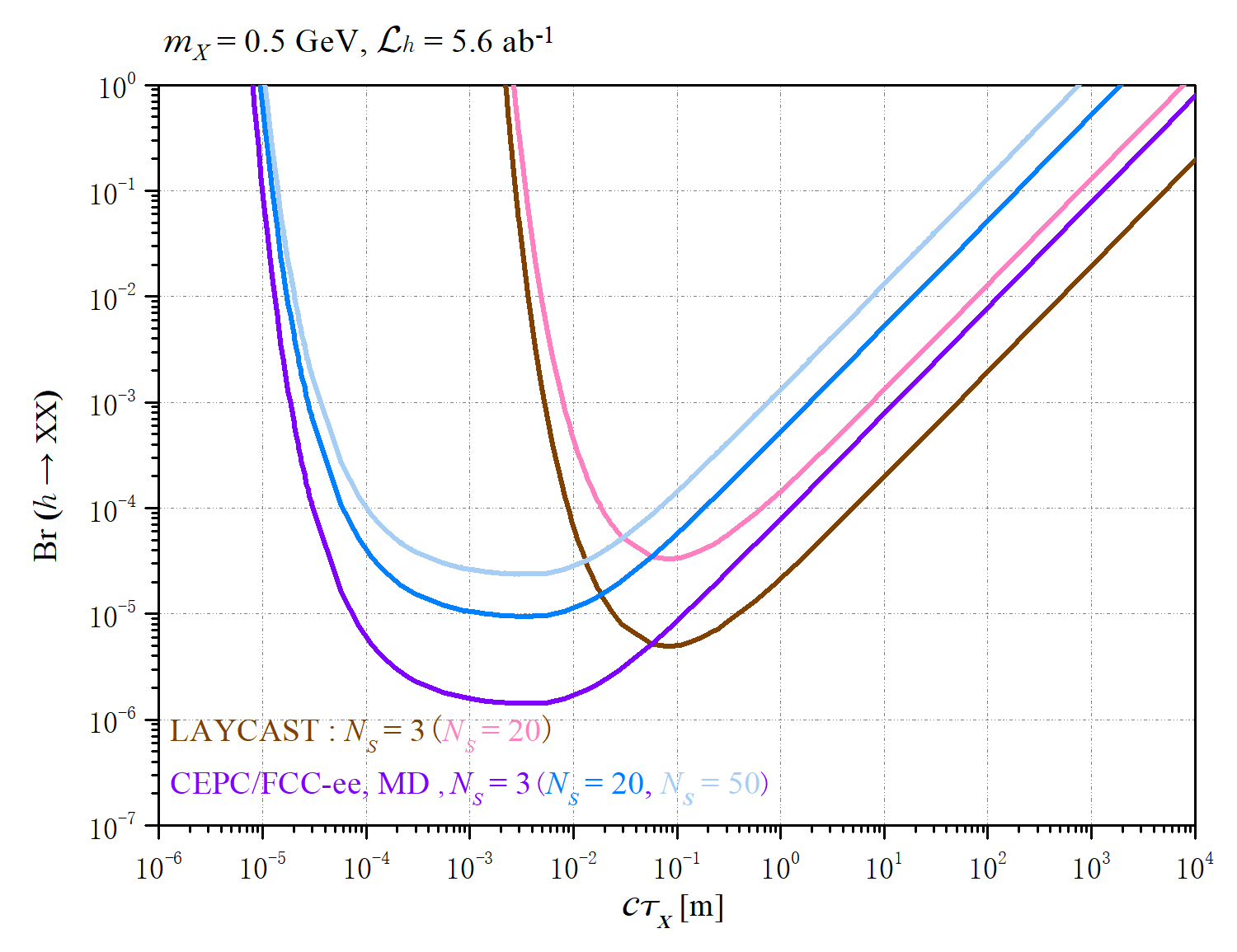}
\includegraphics[width=\columnwidth]{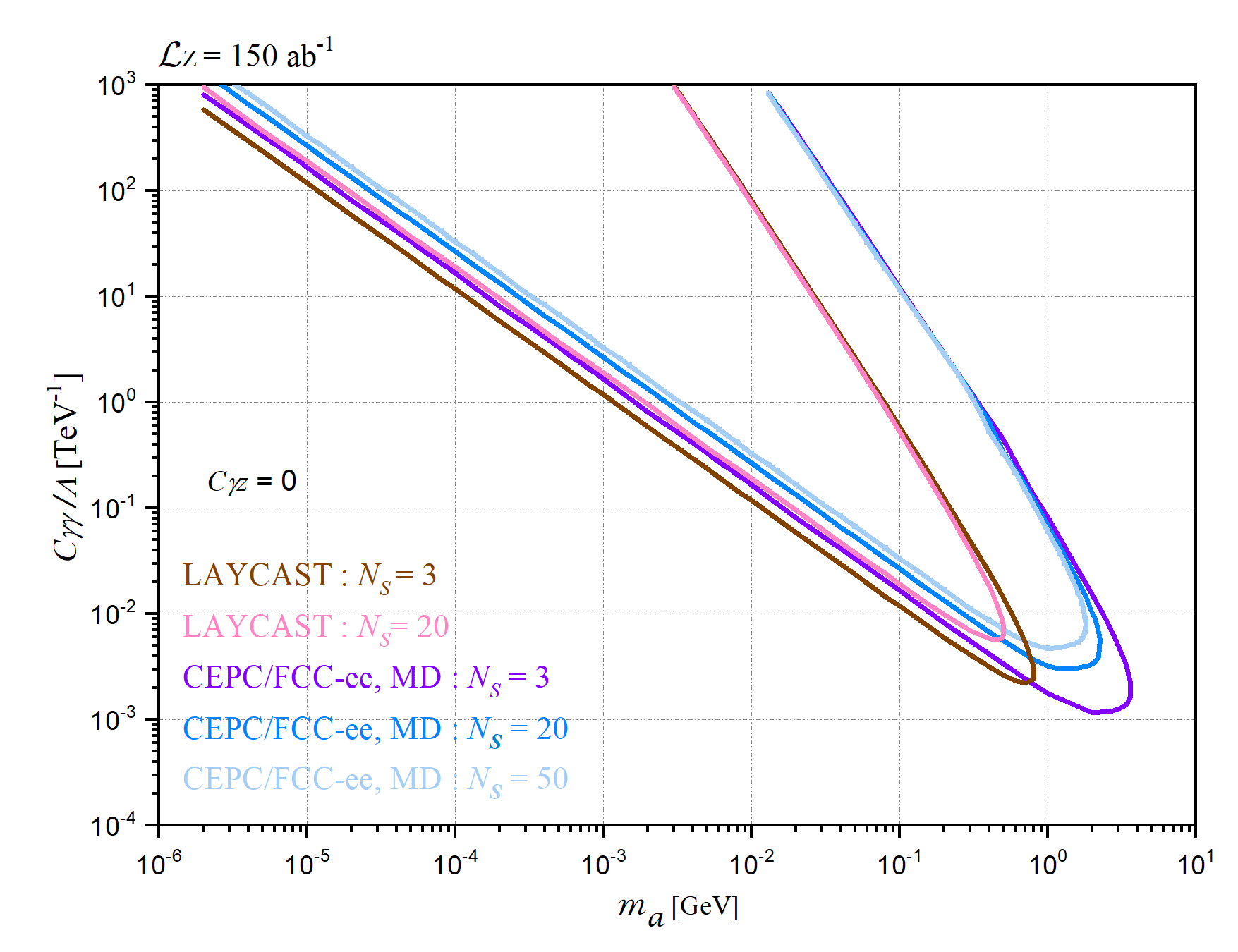}
\caption{\textit{Upper panel}: comparison of the sensitivity reaches of CEPC/FCC-ee's main detector with those of LAYCAST, for different number of light-scalar signal events, shown in the Br($h \rightarrow X X$) vs.~$c\tau_X$ plane for $m_X = 0.5$ GeV.
\textit{Lower panel}: the same comparison but for the ALP scenario with $\mathcal{L}_Z$ = 150 ab$^{-1}$ and $C_{\gamma Z}=0$, in the $C_{\gamma\gamma}/\Lambda$ vs. $m_a$ plane.}
\label{fig:main_detector_NS_compare}
\end{figure}

In the upper plot of Fig.~\ref{fig:main_detector_NS_compare}, we show the comparison of the sensitivity reaches of CEPC/FCC-ee's main detector with that of LAYCAST, for the long-lived light scalar $X$ with $m_X=0.5$ GeV, considering various background levels.
In the prompt regime, we observe that the main detector always shows much better performance, because its fiducial volume cuts on the minimal radial distance according to the $\mathcal{O}(\text{mm})$ beam-pipe radius~\cite{Wang:2019xvx}.
However, with larger samples of background events for the main detector, we find that LAYCAST can be much more sensitive in the large $c\tau_X$ regime.
This leads us to solidify the conclusion that LAYCAST can be complementary to the main detectors to a large extent in the optimal case.

Then in the lower panel of Fig.~\ref{fig:main_detector_NS_compare} we make the same comparison but for the long-lived ALP model with $C_{\gamma Z}$ set to 0, now given in the $C_{\gamma\gamma}/\Lambda$ vs.~$m_a$ plane.
We find that when background events are present for the main detector, the corresponding discovery potentials are weakened and thus LAYCAST shows some advantage in probing $C_{\gamma \gamma}/\Lambda$ for $m_a \lesssim 1$ GeV.
Consistent with the findings in the upper plot of Fig.~\ref{fig:main_detector_NS_compare}, we conclude that in the presence of background events for the CEPC/FCC-ee's main detector, the LAYCAST experiment can probe new parameter space for larger $c\tau$ values.

\begin{acknowledgments}
We thank Marco Drewes, Jan Hajer, Manqi Ruan, and Xiaolong Wang for useful discussions.
Y.L. and K.W. are supported by the National Natural Science Foundation of China under grant no.~11905162, the Excellent Young Talents Program of the Wuhan University of Technology under grant no.~40122102, and the research program of the Wuhan University of Technology under grant no. 3120625397 and 2020IB024.
Y.N.M.~is supported by the National Natural Science Foundation of China under grant no.~12205227 and the Fundamental Research Funds for the Central Universities (WUT: 2022IVA052).
The simulation and analysis work of this paper was completed with the computational cluster provided by the Theoretical Physics Group at the Department of Physics, School of Sciences, Wuhan University of Technology.
Z.S.W.~was supported by the National Natural Science Foundation of China under grant Nos.~12475106 and~12505120, and the Fundamental Research Funds for the Central Universities under Grant No.~JZ2025HGTG0252.
\end{acknowledgments}

\bigskip
\noindent
\textbf{Note added.}
Following standard practice in high-energy physics, authors are listed in strict alphabetical order by surname. 
This ordering should not be interpreted as indicating any ranking of contribution, seniority, or leadership.
All authors contributed equally to this work.

\bibliography{Refs}

\bibliographystyle{h-physrev5}

\end{document}